\documentclass[a4paper,11pt]{article}
\usepackage[skins]{tcolorbox}
\usepackage{mathtools,slashed}
\usepackage[T1]{fontenc}
\usepackage{slashed,verbatim,subfigure}
\usepackage[numbers,sort&compress]{natbib}
\usepackage{amsmath}
\usepackage{mathrsfs}
\usepackage{amsbsy}
\usepackage{amssymb}
\usepackage{appendix}
\usepackage{graphicx}
\usepackage[final]{pdfpages}
\usepackage{dcolumn}
\usepackage{bm}
\usepackage[colorlinks=true,linktocpage=true,
linkcolor=blue,citecolor=blue]{hyperref}
\usepackage{url}
\definecolor{db5}{cmyk}{0.5,0.5,0,0.5}
\definecolor{mauve}{cmyk}{0.3,0.7,0.1,0.3}
\definecolor{palemauve}{cmyk}{0.3,0.7,0.1,0.0}
\definecolor{pb}{cmyk}{0.4,0.1,0,0.1}
\definecolor{pgreen}{cmyk}{0.4,0.0,0.3,0.0}
\definecolor{pink}{cmyk}{0.0,0.5,0.3,0.0}
\newcommand{\bs}{\begin{slide}}
\newcommand{\es}{\end{slide}}
\newcommand{\tcb}{\textcolor {blue}}
\usepackage[a4paper]{geometry}
\catcode`@11
\def\seceqaa{\@addtoreset{equation}{section}
	\def\theequation{A\arabic{equation}}}
\def\seceqbb{\@addtoreset{equation}{section}
	\def\theequation{B\arabic{equation}}}
\def\seceqcc{\@addtoreset{equation}{section}
	\def\theequation{C\arabic{equation}}}
\def\seceqdd{\@addtoreset{equation}{section}
	\def\theequation{D\arabic{equation}}}
\def\seceqee{\@addtoreset{equation}{section}
	\def\theequation{E\arabic{equation}}}
\catcode`@11
\newcommand{\be}{\begin{eqnarray}}
\newcommand{\ee}{\end{eqnarray}}
\begin{document}
\large
\title{Bulk Viscosity, Speed of Sound and Contact Structure at Intermediate Coupling}
\author{Shivam Singh Kushwah\footnote{email-shivams\_kushwah@ph.iitr.ac.in}~ and ~Aalok Misra\footnote{email- aalok.misra@ph.iitr.ac.in}\vspace{0.1in}\\
Department of Physics,\\
Indian Institute of Technology Roorkee, Roorkee 247667, Uttarakhand, India}
\date{}
\maketitle
\begin{abstract}
Thermal QCD systems like strongly coupled QGP require not only a large 't-Hooft coupling but also a finite gauge coupling \cite{Natsuume:2007qq} in the large-$N$ gravity dual. Unlike almost all top-down holographic models in the literature, holographic large-$N$ thermal QCD models based on this assumption, therefore necessarily require addressing this limit from ${\cal M}$ theory. In the context of strongly coupled (very large `t-Hooft coupling) thermal QCD-like theories, the bulk viscosity($\zeta$)-to-shear viscosity($\eta$) ratio using the type IIA-theory dual of thermal QCD-like theories was shown to vary like $\frac{1}{3} - c_s^2$ \cite{IITR-McGill-bulk-viscosity} ($c_s$ being the speed of sound), and the same ($\frac{\zeta}{\eta}$) at weak coupling using kinetic theory and finite temperature field theory, was shown and is known to vary like $\left(\frac{1}{3} - c_s^2\right)^2$ (\cite{IITR-McGill-bulk-viscosity} and references therein). The novelties of the results of this paper are that we not only show for the first time from ${\cal M}$ theory that at intermediate coupling, with the inclusion of the ${\cal O}(R^4)$ corrections, the result obtained for $\frac{\zeta}{\eta}$ interpolates between the strong and weak coupling results in a way consistent with lattice computations of $SU(3)$ Gluodynamics \cite{lattice-SU3_Glue} within statistical errors (and within the temperature range permissible by our ${\cal M}$-theory uplift), but also observe that this behavior is related to the existence of Contact 3-Structures that exist only at intermediate coupling effected by the intermediate-$N$ "MQGP" limit of \cite{ACMS}. We also obtain an explicit dependence of $\frac{\zeta}{\eta}$ on (fractional powers of) the temperature-dependent/running gauge coupling (and its temperature derivative), and verify that the weak-coupling result dominates at large temperatures. We further conjecture that the aforementioned fractional-power-dependence of $\frac{\zeta}{\eta}$ on the gauge coupling is related to the lack of ``$N$-connectedness'' in the parameter space of Contact 3-Structures (as shown in \cite{ACMS}). 
\end{abstract}
\newpage
\tableofcontents

\section{Introduction}
QCD as a theory of strong interaction has a very interesting phase diagram. Below the deconfinement temperature $T<T_c$ -  known as the confining/hadronic phase where the degrees of freedom are hadrons. Above the deconfinement temperature, $T>T_c$, one has the Quark Gluon plasma phase where the degrees of freedom are quarks and gluons. Very recently in Relativistic Heavy Ion Collision (RHIC)  experiments, it was revealed that QGP behaves like an ideal fluid \cite{STAR:2005gfr} - in fact, the most ideal fluid found in any experiments so far, and hence often called as the most perfect fluid of nature. These findings make the hydrodynamic exploration of Quark Gluon Plasma (QGP) to be an obviously interesting enterprise.

Hydrodynamics explores the long wavelength or short frequency behavior of the fluid. The gauge/gravity duality \cite{Maldacena:1997re,Gubser:1998bc,Witten:1998qj,Polchinski:2000uf,Aharony:1999ti} plays a central role in providing the possible tools to compute the transport coefficients for the strongly coupled large-$N$ gauge theory plasmas via its gravity dual containing a black hole; it is well known that transport coefficients can be expressed as the zero-energy and long-wavelength limit of appropriate correlation functions \cite{Kovtun:2003wp, Policastro:2001yc,Policastro:2002tn,Herzog:2002fn,Herzog:2003ke,Policastro:2002se}. The definitions of bulk viscosity and shear viscosity are given by the variation of the stress-energy tensor of the fluid. The variation of stress-energy tensor from equilibrium can be given as (considering only the spatial part)
\begin{equation}
\delta T^{ij}=\eta \Biggl(\partial^{i} u^{j} +\partial^{j} u^{i}-\frac{2}{3}g^{ij} \partial\cdot u \Biggr)+\zeta g^{ij} \partial\cdot u,
\end{equation}
where $\eta\equiv$ Shear Viscosity, $\zeta\equiv$ Bulk Viscosity and $u^i$ is the fluid velocity.

The trace of the stress-energy tensor carries the information about the pressure. In the rest frame, the fluid pressure, ${\cal{P}}$, and the speed of the sound mode, $c_{s}$, is given by: .
\begin{eqnarray}
{\cal{P}}=-\frac{1}{3}T^{i}_{i}, \quad\quad\quad c_s^2=\frac{\partial\cal{P}}{\partial\epsilon}.
\end{eqnarray}
 In the past couple of decades, some very interesting results were found towards demystifying the behavior of QGP via a top-down holographic approach, such as the ratio of shear viscosity to entropy density follows a lower bound proposed by Kovtun, Son, and Starinets i.e., $\frac{\eta}{s} \geq \frac{1}{4\pi}$\cite{Kovtun:2004de} which is satisfied by a large class of physical systems in nature. The QGP produced in RHIC experiments at the Brookhaven National Laboratory follows the aforementioned shear viscosity bound up to multiplicative numerical factors \cite{Romatschke:2007mq}. The bulk viscosity is an essential feature of non-conformal theories. For conformal theories the bulk viscosity $\zeta=0$, and for non-conformal theories $\zeta \neq 0$. In \citep{Yarom:2009mw} a general formula for the bulk viscosity is derived for softly broken conformal symmetry. In \cite{Buchel:2007mf} another bound referred to as the Buchel bound about which we are concerned here, is also explored. This bound says that the ratio of the bulk viscosity ($\zeta$) to shear viscosity($\eta$)  has a lower bound related to the speed of sound mode ($c_s$). This is a dynamical bound since the speed of sound and the ratio of bulk viscosity to shear viscosity vary with temperature, but the bound is expected to be followed at arbitrary temperature by all gauge theories that possess a gravitational dual. Several backgrounds supporting this bound have been proposed in recent years  \cite{Mas:2007ng,Buchel:2008uu,Springer:2008js,Kanitscheider:2009as,David:2009np}. The bound is saturated in a number of cases, including the compactification of the $p$-dimensional gauge theories on a $k$-dimensional torus \cite{Benincasa:2006ei}, the case of Little String Theory \cite{Parnachev:2005hh}, and the $p+1$-dimensional gauge theory plasma holographically dual to a stack of near-extremal D$3$-branes \cite{Mas:2007ng}.
 
 To the best of our knowledge, \cite{MQGP} is the only  holographic ${\cal M}$-Theory dual of realistic thermal QCD-like theories that is able to:
\begin{itemize}
\item
 yield a deconfinement temperature\footnote{The very high temperature  at and above which the quarks become free, and below which they are ``'confined''.} $T_c$ from a Hawking-Page phase transition \cite{Trace anomaly_AM+CG} (also see \cite{Vikas+Gopal+Aalok} for non-renormalization of the same up to ${\cal O}(R^4)$ in the ${\cal M}$-theory dual); for completeness and the relevance of $T_c$ to this paper, this is summarized in Appendix \ref{Tc-M-theory}

\item      
yield a conformal anomaly variation with temperature compatible with lattice results at  high ($T>T_c$) {\it and} low ($T<T_c$) temperatures \cite{Trace anomaly_AM+CG}

\item
Condensed Matter Physics: inclusive of  the non-conformal corrections, obtain
\begin{enumerate}
\item
 a lattice-compatible shear-viscosity-to-entropy-density ratio (first reference in \cite{bulk+gauge_KS+AM}) {\it compared well with, i.e., within the error bars of QGP-related RHIC data as given in the third reference of \cite{bulk+gauge_KS+AM} for 
 $T\in[T_c, 1.6 T_c]$}.

\item
the temperature variation of a variety of transport coefficients including the bulk-viscosity-to-shear-viscosity ratio,  diffusion coefficient, speed of sound, electrical and thermal conductivity and the Wiedemann-Franz law  \cite{IITR-McGill-bulk-viscosity}, \cite{bulk+gauge_KS+AM};
\end{enumerate}

\item
Particle Phenomenology:  obtain
\begin{enumerate}
\item
 lattice-compatible glueball spectroscopy \cite{Glueball-Roorkee}

\item
Particle Data Group(PDG)-compatible meson spectroscopy (first reference of \cite{Yadav+Misra+Sil-Mesons})\footnote{As noted in the same, we obtain $0^{--}$ (\`{a} la $J^{PC}$-assignment) pseudo-scalar mesons that thus far have not been found in the Particle Data Group (PDG) data further justifying the use of QCD-like theories in our work.}

\item
PDG-compatible glueball-to-meson decay widths (\cite{VA-Glueball-decay})

\item
the values of phenomenology-compatible coupling constants of the NNLO (in a chiral expansion) $\chi$PT Lagrangian in the chiral limit involving the NGBs and $\rho$ meson (and its flavor partners) from the ${\cal M}$-theory /type IIA dual of large-N thermal QCD, inclusive of the ${\cal O}(R^4)$ correction \cite{Vikas+Gopal+Aalok} \footnote{It was also shown that 
(i) ensuring compatibility with phenomenological/lattice results requires a relationship relating the ${\cal O}(R^4)$ corrections and large-$N$ suppression, (ii) at the ${\cal O}(R^4)$ corrections in the UV to the ${\cal M}$-theory uplift of the type IIB dual of large-$N$ thermal QCD-like theories at low temperatures, can be consistently set to be vanishingly small. }

\item
QCD-compatible supermassive inert mesinos \cite{Aalok+Gopal-Mesino} thereby resolving a longstanding problem with the Sakai-Sugimoto type IIA top-down holographic dual \cite{SS} of large-$N$ QCD-like theories of meson-mesino isospectrality and unsuppressed mesino-meson interaction.  

\end{enumerate}

\item
Mathematics:  provide, for the first time,  an $SU(3)$-structure (for type IIB (second reference of \cite{bulk+gauge_KS+AM})/IIA \cite{NPB}, \cite{OR4}   holographic dual), $G_2$-structure  \cite{NPB}, \cite{OR4} and the resultant (Almost) Contact 3-Structures \cite{ACMS}, and ${\rm Spin}(7)$ torsion classes \cite{OR4} of the six-, seven- and eight-folds in the UV-IR interpolating region/UV, relevant to type II string/${\cal M}$-Theory holographic duals of thermal QCD-like theories at high temperatures.

\end{itemize}
 
Recent endeavors in hydrodynamics of ${\cal M}$-theoretic dual of large-$N$ thermal QCD-like theories above deconfinement temperature, $T>T_c$, in the strong and weak coupling limits, have been explored. The results of these calculations for the compactified ${\cal M}$-theoretic eleven dimensional background to five-dimensional non-compact background ($M_5=(S^1 \times \mathbb{R}^3)\times \mathbb{R}_{>0}$) are:
\begin{itemize}
\item In the strong coupling limit, the lower bound on the bulk-to-shear-viscosity ratio $\frac{\zeta}{\eta}$ was found tobe proportional to $(\frac{1}{3}-c_s^2)$ \cite{IITR-McGill-bulk-viscosity,Czajka:2018egm}.
\item In the weak coupling limit, the lower bound on $\frac{\zeta}{\eta}$ was found to be proportional to $(\frac{1}{3}-c_s^2)^2$ \cite{Czajka:2018egm}, mimicking QCD \cite{Moore:2008ws}.
\end{itemize}   

Kubo's formulae are relations used to express the linear response of a quantity  to time-dependent perturbation.  They utilize the two-point correlation function of the stress-energy tensor to derive the desired expression for the transport coefficients such as bulk viscosity, shear viscosity, etc. We can also obtain susceptibilities  using Kubo's relation \cite{Kovtun:2018dvd}. Kubo's formula for bulk viscosity can be written as:
\begin{equation}
\zeta=\frac{1}{2} \lim_{\omega\rightarrow 0} \lim_{k\rightarrow 0} \frac{\rho_{PP}(\omega,k)}{\omega}
\end{equation}
where $\rho_{PP}\equiv$ spectral function of the pressure-pressure correlator where $\omega\equiv$ frequency of the hydrodynamic mode. Here the limits for $\omega$ and $k$ are the hydrodynamic limit for extracting the short frequency 
$(\omega)$, or long wavelength $(\lambda$, with $k=\frac{2\pi}{\lambda}$) behavior of the system. The spectral function is related to the imaginary part of the pressure-pressure retarded correlation function,
\begin{equation*}
\rho_{PP}= 2 ImG_{R}^{PP}(\omega).
\end{equation*}

\subsection*{Organization of the paper}
In Section \ref{review}, on one hand in \ref{M-theory-uplift}, we take a brief tour of the $\cal M$-theory uplift inclusive of $\mathcal{O}(R^4)$ corrections of type IIB string theory dual to thermal QCD-like theories on the basis of literature developed in \citep{metrics,MQGP,NPB,OR4}. On the other hand, in \ref{ACM3S-basics}, we review some relevant results of \cite{ACMS} pertaining to (Almost) Contact 3-Structures arising from $G_2$-structures in the aforementioned ${\cal M}$-theory uplift.

Section \ref{cs-Zs} has to do with obtaining the speed of sound from the EOM for a gauge-invariant combination of scalar modes of metric perturbations. We consider the bosonic part of eleven-dimensional supergravity action inclusive of the $\mathcal{O}(R^{4})$ corrections \cite{OR4}.  Utilizing the metric perturbations of the form (\ref{metric-perturbations}), we can always decompose the equation of motion into the parts which do not contain the $\mathcal{O}(\beta\equiv l_p^6)$ corrections, i.e., $EOM^{\beta^{0}}_{MN}$, and the part which receives $\mathcal{O}(\beta)$ corrections, i.e., $EOM^{\beta}_{MN}$. We write the metric of the non-compact part in the ingoing Eddington-Finkelstein coordinates. Then in the EOM for the gauge-invariant combination of scalar modes of the metric perturbations, we find that $r_h$ turns out to be an irregular singular point. Requiring the horizon to be a regular singular point of the EOMs for the  scalar modes of metric perturbations (\ref{EOMs-LON-LOlogrh}), we find the expression relating the deviation of the square of the speed of sound from its conformal value, $\Biggl(\frac{1}{3}-c_{s}^{2} \Biggr)^{1\ {\rm and}\ 2}$ in terms  of $g_s$ (string coupling), $M$ (the number of fractional $D3$-branes in the type IIB dual of \cite{metrics}), $N$ (the number of color $D3$-branes in the type IIB dual of \cite{metrics}),  $N_f$ (the number of flavor $D7$-branes in the same type IIB dual of \cite{metrics}), and $r_h$ (black-hole horizon radius).

In Section \ref{spectral}, having shown the coupling of metric and gauge perturbations, we consider the DBI action for the $D6$-branes having turned on a $U(1)$(-subgroup of $U(N_f)$) gauge field on the $N_f\ D6$-branes' world volume. Coupling the $U(1)$-gauge field to the pulled-back metric and the pulled-back NS-NS $B$-field, and solving the EOM for the $U(1)$ gauge field  $A_{t}(Z)$ up to a contribution of $\cal{O}(\beta)$ in the UV and IR. Considering only fluctuations in the gauge field $A_{\mu}$ here $\mu={t,x_1,x_2,x_3}$ in the $A_Z=0$-gauge, using the DBI action we obtain the EOM for gauge field fluctuations. Via the gauge-invariant variables, $E_{x_{1}},E_{x_{2}}=E_{x_{3}}=E_{T}$, the EOM corresponding to the fluctuations in the direction $Z, t, x_{1}$ can be converted into a single EOM, and for $x_2$ or $x_{3}$ in terms of new variable $E_{T}$. Since in the limit $q\rightarrow 0$, $E_{T}$ and $E_{x_1}$ are same, and hence a single Schr\"{o}dinger-like equation is obtained for all $E_{\mu}$ with $\mu = t, x_{1}, x_{2}, x_{3}$. Writing the on-shell DBI action in terms of the gauge invariant variables $E_{\mu}$'s, we work out the retarded Green function $G_{x_{1}x_{1}}$ in the zero-momentum limit i.e, $q\rightarrow 0$, the imaginary part of which provides the spectral function $ \mathcal R_{x_{1} x_{1}} $. Using Kubo's formula for the bulk viscosity in terms of the retarded Green function for EM-tensor two-point correlation function, and the relationship between the same and the retarded Green function for current-current two-point correlation function arising from the aforementioned connection between metric and gauge fluctuations, the abovementioned Green/spectral function involving $E_T$ suggests that the bulk viscosity at intermediate coupling is the linear combination of the results at weak (i.e.,$~\left(\frac{1}{3}-c_s^2\right)^2$) and strong  (i.e.,$~\left(\frac{1}{3}-c_s^2\right)$) coupling.

In Section \ref{Eling+Oz}, we use the formalism provided by Eling and Oz\cite{EO} where they derived a generic result for bulk-to-shear viscosity ratio for general backgrounds containing the scalar fields say $\Phi_i$'s, and non-abelian gauge fields in gauge stress energy tensor in $d+1$-dim gravitational action in Einstein frame. In \cite{EO} the formula for the bulk-to-shear viscosity ratio is obtained by establishing the focusing (or Raychaudhuri) equation for the background which contains the holographic dual to thermal states in strongly coupled $d$-dimensional theory: 
\begin{equation}
\label{EO-1}
\frac{\zeta}{\eta} = c_s^4 T^2\sum_i \left.\left(\frac{d\Phi^i}{dT}\right)^2\right|_{r=r_h},
\end{equation}
where,
$\eta\equiv$ shear viscosity, $s\equiv$ entropy density, and  $\Phi_i\equiv$ Scalar fields.  We use (\ref{EO-1}) to derive the relation between the $\frac{\zeta}{\eta}$ ratio and $g_s$, $M$, $N_f$, $N$, and $r_h$. Consequently, we obtain the desired result for the bulk-to-shear viscosity ratio for the theories at intermediate 't-Hooft coupling and find the same to be a linear combination of the results in the strong and weak coupling limits. We obtained a similar result by using $\frac{\zeta}{\eta} = \left(s \frac{d\Phi(r = r_h)}{ds}\right)^2$ \cite{EO}, $s$  being the entropy density and $\Phi$ being a scalar which we take to be the most dominant (in large-$N$ limit) component of the ${\cal M}$-theory potential $C_{\theta_1\theta_2 x^{10}}$.

Section \ref{Summary} has a summary of the results obtained in this paper in \ref{summary} and a discussion on the significance of the results obtained in \ref{significance}. This also includes the relevance of Contact 3-Structures obtained in the intermediate-$N$ MQGP limit (\ref{MQGP_limit}) to obtaining the bulk-to-shear-viscosity ratio at intermediate coupling.

There are four supplementary appendices. Appendix \ref{review-(A)C3S} is on a brief review of (Almost) Contact 3-Structures. Appendix \ref{Tc-M-theory} is a review of our computation of $T_c$ from ${\cal M}$-theory, both, from a semiclassical computation involving a thermal and black-hole backgrounds as the pair of saddle points relevant respectively to $T<T_c$ and $T>T_c$, as well as from the point of view of entanglement entropy involving a spatial interval and its complement. Appendix \ref{details of Sec 5} provides details of the computations of Sec. \ref{Eling+Oz}, in particular the regularization of the (polar) angular integrals that are required to be evaluated when computing $\frac{\zeta}{\eta}$ using the formalism of \cite{EO}.  The coframes $e^{2, 3, 4}\in e^{a=1,...,7}$ diagonalizing $M_7=S^1_{\cal M}\times_w\left(S^1_{\rm thermal}\times_w M_5\right)$, $M_5$ being a non-Einsteinian generalization of $T^{1,1}$, near $\psi=2n\pi, n=0, 1, 2$-coordinate patches, are given, for this paper to be self-contained, in appendix \ref{five-fold-coframes}.
 
\section{A brief review of $\cal{M}$ theory uplift inclusive of ${\cal O}(l_{p}^6$ Corrections, and (Almost) Contact 3-Structures}
\label{review}

\subsection{${\cal M}$-theory Uplift of Thermal QCD-Like Theories Inclusive of ${\cal O}(R^4)$ corrections}
\label{M-theory-uplift}

Here we will give a brief review of the ${\cal M}$-theory uplift (including the ${\cal O}(R^4)$ corrections) of type IIB string theory dual of thermal QCD-like theories. The thermal QCD like theories refer to the equivalence class of theories that are UV conformal, IR confining, and have bi-fundamental quarks\footnote{The quarks transforming under the fundamental representation of color and flavor group are known as the bi-fundamental quark.}. The UV-complete type IIB string dual dual to these large $N$-thermal QCD-like theories was constructed in \cite{metrics}. The $\mathcal{M}$-theory uplift of this set-up without $\mathcal{O}(R^{4})$ terms are in  \cite{MQGP} and with inclusion of $\mathcal{O}(R^{4})$ terms are constructed in \cite{OR4}. We will take a brief tour of this construction here.
\begin{itemize}
\item {\bf Type IIB Brane scenario of \cite{metrics}}: The brane scenario contains:
\begin{itemize}
\item $N$ $D3$-branes filling the space-time located at the tip of the warped resolved conifold.
\item  At the same tip of warped resolved conifold, there are $M$ spacetime-filling $D5$ branes are wrapping the vanishing ``squashed'' $S^2$, that are in addition, located at the North Pole (NP) of the resolved squashed $S^2$ of the radius $a$ (resolution parameter).  There are also spacetime-filling $\overline{D5}$-branes at the tip of the same warped resolved conifold wrapping the earlier mentioned vanishing squashed $S^2(\theta_1,\phi_1)$, and located at the South Pole (SP) of the resolved squashed $S^2(\theta_2,\phi_2)$.
\item  Spacetime-filling flavor $N_f$ $D7$-branes surround the vanishing squashed $S^3(\theta_1,\phi_1,\psi)$ and are situated at the North Pole of the squashed resolved $S^2(\theta_2,\phi_2)$. They continue into the IR to the extent $|\mu_{\rm Ouyang}|^{\frac{2}{3}}$, where $|\mu_{\rm Ouyang}|$ represents the modulus of the Ouyang embedding corresponding with the Ouyang embedding related to the flavour $D7$-branes:
\begin{equation}
\label{Ouyang-definition}
\left(9 a^2 r^4 + r^6\right)^{1/4}e^{\frac{i}{2}(\psi - \phi_1-\phi_2)}\sin\frac{\theta_1}{2} \sin\frac{\theta_2}{2}=\mu_{\rm Ouyang}.
\end{equation}
The same number of flavor $\overline{D7}$-branes is located at the South Pole of the blown-up $S^2(\theta_2,\phi_2)$ wrapping the vanishing squashed $S^3(\theta_1,\phi_1,\psi)$; the same number of respetively $D5/D7$-branes and $\overline{D5}/\overline{D7}$-branes,  ensure the UV conformal behaviour of the gauge/string coupling.

\item The $N_f$ flavor $D7$ and $\overline{D7}$-branes correspond to the flavor gauge group $SU(N_f) \times SU(N_f)$ in the UV. Since there are no  $\overline{D7}$-branes in the IR region then the symmetry group $SU(N_f) \times SU(N_f)$ in the UV breaks down to $SU(N_f)$ in the IR region. This is equivalent to the chiral symmetry breaking.

\item
Let us elaborate on the choice of $M$ and and $N_f$. As explained in \cite{Trace anomaly_AM+CG}, in the IR, at the end of a Seiberg-like duality cascade,  the number of colors $N_c$ gets identified with $M$, which in the `MQGP limit' (\ref{MQGP_limit})  can be tuned to equal 3. This is briefly summarized here. One can identify $N_c$ with the sum of effective number $N_{\rm eff}$ of $D3$-branes and the effective number of $D5$-branes $M_{\rm eff}$: $N_c = N_{\rm eff}(r) + M_{\rm eff}(r)$.  As $N_{\rm eff}$  varies between $N\gg1$ in the UV and 0 (due to ensure conformal invariance) in the deep IR, and  $M_{\rm eff}$  varies between 0 in the UV and $M$ in the deep IR, $N_c$ varies between $M$ in the deep IR and a large value [in the MQGP limit of (\ref{MQGP_limit}) for a large value of $N$] in the UV.  Therefore, at very low energies, the number of colors $N_c$ can be approximated by $M$, which in the MQGP limit is  finite and can hence be taken to be equal to three. Similarly, the effective number of flavor $D7$-branes $N_f^{\rm eff}$, varies from 0 in the UV (once again to ensure conformal invariance) and $N_f$, which one can set as $N_f = 2(u/d) + 1(s) = 3$ in the MQGP limit of (\ref{MQGP_limit}), in the IR. This is hence summarized in Table \tcb{1} below.
\begin{table}[h]
\label{Parameters-real-QCD}
\begin{center}
\begin{tabular}{|c|c|c|c|} \hline
S. No. & Parameterc & Value chosen consistent with (\ref{MQGP_limit}) & Physics reason \\ \hline
1. & $g_s$ & 0.1 & QCD fine structure constant \\ 
&&& at EW scale \\ \hline
2. & $M$ & 3 & Number of colors after a \\ 
&&& Seiberg-like duality cascade \\
&&& to match real QCD \\ \hline
3. & $N_f$ & 2 or 3 & Number of light quarks in real QCD \\ \hline
\end{tabular}
\end{center}
\caption{QCD-inspired values of $g_s, N_f, M$}
\end{table}

\end{itemize} 
\item {\bf Type IIB Bulk picture of \cite{metrics}}: Here we give a brief tour of the dual bulk picture of the type IIB set-up of \cite{metrics} for thermal QCD like theories:
\begin{itemize}
\item Like the Klebanov-Strassler model, in the gravity dual, the IR confinement is effected via the deformation of the vanishing squashed $S^3$ in the conifold. The high-temperature (i.e., $T>T_c$) regime of the QCD-like theories corresponds to black hole background in the gravity dual.
\item Due to the finite temperature and finite separation between $D5$ and $\overline{D5}$-branes on the brane side, the conifold also requires a $S^2$-blow-up/resolution (with radius/resolution parameter $a$).
\item As a consequence, we conclude that the string dual associated with thermal QCD-like theories includes a warped resolved deformed conifold in the large-$N$ limit.
\item Furthermore, the ten-dimensional warp factor and fluxes take the back-reaction of the $D5$- and $D7$-branes into account.
\item There are several advantages of the type IIB dual of \cite{metrics} under the the intermediate-$N$ MQGP limit \footnote{The very-large-$N$ MQGP limit with $N\gg1, \frac{\left(g_s M^2\right)^{m_1}\left(g_s N_f\right)^{m_2}}{N}\ll1,\ m_{1,2}\in\mathbb{Z}^+\cup\{0\}$, was actually the MQGP limit that was originally suggested in \cite{MQGP}.}\cite{MQGP}, \cite{ACMS},
\begin{equation}
\label{MQGP_limit}
g_s\sim\frac{1}{{\cal O}(1)}, M, N_f \equiv {\cal O}(1), N>1,  \frac{\left(g_s M^2\right)^{m_1}\left(g_s N_f\right)^{m_2}}{N}<1,\ m_{1,2}\in\mathbb{Z}^+\cup\{0\},
\end{equation}
 as the Seiberg-like duality cascade comes to an end, the number of colors $N_c$ in the IR equals $M$. 
 \item To obtain  Contact (3-) structure(s) and  corresponding transverse $SU(3)$ transverse 3-structures, $N=100\pm{\cal O}(1)$ was explicitly picked out in \cite{ACMS}. Other set of values of $(g_s, M, N_f)$ will select another intermediate $N$.
\item The lightest quark flavours corresponds when addressing the embedding of the flavour $D7$-branes in the vanishing-Ouyang-modulus limit ($|\mu_{\rm Ouyang}|\ll1$ in (\ref{Ouyang-definition})), with $N_f =$ 2 or 3 \cite{Vikas+Gopal+Aalok}.

\item
From \cite{metrics} the form of ten-dimensional type IIB supergravity solution is as follows:
\begin{equation}
\label{metric}
ds^2 = \frac{1}{\sqrt{h}}
\left(-g_1 dt^2+dx_1^2+dx_2^2+dx_3^2\right)+\sqrt{h}\biggl[g_2^{-1}dr^2+r^2 d{\cal M}_5^2\biggr],
\end{equation}
  where 10D warp factor $h$ will be specified later, the black hole functions are denoted by $g_i$ can be given as:
$ g_{1,2}(r,\theta_1,\theta_2)= 1-\frac{r_h^4}{r^4} + {\cal O}\left(\frac{g_sM^2}{N}\right)$
with ${\cal O}\left(\frac{g_sM^2}{N}\right)$ corrections are the source of ($\theta_1, \theta_2$), where $r_h$ represents the black hole horizon.
The compact five-dimensional metric  found in (\ref{metric}) can be expressed as:
\begin{eqnarray}
\label{RWDC}
& & d{\cal M}_5^2 =  h_1 (d\psi + {\rm cos}~\theta_1~d\phi_1 + {\rm cos}~\theta_2~d\phi_2)^2 +
h_2 (d\theta_1^2 + {\rm sin}^2 \theta_1 ~d\phi_1^2) +   \nonumber\\
&&  + h_4 (h_3 d\theta_2^2 + {\rm sin}^2 \theta_2 ~d\phi_2^2) + h_5~{\rm cos}~\psi \left(d\theta_1 d\theta_2 -
{\rm sin}~\theta_1 {\rm sin}~\theta_2 d\phi_1 d\phi_2\right) + \nonumber\\
&&  + h_5 ~{\rm sin}~\psi \left({\rm sin}~\theta_1~d\theta_2 d\phi_1 +
{\rm sin}~\theta_2~d\theta_1 d\phi_2\right),
\end{eqnarray}
$r> a, h_5\sim\frac{({\rm deformation\ parameter})^2}{r^3}\ll \frac{a^2}{r^2} \forall r \gg({\rm deformation\ parameter})^{\frac{2}{3}}$.  The squasing factors $h_i$'s that occur in the compact five-dimensional metric (\ref{RWDC}) are:
\begin{eqnarray}
\label{h_i}
& & \hskip -0.45in h_1 = \frac{1}{9} + {\cal O}\left(\frac{g_sM^2}{N}\right),\  h_2 = \frac{1}{6} + {\cal O}\left(\frac{g_sM^2}{N}\right),\ h_4 = h_2 + \frac{a^2}{r^2},\nonumber\\
& & h_3 = 1 + {\cal O}\left(\frac{g_sM^2}{N}\right),\ h_5\neq0,\
\end{eqnarray}
The non-extremal resolved warped deformed conifold implied by equations (\ref{RWDC}) and (\ref{h_i}) includes a $S^2$-blowup (as $h_4 - h_2 = \frac{a^2}{r^2}$), a $S^3$-blowup (as $h_5\neq0$), and squashing of an $S^2$ (as $h_3$ is not specifically unity). In the deep infrared, we obtain a warped squashed $S^2(a)\times S^3(\epsilon)$, where $\epsilon$ is the deformation parameter. The horizon's shape is warped squashed $S^2\times_w S^3$. The ten-dimensional warp factor in the IR, which takes back-reaction into account, is provided by: 
\begin{eqnarray}
\label{h-def}
&& \hskip -0.45in h(r, \theta_{1,2}) =\frac{L^4}{r^4}\Bigg[1+\frac{3g_sM_{\rm eff}^2}{2\pi N}{\rm log}r\left\{1+\frac{3g_sN^{\rm eff}_f}{2\pi}\left({\rm
log}r+\frac{1}{2}\right)+\frac{g_sN^{\rm eff}_f}{4\pi}{\rm log}\left({\rm sin}\frac{\theta_1}{2}
{\rm sin}\frac{\theta_2}{2}\right)\right\}\Biggr],\nonumber\\
& & 
\end{eqnarray}
While $M_{\rm eff}/N_f^{\rm eff}$ does not always equal $M/N_f$, the two are the same in the IR. 
\end{itemize}


\item {\bf Color-Flavor Length Scale's Enhancement}: An Infra-Red (IR) color-flavor enhancement associated with the length scale in the MQGP limit (\ref{MQGP_limit}) with respect to a Planckian length scale in the Klebanov-Strassler(KS)'s model \cite{Klebanov:2000hb} even for ${\cal O}(1)$ $M$ is obtained by incorporating terms higher order in $g_s N_f$ in the RR and NS-NS three-form fluxes as given in \cite{metrics} and the next-to-leading order terms in $N$ within the metric (\ref{h_i}). This guarantees that quantum and stringy corrections are suppressed. This is partly discussed in \cite{NPB}. The following is a summary of the argument, which we also extend to the argue suppression of loop/quantum corrections. 

\begin{itemize}
\item {\bf Stringy/$\alpha^\prime$ correction's Suppression}: In the ten-dimensional warp factor (\ref{h-def}) near, e.g., (\ref{alpha_theta_12}),  the angular part can be ignored (as near, e.g., (\ref{alpha_theta_12}), $\left|\log\left(\sin\left(\frac{\theta_1}{2}\right)\sin\left(\frac{\theta_2}{2}\right)\right)\right|\sim\log N<|\log r|\sim N^{1/3}, r\in{\rm IR}$ \cite{IITR-McGill-bulk-viscosity}):
\begin{equation}
\label{h10d}
h = \frac{4\pi g_s}{r^4}\Biggl[N_{\rm eff}(r) + \frac{9 g_s M^2_{\rm eff} g_s N_f^{\rm eff}}{2\left(2\pi\right)^2}\log r \Biggr],
\end{equation}
where \cite{metrics}
\begin{eqnarray}
\label{NeffMeffNfeff}
& & N_{\rm eff}(r) = \int_{\rm resolved\ warped\ deformed\ conifold\ base\ \mathbb{M}_5}(F_5 + B_2\wedge F_3)\nonumber\\
& &  = N\left[ 1 + \frac{3 g_s M_{\rm eff}^2}{2\pi N}\left(\log r + \frac{3 g_s N_f^{\rm eff}}{2\pi}\left(\log r\right)^2\right)\right],\nonumber\\
& & M_{\rm eff}(r) = \int_{S^3\ {\rm dual\ to}\ e_\psi\wedge(\Omega_{11} - B_1\Omega_{22})}\tilde{F}_3 \nonumber\\
& & = M + \frac{3g_s N_f M}{2\pi}\log r + \sum_{m\geq1}\sum_{n\geq1} N_f^m M^n f_{mn}(r),\nonumber\\
& & N^{\rm eff}_f(r) = \frac{4\pi C_0}{\left(\psi - \phi_1 - \phi_2\right)} = N_f + \sum_{m\geq1}\sum_{n\geq0} N_f^m M^n g_{mn}(r);
\end{eqnarray}
$\Omega_{ii} = \sin\theta_i d\theta_i\wedge d\phi_i, i=1, 2$;$e_\psi = d\psi + \cos\theta_1 d\phi_1 + \cos\theta_2 d\phi_2$. Generally in the IR, $f_{mn}(r)$ and $g_{mn}(r)$ are proportional to positive integral powers of $\log r$. Terms occurring in double summation in $M_{\rm eff}$ in (\ref{NeffMeffNfeff}) have higher order in $g_s N_f$ arising from  the type IIB three-form fluxes \cite{metrics} and the next-to-leading order terms in (\ref{h_i}). By means of $r\rightarrow r e^{-\frac{2\pi}{3g_s M_{\rm eff}}}$, i.e., Seiberg-like duality, $N_{\rm eff}\rightarrow N_{\rm eff} - M_{\rm eff} + N_f^{\rm eff}$, \cite{Ouyang}.  Thus, one anticipates a $r={\cal R}_0: N_{\rm eff}({\cal R}_0)=0$ following a Seiberg-like duality cascade. Thus, the length scale of (\ref{metric}) in the IR is as follows\footnote{Employing $L=\left(4\pi g_s N_{\rm eff}\right)^{\frac{1}{4}}$ and $h \sim \frac{L^4}{r^4}$.}:
\begin{eqnarray}
\label{length-IR}
& & L_{(\ref{metric})}\sim N_f^{\frac{3}{4}}\sqrt{\left(\sum_{m\geq0}\sum_{n\geq0}N_f^mM^nf_{mn}(r_0)\right)}\left(\sum_{l\geq0}\sum_{p\geq0}N_f^lM^p g_{lp}(r_0)\right)^{\frac{1}{4}}\left(\log r_0\right)^{\frac{1}{4}} L_{\rm KS},\nonumber\\
& & 
\end{eqnarray}
\noindent where $L_{\rm KS}=\left({g_s M}\right)^{\frac{1}{4}}\sqrt{\alpha^\prime}$. The length scale for the MQGP limit (\ref{MQGP_limit}), where $M$ and $N_f$ are of ${\cal O}(1)$, may have a color-flavor enhancement in the IR ({arising from the logarithmic IR enhancement in the double series summation in (\ref{length-IR})}) as compared to KS, according to the relationship (\ref{length-IR}). Consequently, $L\gg L_{\rm KS}(\sim L_{\rm Planck})$ within the MQGP limit (\ref{MQGP_limit}) implies that the stringy ($\alpha^\prime$) corrections have been suppressed so we can rely on supergravity calculations with sub-dominant contributions arising from the higher derivative (type II $l_s^6/{\cal M}-{\rm theory}\ l_p^6$) corrections within the IR, regardless of $N_c^{\rm IR}=M=3$ and $N_f=2(u/d)+1(s)$.

\end{itemize}

\item{\bf ${\cal M}$-Theory Uplift of \cite{metrics}}: The ${\cal O}(R^4)$ terms in eleven-dimensional supergravity action have been included in \cite{OR4} to investigate, holographically, thermal QCD-like theories at intermediate coupling. The type IIA Strominger-Yau-Zaslow (SYZ) mirror of the type IIB dual was first constructed and then uplifted to ${\cal M}$-theory \cite{MQGP}, \cite{NPB}. The delocalized SYZ mirror (around some fixed values of $\theta_{1,2}, \psi$) was constructed via a triple T-duality performed along a local special Lagrangian (sLag) three-cycle $T^3(x,y,z)$, where $(x,y,z)$ are the toroidal equivalents of $(\phi_1,\phi_2,\psi)$. We can understand this within large-$N$ limit as the $T^2$-invariant sLag of \cite{M.Ionel and M.Min-OO (2008)} where the limit for large complex structures is effected by guaranteeing the base ${\cal B}(r,\theta_1,\theta_2)$ (of a $T^3(\phi_1,\phi_2,\psi)$-fibration via ${\cal B}(r,\theta_1,\theta_2)$) to be large \cite{MQGP}, \cite{NPB}. T-dualization of all the color and flavor $D$-branes of type IIB setup yields color and flavor $D6$-branes. Analogous to \cite{SYZ-free-delocalization} (which involves $D5$-branes encircling a resolved squashed $S^2$), after freeing the ${\cal M}$-theory uplift (of the delocalized type IIA mirror) of the aforementioned delocalization, this uplift maps to a bona fide $G_2$ structure that satisfies the equations of motion (\cite{SYZ-free-delocalization}, second reference in \cite{bulk+gauge_KS+AM}).  Additionally, working in the vanishing-Ouyang-embedding-modulus limit $|\mu_{\rm Ouyang}|$ (essentially limited to the first-generation quarks[+s quark] as $|\mu_{\rm Ouyang}|$, a measure of the $D3-D7$ string length, is, therefore, a measure of the bi-fundamental quark mass) from (\ref{Ouyang-definition}) makes it evident that one will have to work close to small values of $\theta_{1,2}$. We do, for example, work close to  
   \begin{eqnarray}
\label{alpha_theta_12}
& & (\theta_1, \theta_2) = \left(\frac{\alpha_{\theta_1}}{N^{1/5}}, \frac{\alpha_{\theta_2}}{N^{3/10}}\right),\ \ \ \ \ \ \ \alpha_{\theta_{1,2}}\equiv{\cal O}(1).
\end{eqnarray}
Furthermore, the somewhat different powers of $N$ for the delocalized $\theta_{1,2}$ remind us that the resolved $S^2(\theta_2,\phi_2)$ and vanishing $S^2(\theta_1,\phi_1)$ in the two squashed $S^2$s are not at the ``same footing''. From the point of view of the on-shell action, the results up to ${\cal O}(\frac{1}{N})$ become independent of the de-localization by replacing the ${\cal O}(1)$ delocalization parameters $\alpha_{\theta_{1,2}}$ with $N^{1/5}\sin\theta_1$ or $N^{3/10}\sin\theta_2$, respectively (as demonstrated in \cite{OR4}). 

The  ${\cal M}$-theory uplift  of the type IIB dual \cite{metrics} of thermal QCD-like theories at high temperatures, i.e., for $T>T_c$ is given as \cite{MQGP}, \cite{NPB}, \cite{OR4}:  
\begin{eqnarray}
\label{TypeIIA-from-M-theory-Witten-prescription-T>Tc}
\hskip -0.1in ds_{11}^2 & = & e^{-\frac{2\phi^{\rm IIA}}{3}}\Biggl[\frac{1}{\sqrt{h(r,\theta_{1,2})}}\left(-g(r) dt^2 + \left(dx^1\right)^2 +  \left(dx^2\right)^2 +\left(dx^3\right)^2 \right)
\nonumber\\
& & \hskip -0.1in+ \sqrt{h(r,\theta_{1,2})}\left(\frac{dr^2}{g(r)} + ds^2_{\rm IIA}(r,\theta_{1,2},\phi_{1,2},\psi)\right)
\Biggr] + e^{\frac{4\phi^{\rm IIA}}{3}}\left(dx^{11} + A_{\rm IIA}^{F_1^{\rm IIB} + F_3^{\rm IIB} + F_5^{\rm IIB}}\right)^2,
\end{eqnarray} 
where the type IIA dilaton profile is $\phi^{\rm IIA}$, $A_{\rm IIA}^{F^{\rm IIB}_{i=1,3,5}}$ is type IIA RR one-form potential arising from tyiple T-duality of the type IIB fluxes ($F_{1,3,5}^{\rm IIB}$), and $g(r) = 1 - \frac{r_h^4}{r^4}$. The thermal gravitational dual for low-temperature QCD-like theories, i.e., for $T<T_c$,  is as follows:  
\begin{eqnarray}
\label{TypeIIA-from-M-theory-Witten-prescription-T<Tc}
\hskip -0.1in ds_{11}^2 & = & e^{-\frac{2\phi^{\rm IIA}}{3}}\Biggl[\frac{1}{\sqrt{h(r,\theta_{1,2})}}\left(-dt^2 + \left(dx^1\right)^2 +  \left(dx^2\right)^2 + \tilde{g}(r)\left(dx^3\right)^2 \right)
\nonumber\\
& & \hskip -0.1in+ \sqrt{h(r,\theta_{1,2})}\left(\frac{dr^2}{\tilde{g}(r)} + ds^2_{\rm IIA}(r,\theta_{1,2},\phi_{1,2},\psi)\right)
\Biggr] + e^{\frac{4\phi^{\rm IIA}}{3}}\left(dx^{11} + A_{\rm IIA}^{F_1^{\rm IIB} + F_3^{\rm IIB} + F_5^{\rm IIB}}\right)^2,\nonumber\\
& & 
\end{eqnarray}
with $\tilde{g}(r) = 1 - \frac{r_0^4}{r^4}$ and $h(r,\theta_{1,2})$ is the ten dimensional warp factor\cite{metrics, MQGP}. Note that in (\ref{TypeIIA-from-M-theory-Witten-prescription-T>Tc}), $t\rightarrow x^3,\ x^3\rightarrow t$, and then performing a Double Wick rotation in the new $x^3, t$ coordinates, produces (\ref{TypeIIA-from-M-theory-Witten-prescription-T>Tc}). We can also express this as:  $-g_{tt}^{\rm BH}(r_h\rightarrow r_0) = g_{x^3x^3}\ ^{\rm Thermal}(r_0)$ in the results of \cite{VA-Glueball-decay}, \cite{OR4} (Refer to \cite{Kruczenski et al-2003} regarding Euclidean/black $D4$-branes in type IIA). The solitonic $M3$ brane's spatial component will be taken [locally, it may be seen as a solitonic $M5$-brane wrapping a homologous sum of $S^2_{\rm squashed}$s  \cite{DM-transport-2014}]. In (\ref{TypeIIA-from-M-theory-Witten-prescription-T<Tc}), where $\frac{2r_0}{ L^2}\left[1 + {\cal O}\left(\frac{g_sM^2}{N}\right)\right]$ ($r_0$ denotes the IR cut-off that characterizes the thermal background) provides the very tiny $M_{\rm KK}$. We get the period of  $S^1(x^3)$ from very large $\frac{2\pi}{M_{\rm KK}}$, where $L = \left( 4\pi g_s N\right)^{\frac{1}{4}}$  (see also \cite{Armoni et al-2020}).  Regarding the thermal background associated with $T<T_c$, in the working metric (\ref{TypeIIA-from-M-theory-Witten-prescription-T<Tc}), $\tilde{g}(r)$ is set to unity.

\end{itemize}
We consider here the bosonic part of eleven-dimensional supergravity action inclusive of $\mathcal{O}(R^4)$ corrections\cite{OR4},
\begin{eqnarray}
\label{D=11_O(l_p^6)}
& & \hskip -0.5in S = \frac{1}{2\kappa_{11}^2}\Biggl[\int_{M_{11}}\sqrt{g}R + \int_{\partial M_{11}}\sqrt{h}K -\frac{1}{2}\int_{M_{11}}\sqrt{g}G_4^2
-\frac{1}{6}\int_{M_{11}}C_3\wedge G_4\wedge G_4\nonumber\\
& & \hskip -0.5in  + \frac{\left(4\pi\kappa_{11}^2\right)^{\frac{2}{3}}}{{(2\pi)}^4 3^2.2^{13}}\Biggl(\int_{\cal{M}} d^{11}\!x \sqrt{g}\left(J_0-\frac{1}{2}E_8\right) + 3^2.2^{13}\int C_3 \wedge X_8 + \int t_8 t_8 G^2 R^3 + \cdot \cdot \Biggr)\Biggr],\nonumber\\
\end{eqnarray}
where:
\begin{eqnarray}
\label{J0+E8-definitions}
& & \hskip -0.8inJ_0  =3\cdot 2^8 (R^{HMNK}R_{PMNQ}{R_H}^{RSP}{R^Q}_{RSK}+
{1\over 2} R^{HKMN}R_{PQMN}{R_H}^{RSP}{R^Q}_{RSK}),\nonumber\\
& & \hskip -0.8inE_8  ={ 1\over 3!} \epsilon^{ABCM_1 N_1 \dots M_4 N_4}
\epsilon_{ABCM_1' N_1' \dots M_4' N_4' }{R^{M_1'N_1'}}_{M_1 N_1} \dots
{R^{M_4' N_4'}}_{M_4 N_4},,\nonumber\\
& & \hskip -0.8in t_8t_8G^2R^3 = t_8^{M_1...M_8}t^8_{N_1....N_8}G_{M_1}\ ^{N_1 PQ}G_{M_2}\ ^{N_2}_{\ \ PQ}R_{M_3M_4}^{\ \ \ \ N_3N_4}R_{M_5M_6}^{\ \ \ \ N_5N_6}R_{M_7M_8}^{\ \ \ \ N_7N_8},
\nonumber\\
& & \hskip -0.8in X_8 = {1 \over 192} \left( {\rm tr}\ R^4 -
{1\over 4} ({\rm tr}\ R^2)^2\right),\nonumber\\
& & \hskip -0.8in\kappa_{11}^2 = \frac{(2\pi)^8 l_p^{9}}{2}.
\end{eqnarray}
The holographic renormalization of the eleven-dimensional on-shell bulk action was discussed in \cite{Gopal+Vikas+Aalok}.

Via the variation of action with respect to metric $g_{MN}$, and the three form potential $C_{MNP}$, we obtin the following equations of motion:
\begin{eqnarray}
\label{eoms}
& & {\rm EOM}_{\rm MN}:\ R_{MN} - \frac{1}{2}g_{MN}{\cal R} - \frac{1}{12}\left(G_{MPQR}G_N^{\ PQR} - \frac{g_{MN}}{8}G_{PQRS}G^{PQRS} \right)\nonumber\\
 & &  = - \beta\left[\frac{g_{MN}}{2}\left( J_0 - \frac{1}{2}E_8\right) + \frac{\delta}{\delta g^{MN}}\left( J_0 - \frac{1}{2}E_8\right)\right],\nonumber\\
& & d*G = \frac{1}{2} G\wedge G +3^22^{13} \left(2\pi\right)^{4}\beta X_8,\nonumber\\
& &
\end{eqnarray}
where \cite{Becker:2001pm}:
\begin{equation}
\label{beta-def}
\beta \equiv \frac{\left(2\pi^2\right)^{\frac{1}{3}}\left(\kappa_{11}^2\right)^{\frac{2}{3}}}{\left(2\pi\right)^43^22^{12}} \sim l_p^6.
\end{equation}

 The symbols $R_{MNPQ}, R_{MN}, {\cal R}$ represent the elven-dimensional Riemann curvature tensor, the Ricci tensor, and the Ricci scalar in (\ref{D=11_O(l_p^6)})/(\ref{eoms}). The ansatz created to solve (\ref{eoms}) was as follows:
\begin{eqnarray}
\label{ansaetze}
& & \hskip -0.8ing_{MN} = g_{MN}^{(0)} +\beta g_{MN}^{(1)},\nonumber\\
& & \hskip -0.8inC_{MNP} = C^{(0)}_{MNP} + \beta C_{MNP}^{(1)}.
\end{eqnarray}
Now, $C^{\beta}_{MNP}=0$ has been shown in \cite{OR4} to be a consistent truncation of ${\cal O}(\beta)$ corrections provided ${\cal C}_{zz} - 2 {\cal C}_{\theta_1z} = 0, |{\cal C}_{\theta_1x}|\ll1$,  where $C_{MN}$ are the constants of integration appearing in the solutions to the equations of motion of $h_{MN}$, and the delocalized toroidal coordinates $T^3(x, y, z)$ that corresponds to some $(r, \theta_1, \theta_2) = (\langle r\rangle, \langle\theta_1\rangle, \langle\theta_2\rangle)$ can be expressed as \cite{MQGP}:
\begin{eqnarray}
\label{xyz-defs}
& & dx = \sqrt{\frac{1}{6} + {\cal O}\left(\frac{g_sM^2}{N}\right)}h^{\frac{1}{4}}\Bigl(\langle r\rangle, \langle \theta_{1,2}\rangle\Bigr)\langle r\rangle
\sin\langle\theta_{1}\rangle d\phi_1,\nonumber\\ 
& & dy = \sqrt{\frac{1}{6} + \frac{a^2}{r^2} + {\cal O}\left(\frac{g_sM^2}{N}\right)}h^{\frac{1}{4}}\Bigl(\langle r\rangle, \langle \theta_{1,2}\rangle\Bigr)
\langle r\rangle\sin\langle\theta_{2}\rangle d\phi_2,\nonumber\\  
& & dz =  \sqrt{\frac{1}{6} + {\cal O}\left(\frac{g_sM^2}{N}\right)}h^{\frac{1}{4}}\Bigl(\langle r\rangle, \langle \theta_{1,2}\rangle\Bigr)\langle r\rangle d\psi.
\end{eqnarray}
The 10-D warp factor is given in (\ref{h-def}). Given this, ${\cal O}(R^4)$ corrections are only made to the metric and have the following form:
\begin{eqnarray}
\label{fMN-definitions}
\delta g_{MN} =g^{(1)}_{MN} = g_{MN}^{(0)} f_{MN}(r).
\end{eqnarray}
The ${\cal M}$ theory metric usually has the following form when ${\cal O}(R^4)$ improvements are applied:
\begin{equation}
\label{fMN-def}
g_{MN} = g_{MN}^{(0)}\left(1+\beta f_{MN}(r)\right),
\end{equation}
where \cite{OR4} provides the non-vanishing $f_{MN}(r)$s. The metric components (\ref{fMN-def}) when calculated near the coordinate patches $\psi=2n\pi, \ n=0, 1, 2$ yields $g_{rM}=0, M\neq r$ and decouple the $M_5=(S^1 \times_w \mathbb{R}^3) \times \mathbb{R}_{>0}$ and the unwarped $\tilde{M}_6(S^1_{\cal M} \times_w T^{\rm NE})$ of $SU(3)$ structure, where $S^1_{\cal M}$, and $T^{\rm NE}$ is the ${\cal M}$-theory circle and the non-Einsteinian deformation of $T^{1,1}$ respectively.

We can verify the leading-order-in-$N$ contribution to $J_0$ in the following manner\cite{OR4}:
\begin{eqnarray}
\label{J0-1}
J_0 = \frac{1}{2} R^{\phi_2r \theta_1r}  R_{r \psi\theta_1r}  R_{\phi_2}^{\ \ r \phi_1r}  R^{\psi}_{r \phi_1r} -R^{\phi_2r \theta_1r}
   R_{\phi_1r\theta_1r}  R_{\phi_2}^{\ \ r \phi_1r}  R^{\theta_1}_{\ \ r\theta_1r},
\end{eqnarray}
i.e., independent of the ${\cal M}$-theory circle coordinate. It was shown in \cite{OR4} that $\frac{E_8}{J_0}\sim\frac{1}{N^{\lambda_1}}, \frac{t_8^2 G^2R^3}{E_8}\sim\frac{1}{N^{\lambda_2}}, \lambda_{1,2}>0$. This hence implies a hierarchy: $t_8^2G^2R^3<E_8<J_0$. Consequently, only the ``$J_0$'' term is taken into account amongst the ${\cal O}(R^4)$ terms in (\ref{D=11_O(l_p^6)}) for calculation purposes in this study, which is similar to \cite{Vikas+Gopal+Aalok,Gopal-Tc-Vorticity,Gopal+Vikas+Aalok,ACMS,Aalok+Gopal-Mesino}\footnote{See also \cite{Yadav:2023glu}, where some of the results in the context of holographic thermal QCD at intermediate coupling are discussed.}.

\subsection{Almost Contact 3-Structures Arising from $G_2$ Structure in the ${\cal M}$ Uplift in the Limit (\ref{MQGP_limit})}
\label{ACM3S-basics}

Due to non-trivial ${\cal M}$-theory four-form fluxes, $M_7$ is generically not Ricci-flat and hence does not possess $G_2$ holonomy, but usually possesses $G_2$ structure \footnote{If V is a seven-dimensional real vector space, then a three-form $\Phi$ is said to be positive if it lies in the $GL (7; \mathbb{R})$ orbit of $\Phi_0$, where $\Phi_0$ is a three-form on $\mathbb{R}^7$ which is preserved by $G_2$-subgroup of $GL (7; \mathbb{R})$. The pair $(\Phi; g)$ for a positive 3-form $\Phi$
and corresponding metric $g$ constitute a $G_2$-structure.}. Given that the adjoint of $SO(7)$ decomposes under $G_2$ as ${\bf 21}\rightarrow{\bf 7}\oplus{\bf 14}$ where ${\bf 14}$ is the adjoint representation of $G_2$, one obtains four $G_2$-structure torsion classes:
\begin{equation}
\label{T-G2}
\tau \in \Lambda^1 \otimes g_2^\perp = W_1 \oplus W_7 \oplus W_{14} \oplus W_{27} = \tau_0 \oplus \tau_1 \oplus \tau_2 \oplus \tau_3,
\end{equation}
$g_2^\perp$ being the orthogonal complement of $g_2$, the subscript $a$ in $W_a$ denoting the dimensionality of the torsion class $W_a$, and $p$ in $\tau_p$ denoting the rank of the associated differential form. The four intrinsic $G_2$-structure torsion classes are defined, e.g. in \cite{J. G. J. Held's thesis [2012]}.

The $G_2$-structure torsion classes $\tau_p$'s of the seven-fold $M_7=S^1_{\cal M}\times_w\left(S^1_{\rm thermal}\times_w M_5\right)$, $M_5$ being a non-Einsteinian generalization of $T^{1,1}$, and close to the Ouyang embedding (\ref{Ouyang-definition}) of the flavor $D7$-branes in the parent type IIB dual in the limit of very-small-Ouyang-embedding parameter limit ($|\mu_{\rm Ouyang}|\ll1$) were worked out in \cite{ACMS}:
\begin{equation}
\label{G2-torsion}
\tau\left(M_7\right) = \tau_1\oplus\tau_2\oplus\tau_3.
\end{equation}
It was further shown in \cite{ACMS} that in the $N\gg1$-MQGP limit (footnote \tcb{1}) and the intermediate-$N$ MQGP limit (\ref{MQGP_limit}), the aforementioned closed $M_7$ supports Almost Contact 3-Structures (Lemma 2 of \cite{ACMS}). But $M_7$ supports Contact 3-Structures  only in the latter limit (\ref{MQGP_limit}) [Lemma 4 of \cite{ACMS}].

The main result of \cite{ACMS} is that the four-parameter space ${\cal X}_{G_2}(g_s, M, N_f; N)$ [$g_s\in(0,1)$ and varying continuously; $M_{\rm UV}, N_f^{\rm UV} N$ varying in steps of 1 such that $M, N_f$ are ${\cal O}(1)$ and $\frac{1}{N}\ll1$] of $M_7$ supporting $G_2$ structures and relevant to the aforementioned ${\cal M}$-theory uplift of thermal QCD-like theories, is not $N$-path connected with reference to Contact Structures in the IR, i.e., the $N\gg1$ Almost Contact 3-Structures arising from the $G_2$ structure in the $N\gg1$ MQGP limit (footnote \tcb{1}), do not connect to a Contact 3-Structures (in the IR) which is shown to exist only for an appropriate intermediate $N$ effected by the intermediate-$N$ MQGP limit (\ref{MQGP_limit}) and, e.g., by the QCD-inspired parameters $M_{\rm UV}, N_f^{\rm UV} g_s$ of Table \tcb{1}.

\begin{figure}
\begin{center}
\includegraphics[width=0.80\textwidth]{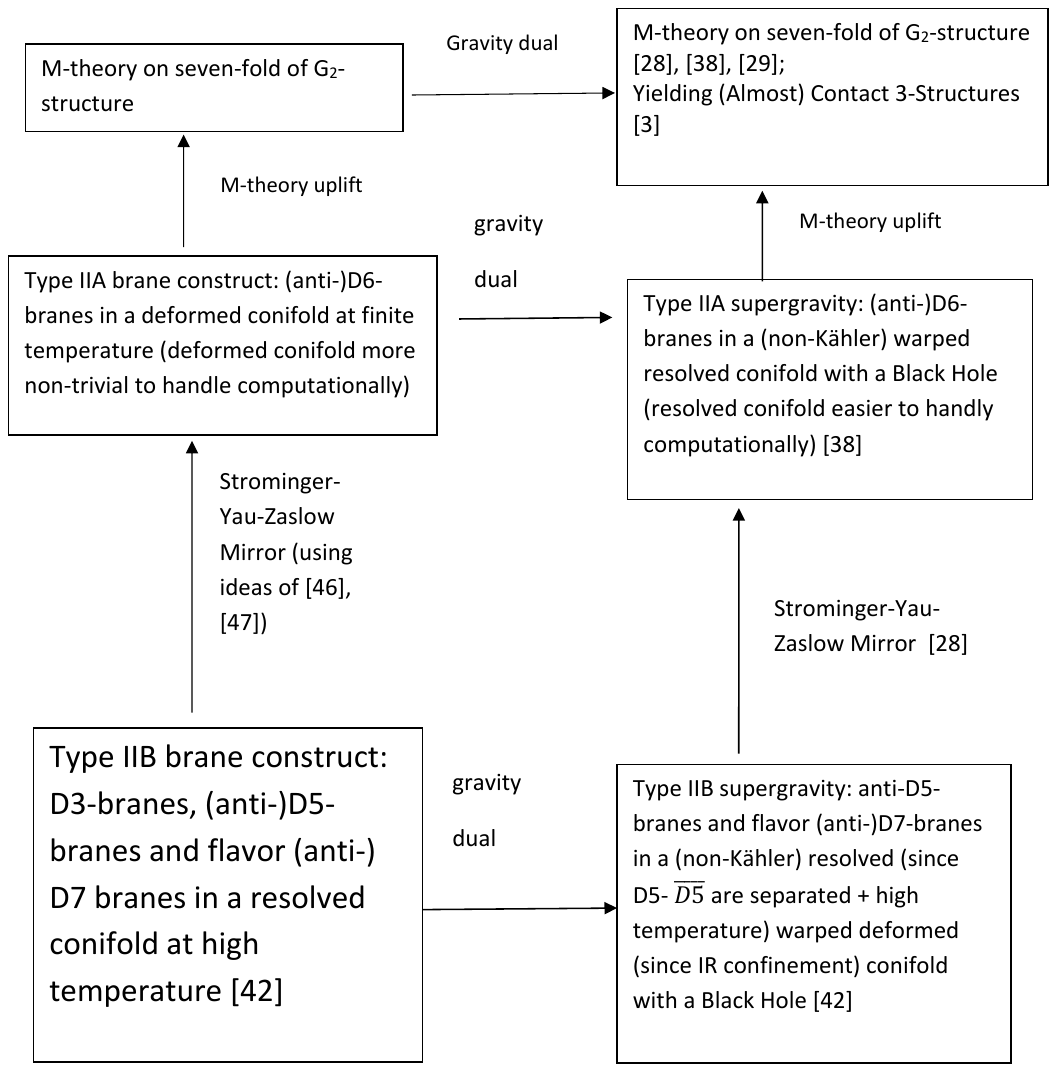}
\end{center}
\caption{String/${\cal M}$-theory dual of thermal QCD-like theories}
\label{flowchart}
\end{figure}

From the point of view of development of a UV-complete top-down holographic QCD program initiated in \cite{metrics}, we are now set to go through Figure \ref{flowchart}.

\section{$\frac{1}{3} - c_s^2$ up to ${\cal O}(l_p^6)$ from Scalar Modes of Metric Perturbations}
\label{cs-Zs}

In this section, using the eleven-dimensional supergravity action inclusive of $\mathcal{O}(R^4)$ corrections corresponding to our $\cal M$-theoretic set-up dual to large-$N$ thermal QCD-like theories, and utilizing the equations of motion corresponding to the metric variation of the action, we obtain the deviation of the square of the speed of sound $c_s$ from its conformal value (1/3), i.e., 
$\Biggl(\frac{1}{3}-c_s^2\Biggr)$ in terms of $g_s$, $M$, $N$,  $N_f$, and $r_h$.

As discussed towards the end of Sec. \ref{review}, the $\mathcal{O}(R^{4})$ terms in the action (\ref{D=11_O(l_p^6)}) follow a hierarchy in large-$N$ limit. Following that hierarchy, the contribution of $J_0$ term is dominant over $E_{8}$ and $t_8^2G^2R^3$ terms in the large-$N$ limit \cite{OR4}. So we retain only the $J_0$ contribution from the equations of motion (\ref{eoms}), and hence
\begin{eqnarray}
\label{EOM-D=11}
R_{MN}-\frac{1}{2}g_{MN}R-\frac{1}{12}\left(G_{MPQR}G_N^{\ PQR}-\frac{g_{MN}}{8}G_4^2\right)=-\beta \Biggl[\frac{g_{MN}}{2}J_0+\frac{\delta J_0}{\delta g^{MN}}\Biggr].
\end{eqnarray}
For generic spacetime with dimension $D$, the trace of the above equation can be given as: 
\begin{eqnarray}
\label{trace-EOM-D=11}
& &
R\left(1-\frac{D}{2}\right)-\frac{G_4^2}{12}\left(1-\frac{D}{8}\right)=\beta \Biggl[\frac{D}{2}J_0+g^{MN}\frac{\delta J_0}{\delta g^{MN}}\Biggr].
\end{eqnarray}

As a result of metric upto $\mathcal{O}(R^4)$ (\ref{ansaetze}) we have, $R=R^{\beta^0}+\beta R^{\beta}$, $G_4^2=\left(G_4^2\right)^{\beta^0}$, $J_0=J_0^{\beta^0}$, and $g^{MN}\frac{\delta J_0}{\delta g^{MN}}=\left(g^{MN}\frac{\delta J_0}{\delta g^{MN}}\right)^{\beta^0}$ the following relationship is implied for $D=11$ by equation (\ref{trace-EOM-D=11}) from the ${\cal O}(\beta^0)$ and ${\cal O}(\beta)$ terms:
\begin{eqnarray}
\label{trace-EOM-D=11-i}
& &
R^{\beta^0}=\frac{G_4^2}{144},\nonumber\\
& & J_0=-\frac{2}{11}\left(R^\beta+g^{MN}\frac{\delta J_0}{\delta g^{MN}}\right).
\end{eqnarray}
 Now it is possible to express the equation of motion at ${\cal O}(\beta^0)$ and ${\cal O}(\beta)$ resulting from the substitution of  (\ref{trace-EOM-D=11-i}) within the equation  (\ref{EOM-D=11}) as:
\begin{eqnarray}
\label{EOMs-beta0-beta}
& & {\bf {EOM_{MN}^{\beta^0}}}: \ \ \ 
R_{MN}^{\beta^0}-\frac{1}{2}g_{MN}^{\beta^0}R^{\beta^0}-\frac{1}{12}\left(G_{MPQR}G_N^{\ PQR}\right)^{\beta^0}+\left(\frac{144}{196}\right)g_{MN}^{\beta^0}R^{\beta^0}=0, \nonumber\\
& & {\bf {EOM_{MN}^{\beta}}}: \ \ \ R_{MN}^\beta + g_{MN}^\beta R^{\beta^0}-\frac{13}{22}g_{MN}^{\beta^0} R^{\beta}-\frac{1}{4}\left(G_{MPQR}G_{N\tilde{P}\tilde{Q}\tilde{R}}g^{P\tilde{P}, \beta}g^{Q\tilde{Q},\beta^0}g^{R\tilde{R},\beta^0}\right)\nonumber\\
& & =- \Biggl[-\frac{g_{MN}^{\beta^0}}{11}\left(g^{PQ}\frac{\delta J_0}{\delta g^{PQ}}\right)^{\beta^0}+\left(\frac{\delta J_0}{\delta g^{MN}}\right)^{\beta^0} \Biggr].
\end{eqnarray} 
To calculate the ${\cal O}(R^4)$ contribution to the linearized Einstein's equation in ${\rm EOM_{MN}^{\beta}}$, the following will be used:
{\footnotesize
\begin{eqnarray}
\label{delta J_0}
 \delta J_0 & \stackrel{\rm MQGP,\ IR}{\xrightarrow{\hspace*{1.5cm}}} & 3\times 2^8 \delta R^{HMNK} R_H^{\ RSP}\Biggl(R_{PQNK}R^Q_{\ RSM} + R_{PSQK}R^Q_{\ MNR}  \nonumber\\
 & & + 2\left[R_{PMNQ}R^Q_{\ RSK} + R_{PNMQ}R^Q_{\ SRK}\right]\Biggr) \nonumber\\
& &  \equiv 3\times 2^8 \delta R^{HMNK}\chi_{HMNK}\nonumber\\
& & = -\delta g_{\tilde{M}\tilde{N}}\Biggl[ g^{M\tilde{N}} R^{H\tilde{N}NK}\chi_{HMNK}
+ g^{N\tilde{N}} R^{HM\tilde{M}K}\chi_{HMNK} + g^{K\tilde{M}}R^{HMN\tilde{N}}\chi_{HMNK}
\nonumber\\
& & + \frac{1}{2}\Biggl(g^{H\tilde{N}}[D_{K_1},D_{N_1}]\chi_H^{\tilde{M}N_1K_1} +
g^{H\tilde{N}}D_{M_1}D_{N_1} \chi_H^{M_1[{N}_1\tilde{M}]}  - g^{H\tilde{H}} D_{\tilde{H}}D_{N_1}\chi_H^{\tilde{N}[N_1\tilde{M}]}\Biggr)\Biggr],\nonumber\\
& &
\end{eqnarray}
}
where:
{\footnotesize
\begin{eqnarray}
\label{chi-def}
& & \chi_{HMNK} \equiv R_H^{\ \ RSP}\left[R_{PQNK} R^Q_{\ \ RSM}
+ R_{PSQK} R^Q_{\ \ MNR} + 2\left(R_{PMNQ} R^Q_{\ \ RSK} + R_{PNMQ}R^Q_{\ \ SRK}\right)\right].
\end{eqnarray}
}

The linearized equations of motion of (\ref{EOMs-beta0-beta}) for the metric perturbation defined as:
\begin{eqnarray}
\label{metric-perturbations}
& & 
g_{MN}^{\beta^0}=\left(g_{MN}^{\beta^0}\right)^{(0)}+\eta  h_{MN}^{\beta^0}; \ \ \ g_{MN}^{\beta}=\left(g_{MN}^{\beta}\right)^{(0)}+ \eta  h_{MN}^{\beta}, \nonumber\\
& &
R_{MN}^{\beta^0}=\left(R_{MN}^{\beta^0}\right)^{(0)}+\eta  \delta R_{MN}^{\beta^0}; \ \ \ R_{MN}^{\beta}=\left(R_{MN}^{\beta}\right)^{(0)}+\eta  \delta R_{MN}^{\beta}
\end{eqnarray}
where $h_{MN}^{\beta^0} \neq h_{MN}^{\beta}$, $\delta R_{MN}^{\beta^0} \neq \delta R_{MN}^{\beta}$ and $g_{MN}^{(1)}$ according to the definition in (\ref{ansaetze}) is a synonym  for $g_{MN}^\beta$. To denote the linear order perturbation in metric, Ricci tensor, and Ricci scalar we added a parameter $\eta$. Then $\left(g_{MN}^{\beta^0}\right)^{(0)}$, $\left(g_{MN}^{\beta}\right)^{(0)}$, $\left(R_{MN}^{\beta^0}\right)^{(0)}$, and $\left(R_{MN}^{\beta}\right)^{(0)}$ are metric, Ricci tensor without perturbation whereas $h_{MN}^{\beta^0}$, $h_{MN}^{\beta}$, $\delta R_{MN}^{\beta^0}$ and $\delta R_{MN}^{\beta}$ are the terms with linear order perturbation in metric, and Ricci tensor respectively. Via this parameter, we can identify Einstein's perturbed and unperturbed equations of motion in the following way: In order to calculate the linearized equations of motion for ${EOM_{MN}^{\beta^0}}$ in (\ref{EOMs-beta0-beta}), we must first compute $R_{MN}^{\beta^0}$ and $R^{\beta^0}$ via the metric $g_{MN}^{\beta^0}=\left(g_{MN}^{\beta^0}\right)^{(0)}+\eta h_{MN}^{\beta^0}$. These should then be substituted in ${EOM_{MN}^{\beta^0}}$ of (\ref{EOMs-beta0-beta}) in order to obtain the coefficient of $\eta$. Similar to this, we may use (\ref{metric-perturbations}) to extract the linearized equations of motion from ${EOM_{MN}^{\beta}}$ that appear in (\ref{EOMs-beta0-beta}) and then extract the coefficient of $\eta$.

The $h_{MN}^{\beta^0/\beta}$ has a form that is analogous to \cite{Sil:2020jhr}\footnote{Generically, we can always decompose the metric into three modes on the basis of rotational symmetry in $x_2$-$x_3$ plane \cite{Policastro:2002se}: i) Scalar modes ($h_{vv}, h_{v x_1}, h_{vr}, h_{x_1,x_1}, h_{x_1 r}, h_{x_2 x_2}=h_{x_3 x_3}, h_{rr}$), ii) Vector modes ($h_{v x_2}, h_{r x_2}, h_{x_1 x_2}$), and iii) Tensor mode ($h_{x_2 x_3}$). All three modes are studied separately in thermal QCD context see \cite{Sil:2016jmc}}.
\begin{eqnarray}
& & h_{vv}^{\beta^0/\beta}(r)=e^{-i w v+iq x^1} g_{vv}(r) H_{vv}^{\beta^0/\beta}(r), \ \ \ \ \  h_{v x^1}^{\beta^0/\beta}(r)=e^{-i w v+iq x^1} g_{x^1 x^1}(r) H_{v x^1}^{\beta^0/\beta}(r), \nonumber\\
& & h_{x^1 x^1}^{\beta^0/\beta}(r)=e^{-i w v+iq x^1} g_{x^1 x^1}(r) H_{x^1 x^1}^{\beta^0/\beta}(r), \ \ \ h_{x^2 x^2}^{\beta^0/\beta}(r)=e^{-i w v+iq x^1} g_{x^2 x^2}(r) H_{x^2 x^2}^{\beta^0/\beta}(r),
\end{eqnarray}
where 
\begin{eqnarray}
\label{h-H}
& & \hskip -0.5in H_{vv}^{\beta^0/\beta}(r)=\frac{h_{vv}^{\beta^0/\beta}(r)}{g_{vv}(r)}; \ \ H_{v x^1}^{\beta^0/\beta}(r)=\frac{h_{v x^1}^{\beta^0/\beta}(r)}{g_{x^1 x^1}(r)}; \ \ H_{x^1 x^1}^{\beta^0/\beta}(r)=\frac{h_{x^1 x^1}^{\beta^0/\beta}(r)}{g_{x^1 x^1}(r)}; \ \ H_{x^2 x^2}^{\beta^0/\beta}(r)=\frac{h_{x^2 x^2}^{\beta^0/\beta}(r)}{g_{x^2 x^2}(r)},
\end{eqnarray}
where $g_{vv}(r), \ g_{v r}(r), g_{x_1 x_1}(r)=g_{x_2 x_2}(r)=g_{x_3 x_3}(r)$, we have used the Eddington-Finkelstein ingoing coordinates to define these metric components which can be given as:
\begin{eqnarray}
\label{v-rstar}
& & v=t+r_{*}, \ \ \ \ \ \ r_{*}=\int \sqrt{\frac{G_{rr}^{\cal M}(r)}{G_{tt}^{\cal M}(r)}} dr.
\end{eqnarray} 
Near $\psi=2n\pi$, $n=0,1,2$- coordinate patches the $11$-dimensional metric is worked out where $G_{rM}^{\cal M}=0, M\neq r$. Defining the two sub-manifolds as,
\begin{enumerate}
\item the non-compact one $M_5=(S^1 \times_w \mathbb{R}^3) \times \mathbb{R}_{>0}$,
\item the compact one, the unwarped $\tilde{M}_6(S^1_{\cal M} \times_w T^{\rm NE})$ of $SU(3)$ structure wherein $S^1_{\cal M}$ is the ${\cal M}$-theory circle and $T^{\rm NE}$ is the non-Einsteinian deformation of $T^{1,1}$
\end{enumerate}  
and decouple them. The non-compact five-dimensional part ($M_5$) of the metric (\ref{fMN-def}) using Eddington-Finkelstein ingoing coordinates (\ref{v-rstar}) can be written as:

\begin{eqnarray}
\label{5D-metric-EF}
ds^2=g_{vv} dv^2+g_{vr} dv dr+g_{x_1 x_1} dx_1^2+g_{x_2 x_2} dx_2^2+g_{x_3 x_3} dx_3^2,
\end{eqnarray}
on the other hand, close to the type IIA $D6$ flavor branes \cite{MQGP-meson-spectroscopy}, (\ref{alpha_theta_12}) and $(\tilde{x},\tilde{z}) = (0, {\rm constant})$ - $(\tilde{x}, \tilde{y}, \tilde{z})$ diagonalizing the $T^3$ sLag used to build the type IIA SYZ mirror, up to leading order in $N$:
\begin{eqnarray}
\label{gvv-etc}
& & g_{vv}(r)=-\frac{3^{2/3} \sqrt{\frac{1}{N}} N_f^{2/3} r^2 \left(1-\frac{r_h^4}{r^4}\right) (2 \log(N)-6
   \log (r))^{2/3}}{8 \pi ^{7/6} \sqrt{g_s}} -\beta \Biggl(\frac{b^8 \left(9 b^2+1\right)^3 \Sigma _1 \left(6 a^2+r_h^2\right) }{72 \sqrt[3]{3} \pi ^{13/6} \left(18 b^4-3
   b^2-1\right)^5 \sqrt{g_s} }\nonumber\\
   & & \times \frac{\left(4374 b^6+1035 b^4+9 b^2-4\right) M \left(\frac{1}{N}\right)^{11/4} r^2
    \left(1-\frac{r_h^4}{r^4}\right) \log
   (r_h) (2 \log(N) N_f-6 N_f \log (r))^{2/3}(r-r_h)^2}{\log(N)^2 N_f r_h^2 \alpha _{\theta _2}^3 \left(9
   a^2+r_h^2\right)}\Biggr), \nonumber\\
& & g_{v r}(r)=\frac{1}{4} \left(\frac{3}{\pi }\right)^{2/3} N_f^{2/3} \sqrt{\frac{6 a^2+r^2}{9 a^2+r^2}} (2
   \log(N)-6 \log (r))^{2/3}\left(1+\frac{\beta}{2}\left({\cal{C}}_{zz}-2 {\cal{C}}_{\theta_1 z}+2 {\cal{C}}_{\theta_1 x}\right) \right), \nonumber\\
& & g_{x_1 x_1}(r)=g_{x_2 x_2}(r)=g_{x_3 x_3}(r)=\frac{3^{2/3} \sqrt{\frac{1}{N}} N_f^{2/3} r^2 (2 \log(N)-6 \log (r))^{2/3}}{8 \pi ^{7/6}
   \sqrt{g_s}}\nonumber\\
   & & -\beta \Biggl(\frac{b^8 \left(9 b^2+1\right)^4 \left(39 b^2-4\right) M \left(\frac{1}{N}\right)^{11/4} r^2 \Sigma _1 \left(6
   a^2+r_h^2\right) (r-r_h)^2 \log (r_h) (N_f (\log(N)-3 \log (r)))^{2/3}}{12
   \sqrt[3]{6} \pi ^{13/6} \left(3 b^2-1\right)^5 \left(6 b^2+1\right)^4 \sqrt{g_s} \log(N)^2
   N_f r_h^2 \alpha _{\theta _2}^3 \left(9 a^2+r_h^2\right)}\Biggr),\nonumber\\
   & &
\end{eqnarray}
where $b=\frac{1}{\sqrt{3}}+\epsilon$, and
\begin{eqnarray}
\label{Sigma_1-3-def}
& & \Sigma_1 \equiv 19683
   \sqrt{6} \alpha _{\theta _1}^6+6642 \alpha _{\theta _2}^2 \alpha _{\theta _1}^3-40 \sqrt{6} \alpha _{\theta _2}^4.
\end{eqnarray} 
The $C^{(1)}_{MNP}=0$-truncation of (\ref{ansaetze}), as demonstrated in \cite{OR4}, produces:
\begin{eqnarray}
\label{C1=0-truncation}
& & {\cal{C}}_{zz}-2 {\cal{C}}_{\theta_1 z} = 0,\nonumber\\
& & |{\cal{C}}_{\theta_1 x}|\ll 1.
\end{eqnarray}
Therefore, there is no ${\cal O}(\beta)$ correction for $g_{vr}$.

Via employing the metric perturbation defined in (\ref{metric-perturbations}), we can get the linearized equations of motion ``${\rm EOM_{MN}^{\beta^0}}$'' and ``${\rm EOM_{MN}^{\beta}}$''\footnote{${\rm EOM_{MN}^{\beta^0}}$ represents the equation of motion without $\cal{O}(\beta)$ correction, and ${\rm EOM_{MN}^{\beta}}$ are the equation of motion with $\cal{O}(\beta)$ correction where the indices  M, N runs over $v,r,x_1,x_2,x_3$.} at  ${\cal O}(\beta^0)$ and ${\cal O}(\beta)$ respectively appearing in (\ref{EOMs-beta0-beta}). For this utilize the unperturbed metric given in (\ref{gvv-etc}). The linearzied equations of motion are the ${\cal O}(\eta )$ contribution of the ``${\rm EOM_{MN}^{\beta^0}}$'' and ${\cal O}(\eta )$ contribution of the ``${\rm  EOM_{MN}^{\beta}}$'' and ${\cal O}(\beta)$ respectively.

The linearized equations of motion at ${\cal O}(\beta^0)$ for the metric perturbations specified in (\ref{metric-perturbations}) are derived by reducing the complexity of ``${\rm EOM}_{MN}^{\beta^0}$'' provided in (\ref{EOMs-beta0-beta}) while keeping the terms of ${\cal O}(\eta)$. Therefore, using (\ref{EOMs-beta0-beta}), (\ref{metric-perturbations}) and (\ref{gvv-etc})\footnote{For ``${\rm EOM}_{MN}^{\beta^0}$'', we use ${\cal O}(\beta^0)$ involvement of the metric (\ref{gvv-etc}) whereas for ``${\rm EOM}_{MN}^{\beta}$'', we use ${\cal O}(\beta^0)$ plus ${\cal O}(\beta)$ correction of the metric (\ref{gvv-etc}) as per the need since ``${\rm EOM}_{MN}^{\beta}$'' utilises $R^{\beta^0}, R^\beta$ etc. and in generic metric (\ref{gvv-etc}) has the form: $g_{MN}=g_{MN}^{\beta^0}+\beta g_{MN}^{\beta}$.}, the linearized equations of motion\footnote{Additionally we made the use of:
\begin{eqnarray*}
\label{flux-beta0-subdominant}
& &  G_{rMNP}G_r^{\ \ MNP} = \left.\frac{\sqrt{\frac{2}{3}} {g_s}^{7/4} M N^{3/5}
   \sqrt[4]{g_s} \left(108 a^2
   {N_f} \log ^2(r)+72 a^2 {N_f} \log
   (r)+{N_f} r+4 {N_f} r \log (r)\right)}{27
   \pi  \alpha ^2 r^3 \alpha _{\theta _1}^2 \alpha
   _{\theta _2}^3 {b_1}(r)^2}\right|_{(\ref{r_h})} \nonumber\\
   & & \hskip 0.9in \sim\frac{1}{r^2N^{2/5}}<\left.R_{rr}^{\beta^0}\right|_{(\ref{r_h})}\sim\frac{1}{N^{1/3}r^2},
\end{eqnarray*}
where $b_1(r) = \frac{3 {g_s}^{7/4} M  {N_f} \left(r^2-3
   a^2\right) \log ^2(r)}{\sqrt{2} \pi ^{5/4} r^2
   \alpha _{\theta _1} \alpha _{\theta _2}^2}$.} can be written into a single equation of motion via constructing the gauge invariant combination of metric perturbation \cite{Kovtun:2005ev}.
     
In accordance with the rotational symmetry $SO(2)$ in the $x_2-x_3$ plane, we can write the gauge-invariant combination for scalar mode perturbations, as in our case \cite{Sil:2020jhr}:
{\footnotesize
\begin{eqnarray}
\label{Zs}
& &  {Z_s}(r)=-2 {H_{x_2x_2}}(r) \left(\frac{q^2 {g_{vv}}'(r)}{{g^{x_1x_1}}(r)
   {g_{x_1x_1}}'(r)}+w^2 {g_{x_1x_1}}(r)\right)+4 q w {g_{x_1x_1}}(r) {H_{vx_1}}(r)+2 w^2
   {g_{x_1x_1}}(r) {H_{x_1x_1}}(r)\nonumber\\
   & & \hskip 0.6in +2 q^2 {g_{vv}}(r) {H_{vv}}(r).
\end{eqnarray}
}
One can show that $Z_s(r)$ satisfies \cite{MChaos}:
\begin{equation}
\label{Zs-EOM}
Z_s''(r) = l(r) Z_s'(r) + m(r) Z_s(r).
\end{equation}
where:
{\footnotesize
\begin{eqnarray}
\label{l+m}
& &  l(r) =\nonumber\\
& & -\frac{i \left(135424 i 2^{2/3} \sqrt[3]{3} \pi ^{4/3} \sqrt{{g_s}}
   \sqrt{\frac{1}{N}} \sqrt[3]{{N_f}} q^2 w \left(2 q^2-3 w^2\right)-{\cal L}_1-81 \sqrt{3 \pi } {g_s} {N_f}^{5/3} {r_h}
   w^2 \left(529 q^2+36 w^2\right) \sqrt[3]{-\log (r)}\right)}{8 \sqrt{\frac{1}{N}}
   \sqrt[3]{{N_f}} {r_h} w (r-{r_h})^2 \left(-81 \sqrt{{g_s}}
   {N_f}^{4/3} {r_h} \left(529 q^2+36 w^2\right) (-\log (r))^{4/3}+33856 i
   2^{2/3} (3 \pi )^{5/6} \sqrt{\frac{1}{N}} q^2 w\right)};\nonumber\\
& & m(r) = -\frac{1}{8464 \left(\frac{1}{N}\right)^{3/2} {r_h}^2
   w^2 \left(36 w^2-529 q^2\right) (r-{r_h})^4 \log ^2(r) \left(8 \pi ^2 N-9
   {g_s}^2 M^2 {N_f} \log ^2(r)\right)}\Biggl\{\pi ^{5/3} \sqrt{{g_s}}\nonumber\\
& & \times \Biggl(-81 \sqrt[3]{2} 3^{2/3} \sqrt{{g_s}}
   \sqrt{N} {N_f}^{4/3} \left(1587 q^2-36 w^2\right) \left(36 w^2-529 q^2\right)
   (r-{r_h})^4 (-\log (r))^{10/3}\nonumber\\
& & +\frac{1}{33856\ 2^{2/3} (3 \pi )^{5/6} q^2 w+81 i {N_f}^{4/3}
   {r_h} \sqrt{{g_s} N} \left(529 q^2+36 w^2\right) (-\log
   (r))^{4/3}}\Biggl\{1058 \pi ^{5/6} N \log (r) \biggl(\sqrt{3
   \pi } \sqrt{{g_s}} w^2 \left(36 w^2-529 q^2\right)\nonumber\\
& & +\log (r) \left(\sqrt{3 \pi
   } \sqrt{{g_s}} \left(-1659 q^2 w^2+1058 q^4+108 w^4\right)+576 i
   \sqrt{\frac{1}{N}} {r_h} w^3 ({r_h}-r)\right)\biggr)\nonumber\\
& & \times \Biggl[\sqrt[3]{3} w
   \left(135424\ 2^{2/3} \pi ^{5/6} \sqrt{\frac{1}{N}} q^2 \left(2 q^2-3
   w^2\right)+81 i \sqrt[6]{3} \sqrt{{g_s}} {N_f}^{4/3} {r_h} w \left(529
   q^2+36 w^2\right) \sqrt[3]{-\log (r)}\right)\nonumber\\
& & +81 i \sqrt{3} \sqrt{{g_s}}
   {N_f}^{4/3} {r_h} \left(-1515 q^2 w^2+1058 q^4-108 w^4\right) (-\log
   (r))^{4/3}\Biggr]\Biggr\}+8954912 i \sqrt{3} \pi ^{5/6} q^2 {r_h} w \left(2 q^2-3 w^2\right)
   (r-{r_h}) \log ^2(r)\Biggr)\Biggr\},\nonumber\\
& & 
\end{eqnarray}
}
where ${\cal L}_1 \equiv 81 \sqrt{3
   \pi } {g_s} {N_f}^{5/3} {r_h} \left(-1515 q^2 w^2+1058 q^4-108
   w^4\right) (-\log (r))^{4/3}$. 
We will substitute the dispersion relation \cite{IITR-McGill-bulk-viscosity},
\begin{equation}
\label{wuptoqsquared}
w_3 = \left(\frac{1}{\sqrt{3}} + \alpha \frac{g_s M^2}{N} + \beta \alpha_\beta\right) q_3 + \left(-\frac{i}{6}+ \beta \frac{g_s M^2}{N} + \beta \beta_\beta\right)q^2_3,
\end{equation}
where $w_3\equiv\frac{w}{2\pi T}, q_3\equiv\frac{q}{2\pi T}$, into (\ref{Zs-EOM})-(\ref{l+m}).

Working up to ${\cal O}(q)$, one can show that $l(r) = l_1(r) q + {\cal O}(q^2),
\ m(r) = m_0(r) + q m_1(r)$ where:
{
\begin{eqnarray}
\label{l1m0m1}
& & l_1(r) = -\frac{i \sqrt{3 \pi } \sqrt{{g_s}} \sqrt{\frac{1}{N}} \left(3 \alpha ^2
   {g_s}^2 M^4+2 \sqrt{3} \alpha  {g_s} M^2 N+N^2\right)}{8 {r_h}^2
   (r-{r_h}) \log ^2({r_h}) \left(3 \alpha  {g_s} M^2+\sqrt{3}
   N\right)}\nonumber\\
& & -\frac{i \sqrt{3 \pi } \sqrt{{g_s}} \sqrt{\frac{1}{N}} \left(\log
   ({r_h}) \left(-9 \alpha ^2 {g_s}^2 M^4-6 \sqrt{3} \alpha  {g_s} M^2
   N+3 N^2\right)-3 \alpha ^2 {g_s}^2 M^4-2 \sqrt{3} \alpha  {g_s} M^2
   N-N^2\right)}{8 {r_h} (r-{r_h})^2 \log ({r_h}) \left(3 \alpha 
   {g_s} M^2+\sqrt{3} N\right)}\nonumber\\
& & +\frac{i \sqrt{3 \pi } \sqrt{{g_s}}
   \sqrt{\frac{1}{N}} (\log ({r_h})+2) \left(3 \alpha ^2 {g_s}^2 M^4+2
   \sqrt{3} \alpha  {g_s} M^2 N+N^2\right)}{16 {r_h}^3 \log ^3({r_h})
   \left(3 \alpha  {g_s} M^2+\sqrt{3} N\right)};\nonumber\\
& &  m_0(r) = \frac{382725 \left(\frac{3}{2}\right)^{2/3} {g_s} N {N_f}^{4/3} (-\log
   ({r_h}))^{4/3}}{33856 \sqrt[3]{\pi } {r_h}^2}\nonumber\\
& & -\frac{729
   \left(\frac{3}{2}\right)^{2/3} {g_s}^2 M^2 {N_f}^{4/3} (-\log
   ({r_h}))^{4/3} \left(8464 \sqrt{3} \pi ^2 \alpha -4725 {g_s} {N_f}
   \log ^2({r_h})\right)}{270848 \pi ^{7/3} {r_h}^2};\nonumber\\
& &  m_1(r) = \frac{\sqrt[6]{\pi } \sqrt{{g_s}} \sqrt{N} \left(-1130679 \sqrt[3]{2} \sqrt[6]{3}
   {g_s}^2 M^2 {N_f}^{4/3} {r_h} \left(2 \sqrt{3} \beta +3 i \alpha
   \right) (-\log ({r_h}))^{10/3}-256 i \sqrt[3]{\pi }\right)}{132352 {r_h}^4
   \log ^2({r_h})}\nonumber\\
& & +\frac{27 {g_s}^2 M^2 \left(267289 i {g_s} {N_f}
   \log ^3({r_h})-2800 i \sqrt{3} \pi ^2 \alpha \right)}{4276624 \pi ^{3/2}
   {r_h}^2 \sqrt{{g_s} N} (r-{r_h})^2 \log ^2({r_h})}\nonumber\\
& & -\frac{27
   {g_s}^2 M^2 \left(-2800 i \sqrt{3} \pi ^2 \alpha  \log ({r_h})-5600 i
   \sqrt{3} \pi ^2 \alpha +267289 i {g_s} {N_f} \log
   ^4({r_h})\right)}{8553248 \pi ^{3/2} {r_h}^3 \sqrt{{g_s} N}
   (r-{r_h}) \log ^3({r_h})}\nonumber\\
& & -\frac{3 i \sqrt{\pi } \sqrt{{g_s} N}}{517
   {r_h}^2 (r-{r_h})^2 \log ^2({r_h})}+\frac{3 i \sqrt{\pi }
   \sqrt{{g_s} N}}{1034 {r_h}^3 (r-{r_h}) \log ^2({r_h})}+\frac{3 i
   \sqrt{\pi } \sqrt{{g_s} N}}{4 {r_h} (r-{r_h})^3}.\nonumber\\
& &
\end{eqnarray}
}
As $r=r_h$ turns out to be an irregular singular point, make an ansatz: $Z_s(r) = e^{S(r)}: |S'(r)|^2>S''(r)$ near $r=r_h$. Hence, $S'(r) = \frac{m(r)\pm\sqrt{(m(r))^2 + 4l(r)}}{2} = m_0+\frac{\left(l_1+m_0m_1\right)}{m_0}q+{\cal O}(q^2), \frac{l_1(r)}{m_0(r)}q + {\cal O}(q^2)$. We will consider the first root. Thus,
{\footnotesize
\begin{eqnarray}
\label{root1-i}
& & \hskip -0.8in S'(r) = \frac{3i\sqrt{\pi g_s N}q}{4(r-r_h)^3r_h} -\frac{3 i \sqrt{\pi } \sqrt{N} q \sqrt{{g_s} N}}{517 \sqrt{N} {r_h}^2
   (r-{r_h})^2 \log ^2({r_h})}\nonumber\\
& & \hskip -0.8in -\frac{i  q \left(0.09
   \sqrt{3} \pi ^2 \alpha  {g_s}^2 M^2 {N_f}^{4/3} \log
   ({r_h})-8.3 {g_s}^3 M^2 {N_f}^{7/3} \log
   ^4({r_h})-0.05\times 2^{2/3} \sqrt[3]{3} \pi ^{7/3} {r_h}^3 (-\log
   ({r_h}))^{5/3}-0.02\times 2^{2/3} \sqrt[3]{3} \pi ^{7/3}
   {r_h}^3 (-\log ({r_h}))^{2/3}\right)}{4.9\times \pi ^{3/2}
   \sqrt{{g_s}} {N_f}^{4/3}\sqrt{N} {r_h}^2 (r-{r_h})^2 \log
   ^3({r_h})}\nonumber\\
& &\hskip -0.8in +   \frac{3 i {g_s} q \left(16 \pi ^2 \log ({r_h}) \left(1575 \sqrt{3} \alpha 
   {g_s} M^2+517 N\right)+50400 \sqrt{3} \pi ^2 \alpha  {g_s} M^2-2405601
   {g_s}^2 M^2 {N_f} \log ^4({r_h})\right)}{8553248 \pi ^{3/2}
   {r_h}^3 \sqrt{{g_s} N} (r-{r_h}) \log ^3({r_h})}\nonumber\\
& & \hskip -0.8in \equiv \frac{\lambda_1}{r_h^3(r-r_h)} + \frac{\lambda_2}{r_h^2(r-r_h)^2} + \frac{\lambda_3}{r_h(r-r_h)^3}.
\end{eqnarray}
}
Near $N=100, g_s=0.1, N_f=3,$ \cite{IITR-McGill-bulk-viscosity}
\begin{equation}
\label{r_h}
r_h=e^{-\kappa_{r_h}N^{1/3}}, \kappa_{r_h}=0.3,
\end{equation}
 one notes that for a generic $\alpha={\cal O}(1)$, near $r=r_h, \frac{\lambda_1}{r_h^3(r-r_h)} \sim \frac{\lambda_2}{r_h^2(r-r_h)^2} \sim \frac{\lambda_3}{r_h(r-r_h)^3}$ for $M={\cal O}(1)$. Hence, $S'(r\sim r_h)\sim\frac{\lambda_1}{(r-r_h)}$, implying $r\sim r_h$ is a regular singular point, which is not meaningful due to the $\frac{1}{(r-r_h)^3}$ term in $m(r)$. We will hence, choose $\alpha$ for which the residue of the $\frac{1}{(r-r_h)}$-term, i.e., $\lambda_1=0$. This is effected via:
\begin{equation}
\label{alpha}
\alpha =  -\frac{517(g_s N_f)(16 N\pi^2\log r_h - 4653 g_s^2 M^2 N_f (\log r_h)^4)}{1,200 N\pi^2(2 + \log r_h)}.
\end{equation}

Given the decoupling of $S^1_t\times_w\mathbb{R}^3\times\mathbb{R}_{>0}$-metric and the unwarped $S^1_{\cal M}\times_w T^{\rm NE}$-metric, in (\ref{EOMs-beta0-beta}), as discussed in \cite{MChaos},:
\begin{eqnarray}
\label{NC-sufficient-i}
\left(g^{MN}\frac{\delta J_0}{\delta g^{MN}}\right)^{\beta^0} = \left(g^{\mu\nu}\frac{\delta J_0}{\delta g^{\mu\nu}}\right)^{\beta^0}
+ \left(g^{mn}\frac{\delta J_0}{\delta g^{mn}}\right)^{\beta^0},
\end{eqnarray}
where $\mu, \nu = v, x^{1, 2, 3}, r$ and $m, n = \theta_{1,2}, x/\phi_1, y/\phi_2, z/\psi, x^{10}$. Further, in the intermediate-$N$ MQGP limit (\ref{MQGP_limit}) restricted to (\ref{Ouyang-definition}), it was observed in \cite{MChaos} that:
\begin{eqnarray}
\label{NC-sufficient-iii}
& & \hskip -0.8in  \left.\left(g_{\mu\nu}g^{\tilde{m}\tilde{n}}\frac{\delta J_0}{\delta g^{\tilde{m}\tilde{n}}}\right)^{\beta^0}\right|_{(\ref{MQGP_limit})\cap (\ref{Ouyang-definition}) } <  \left.\left(g_{\mu\nu}g^{\tilde{\mu}\tilde{\nu}}\frac{\delta J_0}{\delta g^{\tilde{\mu}\tilde{\nu}}}\right)^{\beta^0}\right|_{(\ref{MQGP_limit})\cap (\ref{Ouyang-definition}) } < \left.\left(\frac{\delta J_0}{\delta g^{\mu\nu}}\right)^{\beta^0}\right|_{(\ref{MQGP_limit})\cap (\ref{Ouyang-definition}) }.
\end{eqnarray}
To estimate the contribution of the ${\cal O}(R^4)$ terms to the equations of motion of the metric fluctuations $h^\beta_{\mu\nu}$ from (\ref{EOMs-beta0-beta}), as in \cite{MChaos}, one will hence consider only the linear fluctuations in $g_{\mu\nu}g^{\tilde{\mu}\tilde{\nu}}\frac{\delta J_0}{\delta g^{\tilde{\mu}\tilde{\nu}}} +
\frac{\delta J_0}{\delta g^{\mu\nu}}$. 

The equations of motion up to LO in $N$, ${\rm EOM}_{\mu\nu}, \mu,\nu=v, x^{1,2,3}, r$ for $h_{\mu\nu}$s  corresponding to the scalar modes of metric perturbations are:
{\scriptsize
\begin{eqnarray}
\label{EOMs-LON}
&& \hskip -0.8in {\bf EOM}_{x^1x^1}^\beta:\nonumber\\
& & \hskip -0.8in \frac{1}{99\ 3^{5/6} {N_f}^{2/3} {r_h} (-\log
   (r))^{2/3}}\Biggl\{2 i \sqrt[3]{2} \pi ^{2/3} \beta  q \Biggl(-93 {\alpha_\beta} {h^{\beta^0}_{x^1x^1}}(r)+12 {\alpha_\beta} {r_h}
   {h^\beta_{x^1x^1}}'(r)+156 {\alpha_\beta} {r_h} {h^{\beta^0}_{x^2x^2}}'(r) +12 {\alpha_\beta} {h^{\beta^0}_{x^2x^2}}(r)-12 {r_h}
   {h^\beta_{vx^1}} '(r)+27 {h^\beta_{vx^1}} (r)\nonumber\\
& & \hskip -0.8in -4 {r_h} {h^\beta_{x^1x^1}}'(r)+31 {h^\beta_{x^1x^1}}(r)-52 {r_h}
   {h^\beta_{x^2x^2}} '(r)-4 {h^\beta_{x^2x^2}} (r)\Biggr)\Biggr\} =\frac{23680 \sqrt[3]{\frac{2}{3}} \pi ^{8/3} \beta  (2 q {h^{\beta^0}_{vx^1}} (r)+15 {h^{\beta^0}_{vv}} (r))}{6561 {N_f}^{8/3} {r_h}^6 (-\log (r))^{8/3}};\nonumber\\
&& \hskip -0.8in {\bf EOM}_{x^2x^2}^\beta: \frac{1}{99\ 3^{5/6} {N_f}^{2/3} {r_h} (-\log
   ({r_h}))^{2/3}}\Biggl\{2 i \sqrt[3]{2} \pi ^{2/3} \beta  q \Biggl(78 {\alpha_\beta} {r_h} {h^{\beta^0}_{x^1x^1}}'(r)+6 {\alpha_\beta}
   {h^{\beta^0}_{x^1x^1}}(r)+90 {\alpha_\beta} {r_h} {h^{\beta^0}_{x^2x^2}}'(r)-87 {\alpha_\beta} {h^{\beta^0}_{x^2x^2}}(r)-78 {r_h}
   {h^\beta_{vx^1}} '(r)\nonumber\\
& & \hskip -0.8in +27 {h^\beta_{vx^1}} (r) -26 {r_h} {h^\beta_{x^1x^1}}'(r)-2 {h^\beta_{x^1x^1}}(r)-30 {r_h}
   {h^\beta_{x^2x^2}} '(r)+29 {h^\beta_{x^2x^2}} (r)\Biggr)\Biggr\}=\frac{4736 \sqrt[3]{\frac{2}{3}} \pi ^{8/3} \beta  ({r_h}-r) \left(2 q \log ^6({r_h}) {h^{\beta^0}_{vx^1}} (r)+9 {h^{\beta^0}_{vv}} (r) (-\log ({r_h}))^{2/3}\right)}{59049
   {N_f}^{8/3} {r_h}^7 (-\log ({r_h}))^{26/3}};\nonumber\\
& & \hskip -0.8in {\bf EOM}_{x^3x^3}^\beta: -\frac{1}{99\ 3^{5/6}
   {N_f}^{2/3}{r_h} (-\log ({r_h}))^{2/3}}\Biggl\{2 i \sqrt[3]{2} \pi ^{2/3} \beta  q \Biggl(78 {\alpha_\beta} {r_h}
   {h_{x^1x^1}^{\beta^0}}'(r)+6 {\alpha_\beta} {h_{x^1x^1}^{\beta^0}}(r)+90 {\alpha_\beta}
  {r_h} {h_{x^2x^2}^{\beta^0}}'(r)-87 {\alpha_\beta} {h_{x^2x^2}^{\beta^0}}(r)+78 {r_h}
   {h_{vx^1}^\beta}'(r)\nonumber\\
& & \hskip -0.8in -27 {h_{vx^1}^\beta}(r)+26 {r_h}
   {h_{x^1x^1}^\beta}'(r)+2{h_{x^1x^1}^\beta}(r)+30 {r_h}
   {h_{x^2x^2}^\beta}'(r)-29 {h_{x^2x^2}^\beta}(r)\Biggr)\Biggr\}=\frac{4736
   \sqrt[3]{\frac{2}{3}} \pi ^{8/3} (r-{r_h}) (3 {h_{vv}^{\beta^0}}(r)+2 q
   {h_{vx^1}^{\beta^0}}(r))}{59049 {N_f}^{8/3} {r_h}^7 (-\log
   ({r_h}))^{8/3}};\nonumber\\
&& \hskip -0.8in {\bf EOM}_{rr}^\beta:
\frac{\pi ^{7/6} \sqrt{{g_s}} \sqrt{N} \left(12 q^2 {r_h}^2 {h^\beta_{rr}} (r)-12 {r_h}^2
   {h^\beta_{x^1x^1}}''(r)+49 {r_h} {h^\beta_{x^1x^1}}'(r)-52 {h^\beta_{x^1x^1}}(r)-24 {r_h}^2
   {h^\beta_{x^2x^2}} ''(r)+98 {r_h} {h^\beta_{x^2x^2}} '(r)-104 {h^\beta_{x^2x^2}} (r)\right)}{9\ 2^{2/3} \sqrt[3]{3}
   {N_f}^{2/3} {r_h}^4 (-\log ({r_h}))^{2/3}}\nonumber\\
& & \hskip -0.8in = -\frac{32 \sqrt[3]{\frac{2}{3}} \pi ^{8/3} \beta  \left({h^{\beta^0}_{vv}} (r) (484242 {r_h}-78 r) - 52 q (r - 1378{r_h})
   {h^{\beta^0}_{vx^1}} (r)+117936 {r_h}
   ({h^{\beta^0}_{x^2x^2}} (r)+ {h^{\beta^0}_{x^3x^3}}(r))\right)}{177147 {N_f}^{8/3} {r_h}^7 (-\log
   ({r_h}))^{8/3}};\nonumber\\
&& \hskip -0.8in {\bf EOM}_{vv}^\beta: \frac{1}{99\ 3^{5/6}
   (-\log r )^{2/3} {N_f}^{2/3} {r_h}}\Biggl\{2 i \sqrt[3]{2} \pi ^{2/3} \beta  q \Biggl(99 {\alpha_\beta}
   {h^{\beta^0}_{vv}}(r)+2 \Biggl[156 {\alpha_\beta} {h^{\beta^0}_{x^1x^1}}'(r)+33
   {\alpha_\beta} {h^{\beta^0}_{x^1x^1}}(r)+312 {\alpha_\beta}
   {h^{\beta^0}_{x^2x^2}}'(r)\nonumber\\
& & \hskip -0.8in -66 {\alpha_\beta} {h^{\beta^0}_{x^2x^2}}(r)+156
   {h^\beta_{vx^1}} '(r)-66 {h^\beta_{vx^1}} (r)+52 {h^\beta_{x^1x^1}}'(r)+11
   {h^\beta_{x^1x^1}}(r)+104 {h^\beta_{x^2x^2}} '(r)-22
   {h^\beta_{x^2x^2}} (r)\Biggr]+33 {h^\beta_{vv}} (r)\Biggr)\Biggr\}\nonumber\\
& & \hskip -0.8in=\frac{208 \sqrt[3]{2/3}
   \pi ^{8/3} \beta  \left(3 h_{vv}^{\beta^0}(r) + 2 q h_{vx^1}^{\beta^0}(r)\right)}{59049
   {N_f}^{8/3} {r_h}^6 (-\log ({r_h}))^{8/3}} ;\nonumber\\
& & \hskip -0.8in {\bf EOM}_{x^1r}^\beta: \frac{1}{9\ 3^{5/6} (-\log r )^{5/3}
   {N_f}^{2/3} {r_h}^3}\Biggl\{\left(\frac{\pi }{2}\right)^{2/3} \beta  \Biggl(24 i \sqrt{3 \pi }
   \sqrt{{g_s}} \log r  \sqrt{N} q {h^\beta_{x^1x^1}}(r)+24 i
   \sqrt{3 \pi } \sqrt{{g_s}} \log r  \sqrt{N} q {r_h}
   {h^\beta_{x^2x^2}} '(r)\nonumber\\
& & \hskip -0.8in +48 i \sqrt{3 \pi } \sqrt{{g_s}}
   (-\log r )^{2/3} \sqrt{N} q {h^\beta_{x^2x^2}} (r)+61 \log r 
   {r_h}^2 {h^\beta_{vx^1}} '(r)-12 \log r  {r_h}^3
   {h^\beta_{vx^1}} ''(r)+16 \log r  {r_h}
   {h^\beta_{vx^1}} (r)\Biggr)\Biggr\}\nonumber\\
& & \hskip -0.8in =-\frac{128 \sqrt[3]{2} \pi ^{19/6}
   {\alpha_\beta} \beta  \sqrt{{g_s}} \sqrt{N} q \left(95
   {h^{\beta^0}_{x^2x^2}}(r)+107 {h^{\beta^0}_{x^3x^3}}(r)\right)}{2187\ 3^{5/6} {N_f}^{8/3}
   {r_h}^8 (-\log ({r_h}))^{8/3}};\nonumber\\
& & \hskip -0.8in {\bf EOM}_{vr}^\beta: -\frac{1}{33
   {N_f}^{2/3} {r_h}^3 (-\log ({r_h}))^{2/3}}\Biggl\{2 i \sqrt[3]{\frac{2}{3}} \pi ^{7/6} \sqrt{{g_s}} \sqrt{N} q \Biggl(15 \alpha_\beta  \beta  {r_h}
   {h^{\beta^0}_{x^1x^1}}'(r)+24 \alpha_\beta  \beta  {h^{\beta^0}_{x^1x^1}}(r)+30 \alpha_\beta  \beta  {r_h} {h^{\beta^0}_{x^2x^2}}'(r)+48
   \alpha_\beta  \beta  {h^{\beta^0}_{x^2x^2}}(r)\nonumber\\
& & \hskip -0.8in+15 {r_h} {h^{\beta}_{vx^1}}'(r)+35 {h^{\beta}_{vx^1}}(r)+5 {r_h}
   {h^{\beta}_{x^1x^1}}'(r)+8 {h^{\beta}_{x^1x^1}}(r)+10 {r_h} {h^{\beta}_{x^2x^2}}'(r)+16 {h^{\beta}_{x^2x^2}}(r)\Biggr)\Biggr\}=\frac{3848 \sqrt[3]{2} \pi ^{19/6}\sqrt{{g_s}} \sqrt{N} \beta  \left(3 h_{vv}^{\beta^0}(r) + 2 q h_{vx^1}^{\beta^0}(r)\right)}{59049\ 3^{5/6} (-{\log r_h})^{8/3}
   {N_f}^{8/3} {r_h}^8};\nonumber\\   
& & \hskip -0.8in {\bf EOM}_{vx^1}^\beta: \frac{i \beta  q \left({r_h} \left(8400 \pi ^2 \alpha_\beta  N {h^{\beta^0}_{vx^1}}'(r)+{h^{\beta}_{vx^1}}'(r) \left(267289
   \sqrt{3} {g_s}^2 M^2 {N_f} \log ^3({r_h})+2800 \pi ^2 N\right)+8400 \pi ^2 N
   {h^{\beta}_{vv}}'(r)\right)-25200 \pi ^2 N {h^{\beta}_{vv}}(r)\right)}{6300\ 2^{2/3} 3^{5/6} \pi ^{4/3} N {N_f}^{2/3}
   {r_h} (-\log ({r_h}))^{2/3}}\nonumber\\
& & \hskip -0.8in   =-\frac{512 \sqrt[3]{\frac{2}{3}} \pi ^{8/3} \beta  q \left({h^{\beta^0}_{x^2x^2}}(r) (97
   {r_h}-2 r)+{h^{\beta^0}_{x^3x^3}}(r) (10 r+97 {r_h})\right)}{19683 {N_f}^{8/3} {r_h}^7 (-\log
   ({r_h}))^{8/3}}\nonumber\\                     
\end{eqnarray}
}
Up to LO in $|\log r_h|$, using (\ref{r_h}), the left hand sides of (\ref{EOMs-LON}) are negligible as compared to the right hand sides, and the latter yield:
{\footnotesize
\begin{eqnarray}
\label{EOMs-LON-LOlogrh}
& & {\bf EOM}_{x^1x^1}: 2 q {h^{\beta^0}_{vx^1}} (r)+15 {h^{\beta^0}_{vv}} (r) = q{\cal O}\left(r_h^5 \left(-\log r_h\right)^2\right);\nonumber\\
& & {\bf EOM}_{x^2x^2}: 2 q \log ^6({r_h}) {h^{\beta^0}_{vx^1}} (r)+9 {h^{\beta^0}_{vv}} (r) (-\log ({r_h}))^{2/3} =q {\cal O}\left(r_h^5 \left(-\log r_h\right)^2\right);\nonumber\\
& & {\bf EOM}_{x^3x^3}: 3 {h_{vv}^{\beta^0}}(r)+2 q
   {h_{vx^1}^{\beta^0}}(r) = q {\cal O}\left(r_h^5 \left(-\log r_h\right)^2\right);\nonumber\\
& & {\bf EOM}_{rr}: {h^{\beta^0}_{vv}} (r) (484242 {r_h}-78 r) - 52 q (r - 1378{r_h})
   {h^{\beta^0}_{vx^1}} (r)+117936 {r_h}
   ({h^{\beta^0}_{x^2x^2}} (r)+ {h^{\beta^0}_{x^3x^3}}(r)) = \sqrt{N}{\cal O}\left(r_h^2 \left(-\log r_h\right)^2\right);\nonumber\\
& & {\bf EOM}_{vv/vr}: 3 h_{vv}^{\beta^0}(r) + 2 q h_{vx^1}^{\beta^0}(r) = q {\cal O}\left(r_h^5 \left(-\log r_h\right)^2\right);\nonumber\\
   & & {\bf EOM}_{x^1r}: 95
   {h^{\beta^0}_{x^2x^2}}(r)+107 {h^{\beta^0}_{x^3x^3}}(r) = q{\cal O}\left(r_h^5 \left(-\log r_h\right)^2\right);\nonumber\\
& & {\bf EOM}_{vx^1}: {h^{\beta^0}_{x^2x^2}}(r) (97
   {r_h}-2 r)+{h^{\beta^0}_{x^3x^3}}(r) (10 r+97 {r_h}) = q{\cal O}\left(r_h^5 \left(-\log r_h\right)^2\right) ;\nonumber\\      
\end{eqnarray}
}
One therefore sees that in the $q\rightarrow0$ limit, (\ref{EOMs-LON-LOlogrh}) is solved for
\begin{eqnarray}
\label{solution-alphabeta}
& & h^{\beta^0}_{vv}(r\sim r_h) \sim r^{\mu_{vv}+5}\left(-\log r\right)^{\nu_{vv}};\nonumber\\
& &  h^{\beta^0}_{x^2x^2}(r\sim r_h), h^{\beta^0}_{x^3x^3}(r\sim r_h) \sim r^{\mu_{TT}+1}\left(-\log r\right)^{\nu_{TT}},\nonumber\\
& & \mu_{vv, RR},\ \nu_{vv, T}\in\mathbb{R},
\end{eqnarray}
$T\equiv x^2/x^3, \forall\alpha_\beta={\cal O}(1)$. The values of $\mu_{vv,TT}, \nu_{vv,TT}, h_{x^1x^1}^{\beta^0}, h_{vx^1}^{\beta^0}$ are determined from the EOMs for $h^{\beta^0}_{\mu\nu}$. 
Thus,
\begin{eqnarray}
\label{1over3-cssq}
& & \frac{1}{3} - c_s^2 =  \frac{517\left(g_s M^2\right)(g_s N_f)(16 N\pi^2\log r_h - 4653 g_s^2 M^2 N_f (\log r_h)^4)}{113,400N\pi^2(2 + \log r_h)} - \kappa_{\alpha_\beta}\beta,\nonumber\\
& &  |\kappa_{\alpha_\beta}|\equiv{\cal O}(1),
\end{eqnarray}
implying thereby (dropping terms of ${\cal O}\left(\frac{1}{N^2}\right),\ {\cal O}(\beta^2)$,
\begin{eqnarray}
\label{sq-1over3-cssq}
& & \left(\frac{1}{3} - c_s^2\right)^2\sim \kappa_{\alpha_\beta}\frac{517\left(g_s M^2\right)(g_s N_f)(16 N\pi^2\log r_h - 4653 g_s^2 M^2 N_f (\log r_h)^4)}{113,400N\pi^2(2 + \log r_h)}\beta.
\end{eqnarray}

Using the hierarchy in eleven-dimensional supergravity action as regards the $\mathcal{O}(R^4)$ corrections in it, we solved the equations of motion corresponding the scalar modes of the metric perturbation and obtained the general solution under the limit $q\rightarrow0$, which provided us the required relationship between  $\Biggl(\frac{1}{3}-c_s^2\Biggr)$ and $\Biggl(\frac{1}{3}-c_s^2\Biggr)^2$ in terms of $g_s$, $M$, $N$,  $N_f$, and $r_h$ upto $\mathcal{O}(R^{4})$.

\section{Spectral function computation of $\zeta$ at intermediate coupling}
\label{spectral}

In this section we motivate the computation of $\frac{\zeta}{\eta}$ from a computation of $\zeta$, but from a retarded current-current correlation function instead of the conventional retarded EM tensor-EM tensor correlation function. Let us first justify the same. For starters,  $\frac{\zeta}{s} = {\cal O}\left({g_sM^2\over N}\right)$, and $\frac{\eta}{s}$  is dominated by the conformal result plus an ${\cal O}\left({g_sM^2\over N}\right)$  correction term \cite{IITR-McGill-bulk-viscosity}. This means up to ${\cal O}\left({1\over N}\right)$  the ratio $\zeta/\eta$ would mimic $\zeta/s$.

The gauge and the metric perturbations should, in principle, be considered simultaneously as the same will get coupled once we consider the bulk and boundary actions together \cite{bulk+gauge_KS+AM}. Let us explicitly show that in our type IIA dual of QCD-like theories. In $S_{\rm IIA\ SUGRA}$, in the intermediate-$N$ MQGP limit (\ref{MQGP_limit}), one can show: $e^{-2\phi^{\rm IIA}}R>e^{-2\phi^{\rm IIA}}\partial^M\phi^{\rm IIA}\partial_M\phi^{\rm IIA}, e^{-2\phi^{\rm IIA}}H_{MNP}H^{MNP},$ $F_{MN}^{\rm IIA}F^{MN\ {\rm IIA}}$; one can not obtain $A_3^{\rm IIA}$ from SYZ dual of the type-IIB background  \cite{MQGP}. At ${\cal O}(R^4)$, and in the UV, $J_0 > (\nabla^2\phi^{\rm IIA})Q$  where $ Q = \frac{1}{12(2\pi)^3}\left(R_{IJ}^{\ \ \ KL}R_{KL}^{\ \ \ MN}R_{MN}^{\ \ \ \ IJ}\\ - 2 R_{I\ \ J}^{\ \ K\ \ L}R_{K\ \ L}^{\ \ M\ \ \ N} R_{M\ \ N}^{\ \ \ I\ \ \ J}\right)$ \cite{Q_Beckers+Louis+Haack}
\footnote{One can show that $\left.\nabla^2\phi\right|_{(\ref{alpha_theta_12}), r\in\rm UV} = \kappa_{\beta^0}\frac{N^{7/10}\csc^2\left(\frac{\theta_1}{2}\right)\alpha_{\theta_2}^2}{g_s^2M_{\rm UV}^2N_f^{\rm UV}\alpha_{\theta_1}^2\left(\log r\right)^2} + 
\beta\kappa_\beta\frac{\sqrt{g_s}M_{\rm UV}(6a^2+r_h^2)\Sigma_1r^3}{r_h^4(9a^2+r_h^2)N^{7/4}(\log N)^4\alpha_{\theta_2}^3}
\sim\kappa_{\beta^0}\frac{N^{7/10}\csc^2\left(\frac{\theta_1}{2}\right)\alpha_{\theta_2}^2}{g_s^2M_{\rm UV}^2N_f^{\rm UV}\alpha_{\theta_1}^2\left(\log r\right)^2},$ wherein $M_{\rm UV}\sim\frac{1}{\log N}$ \cite{Gopal+Vikas+Aalok} and $\left.\log r\right|_{r\in{\rm UV}}\sim\log N$
 (as $r_{\rm UV}\stackrel{<}{\sim}(g_s N)^{1/4})$. Further, one can show that $\left.R_{IJ}^{\ \ \ KL}R_{KL}^{\ \ \ MN}R_{MN}^{\ \ \ \ IJ}\right|_{r\in{\rm UV}}\sim\frac{1}{L^6}$ where $L\sim \left(g_s N\right)^{1/4}$, and $\left.R_{I\ \ J}^{\ \ K\ \ L}R_{K\ \ L}^{\ \ M\ \ \ N} R_{M\ \ N}^{\ \ \ I\ \ \ J}\right|_{r\in{\rm UV}}\sim\frac{10^{-12}N^{18/5}}{g_s^{12}M_{\rm UV}^6N_{f,\ {\rm UV}}^6(\log r)^{18}}\sim\frac{10^{-12}N^{18/5}}{g_s^{12}M_{\rm UV}^6N_{f,\ {\rm UV}}^6(\log N)^{18}},$ which will consider to be negligible. Now, in the UV,  $J_0(r\in{\rm UV})\sim\left.\frac{a^{14}M_{\rm UV}\Sigma_1(\theta_1,\theta_2)}{r^{32}\sqrt{g_s}N^{7/4}(\log N)^{5/3}N_{f,\ {\rm UV}}^{5/3}\alpha_{\theta_2}^3}\right|_{r_{\rm UV}\sim \xi N ^{1/4},\ 0<\xi<1}$ \cite{OR4} (From (\ref{J0-1}), we see that up to LO-in-$N$, the ${\cal M}$-theory circle coordinate $x^{10}$ does not appear in the Lorentz indices in the four Riemann curvature tensor components;  the ${\cal M}$-theory background is also independent of $x^{10}$. Near the $\psi=2n\pi, n=0, 1, 2$-coordinate patches and in the MQGP limit (\ref{MQGP_limit}), $G^{\cal M}_{x^{10}M}=0, M\neq x^{10}$. Hence, the $D=11$ $J_0$ term upon dimensional reduction to $D=10$ will generate the $D=10$ $J_0$ term with the understanding that $\frac{\sqrt{G^{\cal M}_{x^{10}x^{10}}}}{\kappa_{11}^2}=\frac{1}{\kappa_{10}^2}$.). For $N=100, \xi=0.3, (\nabla^2\phi^{\rm IIA}Q/J_0)<1$ for $N_f^{\rm UV}\sim g_s^{21/4}e^{-3 N^{1/3}}
= g_s^{21/4} r_h^{10}$. Hence, we consider only the $D=10$ $J_0$ term at ${\cal O}(R^4)$ in this paper.}. 

The first-order metric ($h_{\mu\nu }$)/gauge fluctuations ($\tilde{A}_\mu$) of the EOMs arising from the bulk and DBI action (in $2\pi\alpha^\prime=1$-units): 
{\footnotesize
\begin{equation}
\label{Bulk+Boundary actions}
\hskip -0.5in \int_{S^1\times_w S^2}\int_{\left(S^1_t\times_w\mathbb{R}^3 \right)\times_w S^2\times\mathbb{R}_{\geq0}}\left[e^{-2\phi^{\rm IIA}}\sqrt{-g}(R + \beta J_0)+ T_{D6}e^{-\phi^{\rm IIA}}\sqrt{-{\rm det}\left(i^*(g + B) + F\right)}\delta\left(\tilde{x}\right)\delta\left(\tilde{z}-{\cal C}_{\tilde{z}}\right)\delta\left(\theta_1^{\rm Ouyang}\right)\right],
\end{equation}
}  
$T_{D_6}$ being the $D6$-brane tension, are now discussed. At ${\cal O}(\beta^0)$, one then obtains the following EOMs for  $h_{\mu\nu}^{\beta^0}$ coupled to gauge field fluctuations $\tilde{A}_\mu^{\beta^0}$ (that always appears as $\partial_Z\tilde{A}_\mu$; $h_s = h_{x^2x^2} + h_{x^3x^3} = 2 h_{x^2x^2}, \vartheta_{\mu\nu} = - \delta^{x^1}_\mu\delta^{x^1}_\nu$
):
\begin{eqnarray}
\label{Atilde-H-i}
& & \hskip -0.5in \underbrace{\sum_{\mu, \nu = x^{1,2,3}}f_{\mu\nu}(g_s, M_{\rm UV}, N_f^{\rm UV} N) r_h^{m^{\mu\nu}_{r_h}} q^{m^{\mu\nu}_q} \omega\ ^{m^{\mu\nu}_\omega} \left(|\log r_h|\right)^{\mu^{\mu\nu}_{\log r_h}}h_{\mu\nu}^{(m_{\mu\nu}),\ \beta^0}}_{{\rm From}\ \sqrt{-g}e^{-2\phi^{\rm IIA}}R} \nonumber\\
& & \hskip -0.5in  (m^{\mu\nu}_{r_h} - m_{\mu\nu} + m^{\mu\nu}_q + m^{\mu\nu}_\omega >0;\nonumber\\
& & \hskip -0.5in m^{\mu\nu}_{r_h}=0/1/2/3, m_{\mu\nu}=0/1/2, m^{\mu\nu}_q=0/1/2, m^{\mu\nu}_\omega=0/1/2)\nonumber\\
& & \hskip -0.5in +\sum_{\mu, \nu = x^{1,2,3}} {\cal F}_{\mu\nu}^{(1)}(N, M_{\rm UV}, N_f^{\rm UV} g_s; Z_{\rm UV}; \langle F_{Zt} \rangle)(h_{x^1x^1}^{\beta^0} + \vartheta_{\mu\nu} h_s^{\beta^0}) \nonumber\\
& & \hskip -0.5in \underbrace{+ \sum_{\mu, \nu = x^{1,2,3}}{\cal F}^{(2)}_{\mu\nu}(N, M_{\rm UV}, N_f^{\rm UV} g_s; Z_{\rm UV}; \langle F_{Zt} \rangle)\partial_Z\tilde{A}_t\ ^{\beta^0}}_{{\rm from}\ e^{-\phi^{\rm IIA}}\sqrt{-{\rm det}(i^*(g + B)^{\rm IIA} + F)}}=0,
\end{eqnarray}
where $M_{\rm UV} = M_{\rm eff}(r\in{\rm UV}), N_f^{\rm UV} = N_f^{\rm UV}(r\in{\rm UV})$ and both being vanishingly small due to equal number of $D5/\overline{D5}$-branes and $D7/\overline{D7}$-branes in the UV; $Z_{\rm UV}= \frac{\log N}{{\cal O}(1)}$. At ${\cal O}(\beta)$, using (\ref{EOMs-LON})-(\ref{EOMs-LON-LOlogrh}), the metric fluctuations'$h_{\mu\nu}^\beta$ EOMs that couple to  
gauge-field fluctuations $\tilde{A}_t^\beta$:
\begin{eqnarray}
\label{Atilde-H-ii}
& & \hskip -0.5in \underbrace{\sum_{\mu, \nu = x^{1, 2, 3}}f^{\beta,\ \sqrt{-g}R}_{\mu\nu}(g_s, M_{\rm UV}, N_f^{\rm UV} N) r_h^{m^{\mu\nu,\ \beta}_{r_h}}  \left(|\log r_h|\right)^{\mu^{\mu\nu,\ \beta}_{\log r_h}}h_{\mu\nu}^{(m_{\mu\nu}),\ \beta}}_{{\rm From}\ \sqrt{-g}e^{-2\phi^{\rm IIA}}R} \nonumber\\
& & \hskip -0.5in  + \underbrace{\sum_{\mu, \nu = x^{1, 2, 3}}f^{\beta,\ \sqrt{-g}J_0}_{\mu\nu}(g_s, M_{\rm UV}, N_f^{\rm UV} N) r_h^{m^{\mu\nu,\ \beta^0}_{r_h}}  \left(|\log r_h|\right)^{\mu^{\mu\nu,\ \beta^0}_{\log r_h}}h_{\mu\nu}^{(m_{\mu\nu}),\ \beta^0}}_{{\rm From}\ \sqrt{-g}e^{-2\phi^{\rm IIA}}J_0}\nonumber\\
& & \hskip -0.5in +\sum_{\mu, \nu = x^{1, 2, 3}} {\cal F}_{\mu\nu}^{(1)}(N, M_{\rm UV}, N_f^{\rm UV} g_s; Z_{\rm UV}; \langle F_{Zt} \rangle)(h_{x^1x^1}^{\beta} + \vartheta_{\mu\nu} h_s^{\beta}) \nonumber\\
& & \hskip -0.5in \underbrace{+ \sum_{\mu, \nu = x^{1, 2, 3}}{\cal F}^{(2)}_{\mu\nu}(N, M_{\rm UV}, N_f^{\rm UV} g_s; Z_{\rm UV}; \langle F_{Zt} \rangle)\partial_Z\tilde{A}_t\ ^{\beta}}_{{\rm from}\ e^{-\phi^{\rm IIA}}\sqrt{-{\rm det}(i^*(g + B)^{\rm IIA} + F)}}=0 
\end{eqnarray}
where $m^{\mu\nu,\ \beta^0}_{r_h}\in\mathbb{Z}_-, \mu^{\mu\nu,\ \beta^0}_{\log r_h}<0$; $m^{\mu\nu,\ \beta}_{r_h}\in\mathbb{Z}, \mu^{\mu\nu,\ \beta}_{\log r_h}>0: |m^{\mu\nu,\ \beta^0}_{r_h}|>|m^{\mu\nu,\ \beta}_{r_h}|$.


At ${\cal O}(\beta^0)$, the EOMs for gauge-field fluctuations arising from the first-order fluctuations in the metric and gauge fields in the DBI action for the flavor $D6$-branes:
\begin{eqnarray}
\label{Atilde-H-iii}
& & \hskip -0.5in \omega {\cal L}_1^{\beta^0}(g_s, M_{\rm UV}, N_f^{\rm UV} N; Z_{\rm UV}; r_h; \langle F_{Zt}\rangle)(h_{x^1x^1}^{\beta^0} + h_s^{\beta^0})\nonumber\\
& & \hskip -0.5in + q {\cal L}_2^{\beta^0}(g_s, M_{\rm UV}, N_f^{\rm UV} N; Z_{\rm UV}; r_h; \langle F_{Zt}\rangle)\partial_Z\tilde{A}_{x^1}^{\beta^0}+ \omega {\cal L}_3^{\beta^0}(g_s, M_{\rm UV}, N_f^{\rm UV} N; Z; r_h; \langle F_{Zt}\rangle) \partial_Z\tilde{A}_{t}^{\beta^0} =0;\nonumber\\
& & \hskip -0.5in {\cal L}_4^{\beta^0}(g_s, M_{\rm UV}, N_f^{\rm UV} N; Z_{\rm UV}; r_h; \langle F_{Zt}\rangle)(q\tilde{A}_{t}^{\beta^0} + \omega  \tilde{A}_{x^1}^{\beta^0}) + {\cal L}_5^{\beta^0}(g_s, M_{\rm UV}, N_f^{\rm UV} N; Z_{\rm UV}; r_h; \langle F_{Zt}\rangle)(h_{x^1x^1}^{\beta^0} + h_s^{\beta^0}) = 0;\nonumber\\
& & \hskip -0.5in  \omega q\tilde{A}_{t}^{\beta^0} {\cal L}_6^{\beta^0}(g_s, M_{\rm UV}, N_f^{\rm UV} N; Z_{\rm UV}; r_h; \langle F_{Zt}\rangle)
+ \omega^2  \tilde{A}_{x^1}^{\beta^0}{\cal L}_7^{\beta^0}(g_s, M_{\rm UV}, N_f^{\rm UV} N; Z_{\rm UV}; r_h; \langle F_{Zt}\rangle)\nonumber\\
& & \hskip -0.5in  + {\cal L}_8^{\beta^0}(g_s, M_{\rm UV}, N_f^{\rm UV} N; Z_{\rm UV}; r_h; \langle F_{Zt}\rangle; \langle \partial_Z F_{Zt}\rangle)\partial_Z\tilde{A}_{x^1}^{\beta^0} + {\cal L}_9^{\beta^0}(g_s, M_{\rm UV}, N_f^{\rm UV} N; Z_{\rm UV}; r_h; \langle F_{Zt}\rangle)\partial_Z^2\tilde{A}_{x^1}^{\beta^0}=0;\nonumber\\
& & \hskip -0.5in {\cal L}_{10}^{\beta^0}(g_s, M_{\rm UV}, N_f^{\rm UV} N; Z_{\rm UV}; r_h; \langle F_{Zt}\rangle)\tilde{A}_{x^1}^{\beta^0} +  {\cal L}_{11}^{\beta^0}(g_s, M_{\rm UV}, N_f^{\rm UV} N; Z_{\rm UV}; r_h; \langle F_{Zt}\rangle; \langle \partial_Z F_{Zt}\rangle)\partial_Z\tilde{A}_{x^1}^{\beta^0} = 0.
\end{eqnarray}

Similarly, at ${\cal O}(\beta)$:
\begin{eqnarray}
\label{Atilde-H-iv}
& & \hskip -0.5in q {\cal L}_1^{\beta}(g_s, M_{\rm UV}, N_f^{\rm UV}, N; Z; r_h; \langle F_{Zt}\rangle)\partial_Z\tilde{A}_{x^1}^{\beta}+ \omega {\cal L}_2^{\beta}(g_s, M_{\rm UV}, N_f^{\rm UV} N; Z; r_h; \langle F_{Zt}\rangle) \partial_Z\tilde{A}_{t}^{\beta}\nonumber\\
& & \hskip -0.5in + {\cal L}_3^\beta(g_s, M_{\rm UV}, N_f^{\rm UV}, N; Z; r_h; \langle F_{Zt}\rangle) (h_{x^1x^1}^{\beta} + h_s^{\beta})  = 0;\nonumber\\
& & \hskip -0.5in (q \tilde{A}_t^\beta + \omega \tilde{A}_{x^1}^\beta){\cal L}_4^{\beta}(g_s, M_{\rm UV}, N_f^{\rm UV} N; Z; r_h; \langle F_{Zt}\rangle) + (h_{x^1x^1}^\beta + h_s^\beta){\cal L}_5^{\beta}(g_s, M_{\rm UV}, N_f^{\rm UV} N; Z; r_h; \langle F_{Zt}\rangle) = 0;\nonumber\\
& &\hskip -0.5in  (q \tilde{A}_t^\beta + \omega \tilde{A}_{x^1}^\beta) = 0;\nonumber\\
& & \hskip -0.5in \tilde{A}_{x^1}^{\beta}{\cal L}_6^{\beta}(g_s, M_{\rm UV}, N_f^{\rm UV} N; Z_{\rm UV}; r_h; \langle F_{Zt}\rangle) + \partial_Z\tilde{A}_{x^1}^\beta{\cal L}_7^\beta(g_s, M_{\rm UV}, N_f^{\rm UV} N; Z_{\rm UV}; r_h; \langle F_{Zt}\rangle) = 0.
\end{eqnarray}
Solving (\ref{Bulk+Boundary actions}) - (\ref{Atilde-H-iv}), one obtains:
\begin{equation}
\label{Atilde-H-v}
H_{\mu\nu} = \chi^\lambda_{\mu\nu}(g_s, M_{\rm UV}, N_f^{\rm UV}, N; \langle F_{Zt} \rangle; \langle \partial_Z F_{Zt} \rangle; Z_{\rm UV})\tilde{A}_\lambda,
\end{equation}
which can be inverted to yield
\begin{eqnarray}
\label{Atilde-H-vi}
& & \tilde{A}_\lambda = \tilde{\chi}_{\lambda,\mu\nu}(g_s, M_{\rm UV}, N_f^{\rm UV}, N; \langle F_{Zt} \rangle; \langle \partial_Z F_{Zt} \rangle; Z_{\rm UV})H_{\mu\nu}\nonumber\\
& &  \Rightarrow \frac{\delta}{\delta H_{\mu\nu}} = \tilde{\chi}_{\lambda,\mu\nu}(g_s, M_{\rm UV}, N_f^{\rm UV}, N; \langle F_{Zt} \rangle; \langle \partial_Z F_{Zt} \rangle; Z_{\rm UV})\frac{\delta}{\delta\tilde{A}_\lambda}.
\end{eqnarray}
Now, $S_{\rm on-shell}(u(\equiv \frac{r_h}{r})\sim0)\sim\lim_{u\rightarrow0}\int dw dq A(w, q) Z_s'(u,q)Z_s(u,q)$. Writing $Z_s(u) = \xi_{\mu\nu}H_{\mu\nu}(u)$, one hence obtains: $S_{\rm on-shell}(u\sim0)=\lim_{u\rightarrow0}\int dw dq A(w, q)\xi_{\mu\nu}\xi_{\tilde{\mu}\tilde{\nu}}H'_{\mu\nu}(u)H_{\tilde{\mu}\tilde{\nu}}(u)$. Further, $G_{T_{\mu\nu}, T_{\rho\lambda}}\sim\left.\frac{\delta^2S_{\rm on-shell}}{\delta H_{\mu\nu}(u, q)\delta H_{\rho\lambda}(u, -q)}\right|_{u=0}$. One hence sees that:
\begin{equation}
\label{TT-JJ}
G_{T_{\mu\nu}T_{\rho\lambda}} = \tilde{\chi}_{\alpha,\mu\nu}\tilde{\chi}_{\beta,\rho\lambda}G_{J_\alpha J_\beta}.
\end{equation}
One therefore sees that $h_{x^1x^1}$ or equivalently $h_s$ are related to gauge-field fluctuations $\tilde{A}_{t, x^1}/\tilde{E}_{x^1}$ up to ${\cal O}(\beta)$. Hence, to obtain $\zeta$, one will equivalently be looking at gauge/electric-field fluctuations.

In the remainder of this section, for simplicity, we consider only gauge fluctuations, which based on the prior discussion in this section, is hence expected to capture the essential features of $\zeta$ and hence $\frac{\zeta}{\eta}$ (with the inclusion of higher derivative corrections in the ${\cal M}$-theory uplift) at intermediate coupling. Via the DBI action for $D6$-brane working in the gauge $A_{Z}=0$ with $A_{t}$ as the only non-zero component of the gauge field, we establish the required Sch\"odinger like equation for the perturbed gauge fields $\tilde{A_{\mu}}$ which in the limit $q\rightarrow 0$, can be redefined as a single variable $\tilde{E_{T}}, T\equiv$transverse. Calculating the respective Green function will provide the required spectral function relevant to the computation of bulk viscosity using Kubo's formula. The results obtained in this section of validated by a more rigorous computation of $\frac{\zeta}{\eta}$ in \ref{Eling+Oz} using the results of \cite{EO}.

The DBI action for the $D6$ brane is given as:
\begin{equation}\label{D6DBI}
S_{D6}=-T_{D6}N_f\int d^{7}\xi~ e^{-\phi_{IIA}}\sqrt{\det{i^*\{g+B+F\}}},
\end{equation}
with $2\pi\alpha^{\prime}=1$ and $i:\Sigma_{D6}\hookrightarrow M_{10}$ being the embedding of the $D$-brane world volume in the ten-dimensional type IIA gravity dual involing a non-K\''{a}ler resolved conifold. Here the worldvolume directions of the $D6$ brane are denoted by the coordinates: $\{t,x_1,x_2,x_3,Z,\theta_2,\tilde{y}\}$, with $\{t,x_1,x_2,x_3\}$\footnote{Unlike section \tcb{1}, we will not be using the Eddington-Finkelstein coordinates in sections \tcb{3} and \tcb{4}.} as the usual Minkowski coordinate, $Z$ as the newly defined radial direction and two angular coordinate $\theta_2$, $\tilde{y}$; $Z$ is related the the usual radial coordinate $r$ as $r=r_{h}e^{Z}$ and $\tilde{y}$ is the local value for the angle $\phi_2$ (See sections {\bf 3} and {\bf 4} of \cite{Yadav+Misra+Sil-Mesons} for details).

In the above, $\phi_{IIA}$ is the type IIA dilaton which is the triple T-dual version of type IIB dilaton. The pullback metric and the pullback of the NS-NS $B$ field on the world volume of the $D6$ brane are denoted as $g$ and $B$ in (\ref{D6DBI}). $F$ is the field strength for a $U(1)$ gauge field $A_{\mu}$, where the only nonzero component of the same is the temporal component $A_{t}$. In the gauge $A_{Z}=0$, the only nonzero component of $F$ is $F_{Zt}=-F_{tZ}$. Combining together the symmetric $g$ field and the anti-symmetric $B$ field as $G=g+B$ and expanding the DBI action up to quadratic order in $A$, we get:
\begin{equation}\label{D6DBI2}
S_{D6}=\frac{T_{D6}N_f}{4}\int d^{7}\xi~ e^{-\phi_{IIA}}\sqrt{-G}\left(G^{\mu\alpha}G^{\beta\gamma}F_{\alpha\beta}F_{\gamma\mu}-\frac{1}{2}G^{\mu\nu}G^{\alpha\beta}F_{\mu\nu}F_{\alpha\beta}\right).
\end{equation}
The second term in (\ref{D6DBI2}), is coming because of the anti-symmetric $B$ field in $G$. As none of the fields in the DBI action depends on the angular coordinates $\tilde{y}$, the integrand in equation (\ref{D6DBI2}) is independent of the same. Also, we choose to work around a particular small value of $\theta_2$ in the large $N$ limit as given by: $\theta_2=\frac{\alpha}{N^{3/10}}$, for $\alpha\sim \mathcal{O}(1)$ constant. Hence, the upshot is that the integration over $\theta_2$ and $\tilde{y}$ is trivial and we denote $\Omega_{2}$ as the factor one get after the integration over $\theta_2$ and $\tilde{y}$, such that:
\begin{equation}\label{D6DBI3}
S_{D6}=\frac{T_{D6}}{4}\Omega_{2}\int d^{4}x~dZ~ e^{-\phi_{IIA}}\sqrt{-G}\left(G^{\mu\alpha}G^{\beta\gamma}F_{\alpha\beta}F_{\gamma\mu}-\frac{1}{2}G^{\mu\nu}G^{\alpha\beta}F_{\mu\nu}F_{\alpha\beta}\right).
\end{equation}
The equation of motion for the temporal gauge field $A_{t}(Z)$ as obtained from the above lagrangian in (\ref{D6DBI3}) is given as:
\begin{equation}\label{EOMAt}
\partial_{Z}\left(e^{-\phi_{IIA}}\sqrt{-G}~G^{tt}G^{ZZ}\partial_{Z}A_{t}(Z)\right)=0.
\end{equation}

In the large-$Z$ limit or equivalently the UV, the LO and NLO ${\cal O}(R^4)$ terms are given by:
\begin{eqnarray}
\label{Obeta-LO-large-Z}
& & \hskip -0.5in \beta \frac{45 i 3^{2/3} b^{10} \left(9 b^2+1\right)^3 M_{\rm UV} e^{4 Z} \left(-19683 \sqrt{6} \alpha _{\theta _1}^6-6642
   \alpha _{\theta _2}^2 \alpha _{\theta _1}^3+40 \sqrt{6} \alpha _{\theta _2}^4\right) \log ^3({r_h}) {A_t}'(Z)
   }{16 \sqrt{2}
   \pi ^{5/2} \left(3 b^2-1\right)^5 \left(6 b^2+1\right)^3 {g_s}^{3/2} N^{23/20} N_f^{\rm UV} \alpha _{\theta _2}^5
   \log ^4(N)}\nonumber\\
   & & \hskip -0.5in \times ({g_s} N_f^{\rm UV} \log (N)-3 {g_s} N_f^{\rm UV} \log ({r_h})-3 {g_s} N_f^{\rm UV} Z+4 \pi ),
\end{eqnarray}
and the NLO term is:
\begin{eqnarray}
\label{Obeta-NLO-large-Z}
& & \hskip -0.5in \beta \frac{45 i 3^{2/3} b^{10} \left(9 b^2+1\right)^4 M_{\rm UV} e^{3 Z} \left(-19683 \sqrt{6} \alpha _{\theta _1}^6-6642
   \alpha _{\theta _2}^2 \alpha _{\theta _1}^3+40 \sqrt{6} \alpha _{\theta _2}^4\right) \left(6 a^2+{r_h}^2\right)
   \log ^3({r_h}) {A_t}'(Z) }{8 \sqrt{2} \pi ^{5/2} \left(3 b^2-1\right)^5 \left(6 b^2+1\right)^4 {g_s}^{3/2} N^{23/20}
   N_f^{\rm UV} \alpha _{\theta _2}^5 \left(9 a^2+{r_h}^2\right) \log ^4(N)}\nonumber\\
   & & \hskip -0.5in \times ({g_s} {N_f} (\log N -3 Z+\log (16))-3 {g_s} {N_f} \log
   ({r_h})+4 \pi ).
\end{eqnarray} 
Both are proportional to $\frac{M_{\rm UV}}{N_f^{\rm UV}N^{23/20}}$. We choose the UV-profiles of $M$ and $N_f$ to be such that the same is negligible as both $M_{\rm UV}$ and $N_f^{\rm UV}$ are vanishingly small in the UV. So, it is the NNLO term in a large-$Z$ expansion that figures at ${\cal O}(R^4)$. One thus obtains the following EOM up to ${\cal O}(\beta)$:
   \begin{eqnarray}
   \label{AtEOM-2}
 & &    {r_h}^2 e^{2 Z} \left(A_t^{(0)}\ '(Z) + \beta A_t^{(1)}\ '(Z) \right) (g_s  {N_f} (\log N\ -3 Z)-3 g_s  {N_f} \log ({r_h})+4 \pi ) \nonumber\\
 & &  + {r_h}^2 e^{2 Z}
  A_t^{(0)}\ '(Z) \left({\cal C}_{zz} - 2 {\cal C}_{\theta_1z} + 2 
  {\cal C}_{\theta_1x}\right) (g_s  {N_f} (\log N\ -3 Z)-3 g_s  {N_f} \log ({r_h})+4 \pi )= {\rm constant}\nonumber\\
  & & 
   \end{eqnarray}
   The solution to (\ref{AtEOM-2}) is:
   \begin{eqnarray}
   \label{solution-At-EOM}
& & A_t^{(0)}(Z) + \beta A_t^{(1)}(z) \nonumber\\
& & =   c_1^{\beta^0} + \beta c_1^\beta-\frac{\left(\tilde{C}^{\beta^0} + \beta \tilde{C}^\beta\right) e^{-\frac{2}{3} \left(\frac{4 \pi }{g_s  {N_f}}+\log N\ \right)} {Ei}\left(\frac{2}{3} \left(\log N\ -3 Z-3 \log ({r_h})+\frac{4
   \pi }{g_s  {N_f}}\right)\right)}{6  g_s  {N_f}}\nonumber\\
   & & = \frac{\left(C^{\beta^0} + \beta C^\beta\right) e^{-2 Z}}{ Z}+c_1^{\beta^0} + \beta c_1^\beta + {\cal O}\left(\frac{1}{Z^2}\right),
   \end{eqnarray}
where $g_s, N_f, r_h^2$ have been absorbed into $C^{\beta^0,\ \beta}$.
   
Similarly, the EOM in the small-$Z$ limit or equivalently the IR, can be shown to be:
{\footnotesize
\begin{eqnarray}
\label{At-EOM-IR}
& & \hskip -0.9in \left(\frac{i \sqrt[10]{N} {r_h}^2 {A_t}'(Z) \left({g_s} {N_f} \left(\log N  \left(173 Z^2+184
   Z+96\right)-216 Z (2 Z+1)\right)-3 {g_s} {N_f} \left(173 Z^2+184 Z+96\right) \log ({r_h})+4 \pi 
   \left(173 Z^2+184 Z+96\right)\right)}{192 \sqrt{2} 3^{5/6} \pi ^{3/2} {g_s}^{3/2} \alpha _{\theta _2}^2}\right)\nonumber\\
   & & \hskip -0.8in\times(A_t^{(0)}\ '(Z) + \beta A_t^{(1)}\  '(Z))\nonumber\\
& & \hskip -0.9in -\beta\Biggl(\frac{i \sqrt{9 b^2+1} \sqrt[10]{N} {r_h}^2 \left({\cal C}_{zz} - 2 {\cal C}_{\theta_1z} + 2 
  {\cal C}_{\theta_1x}\right) ({g_s}
   {N_f} \log (N)-3 {g_s} {N_f} \log ({r_h})+4 \pi )}{8 \sqrt[3]{3} \pi ^{3/2} \sqrt{12 b^2+2}
   {g_s}^{3/2} \alpha _{\theta _2}^2}  +\frac{i \sqrt[10]{N} {r_h}^2 \left(6 b^2 \log N +\log N \right)^4 \left({\cal C}_{zz} - 2 {\cal C}_{\theta_1z} + 2 
  {\cal C}_{\theta_1x}\right) }{16 \sqrt{2}
   \sqrt[3]{3} \pi ^{3/2} \left(6 b^2+1\right)^{13/2} \left(9 b^2+1\right)^{7/2} {g_s}^{3/2} \log N ^4 \alpha_{\theta _2}^2}\nonumber\\
   & & \hskip -0.9in \times\Biggl[-\left(11664 b^8+5508 b^6+1161 b^4+114 b^2+4\right) \left(9 b^2+1\right)^2 ({g_s}
   \log N  {N_f}-3 {g_s} {N_f} \log ({r_h})+4 \pi )+18 \left(54 b^5+15 b^3+b\right)^2 {g_s}
   {N_f}\nonumber\\
   & & \hskip -0.8in +6 \left(54 b^4+15 b^2+1\right)^2 \left(108 b^4+27 b^2+2\right) {g_s} {N_f}\Biggr]\Biggr)A_t^{(0)}\  '(Z)  = {\rm constant}.     
\end{eqnarray}  
} 
One hence obtains the following IR-valued $A_t(Z)$:
\begin{eqnarray}
\label{At-IR}
& & A_t(Z\in{\rm IR}) = \frac{11.2 i \sqrt{{g_s}} \alpha _{\theta _2}^2 \left(c_1^{(0)} + \beta c_1^{(1)}\right)}{\sqrt[10]{N}
   {N_f} {r_h}^2 \log ({r_h})}-\frac{23 i \pi ^{3/2} \sqrt{{g_s}} Z^2 \alpha _{\theta _2}^2 \left(c_1^{(0)} + \beta c_1^{(1)}\right)}{6 \sqrt{2} \sqrt[6]{3} \sqrt[10]{N} {N_f} {r_h}^2 \log ({r_h})}\nonumber\\
& & +\frac{2 i
   \sqrt{2} \pi ^{3/2} \sqrt{{g_s}} Z \alpha _{\theta _2}^2 \left(c_1^{(0)} + \beta c_1^{(1)}\right)}{\sqrt[6]{3}
   \sqrt[10]{N} {N_f} {r_h}^2 \log ({r_h})}+c_2^{(0)}+\beta c_2^{(1)}.
\end{eqnarray}   
   
Let's now consider the perturbation of the gauge field $A$ about the nonzero background $A_{t}$ as:
\begin{equation}
\hat{A}_{\mu}(x,Z)=\delta^{t}_{\mu}A_{t}(Z)+\tilde{A_{\mu}}(x,Z),
\end{equation}
where the fluctuation $\tilde{A_{\mu}}$ only exist along the directions $\mu=\{t,x_1,x_2,x_3\}$ due to the particular gauge choice and depends only on the radial variable $Z$.
Including the perturbations in the lagrangian of the DBI action (\ref{D6DBI}), one get:
\begin{equation}
\mathcal{L}=e^{-\phi_{IIA}}\sqrt{\det{i^*\{g+B\}+F+\tilde{F}}},
\end{equation}
with $\tilde{F}$ as the field strength for the gauge field fluctuations. Now defining $\tilde{G}=i^*(g+B)+F$ and again expanding the above lagrangian upto quadratic order in the gauge field fluctuation one get:
\begin{equation}\label{D6DBI4}
\mathcal{L}=-\frac{1}{4}e^{-\phi_{IIA}}\sqrt{-\tilde{G}}\left(\tilde{G}^{\mu\alpha}\tilde{G}^{\beta\gamma}\tilde{F}_{\alpha\beta}\tilde{F}_{\gamma\mu}
-\frac{1}{2}\tilde{G}^{\mu\nu}\tilde{G}^{\alpha\beta}\tilde{F}_{\mu\nu}\tilde{F}_{\alpha\beta}\right).
\end{equation}
Writing the field strength $\tilde{F}$ in terms of the gauge field fluctuation $\tilde{A_{\mu}}$ and after doing some simplifications in terms of the interchange of indices, one can write the above lagrangian as:
\begin{equation}\begin{split}\label{lag}
\mathcal{L}&=e^{-\phi_{IIA}}\sqrt{-\tilde{G}}\left\{(\tilde{G}^{\mu\alpha}\tilde{G}^{\beta\gamma}
-\tilde{G}^{\mu\beta}\tilde{G}^{\alpha\gamma})\partial_{[\gamma}\tilde{A}_{\mu]}
-\frac{1}{2}\tilde{G}^{[\alpha\beta]}\tilde{G}^{\mu\nu}\partial_{[\mu}\tilde{A}_{\nu]}
\right\}(\partial_{\alpha}\tilde{A}_{\beta})
\\ & =\partial_{\alpha}\Biggl[e^{-\phi_{IIA}}\sqrt{-\tilde{G}}\left\{(\tilde{G}^{\mu\alpha}\tilde{G}^{\beta\gamma}
-\tilde{G}^{\mu\beta}\tilde{G}^{\alpha\gamma})\partial_{[\gamma}\tilde{A}_{\mu]}
-\frac{1}{2}\tilde{G}^{[\alpha\beta]}\tilde{G}^{\mu\nu}\partial_{[\mu}\tilde{A}_{\nu]}
\right\}\tilde{A}_{\beta}\Biggr]\\
& -\tilde{A}_{\beta}\partial_{\alpha}\Biggl[e^{-\phi_{IIA}}\sqrt{-\tilde{G}}\left\{(\tilde{G}^{\mu\alpha}\tilde{G}^{\beta\gamma}
-\tilde{G}^{\mu\beta}\tilde{G}^{\alpha\gamma})\partial_{[\gamma}\tilde{A}_{\mu]}
-\frac{1}{2}\tilde{G}^{[\alpha\beta]}\tilde{G}^{\mu\nu}\partial_{[\mu}\tilde{A}_{\nu]}
\right\}\Biggr].
\end{split}
\end{equation}
The second line in equation (\ref{lag}) is a total derivative term and equating the last line to zero for any arbitrary $\tilde{A}_{\beta}$ gives the equation of motion for the gauge field fluctuation:
\begin{equation}\label{EOMtildeA}
\partial_{\alpha}\Biggl[e^{-\phi_{IIA}}\sqrt{-\tilde{G}}\left\{(\tilde{G}^{\mu\alpha}\tilde{G}^{\beta\gamma}
-\tilde{G}^{\mu\beta}\tilde{G}^{\alpha\gamma})\partial_{[\gamma}\tilde{A}_{\mu]}
-\frac{1}{2}\tilde{G}^{[\alpha\beta]}\tilde{G}^{\mu\nu}\partial_{[\mu}\tilde{A}_{\nu]}
\right\}
\end{equation}
To derive the equation of motion for the gauge field fluctuation, we first need to write down the fluctuating field as the following Fourier decomposed form:
\begin{equation}
\tilde{A}_{\mu}(t,x_1,Z)=\int \frac{d^4 k}{(2\pi)^4}e^{-i w t+i q x_1}\tilde{A}_{\mu}(w,q,Z),
\end{equation}
where we assume the fluctuation to have momentum along $x_1$ direction only. Now, the equation (\ref{EOMtildeA}) has a free index $\{\beta\}$ and for $\beta=\{t,x_1,x_2,x_3,Z\}$, one get a total of five equation of motion. Out of these five equations, two corresponding to $\beta=x_2, x_3$ are equivalent. Hence the four independent equations are given as:
{\footnotesize
\begin{equation}\begin{split}\label{EOMtildeA1}
(\rm{For}~\beta=Z)&:~~~ w\tilde{G}^{tt}(\partial_{Z}\tilde{A}_{t})-q\tilde{G}^{x_1x_1}(\partial_{Z}\tilde{A}_{x_1})=0\\
(\rm{For}~\beta=t)&:~~~ \partial_{Z}\left(e^{-\phi_{IIA}}\sqrt{-\tilde{G}}~\tilde{G}^{tt}\tilde{G}^{ZZ}\partial_{Z}\tilde{A}_{t}\right)
-e^{-\phi_{IIA}}\sqrt{-\tilde{G}}~\tilde{G}^{tt}\tilde{G}^{x_1x_1}\left(w q \tilde{A}_{x_1}+q^2\tilde{A}_{t}\right)=0\\
(\rm{For}~\beta=x_1)&:~~~ \partial_{Z}\left(e^{-\phi_{IIA}}\sqrt{-\tilde{G}}~\tilde{G}^{x_1x_1}\tilde{G}^{ZZ}\partial_{Z}\tilde{A}_{x_1}\right)
-e^{-\phi_{IIA}}\sqrt{-\tilde{G}}~\tilde{G}^{tt}\tilde{G}^{x_1x_1}\left(w q \tilde{A}_{t}+w^2\tilde{A}_{x_1}\right)=0\\
(\rm{For}~\beta=x_2~\rm{or}~x_3)&:~~~ \partial_{Z}\left(e^{-\phi_{IIA}}\sqrt{-\tilde{G}}~\tilde{G}^{x_2x_2}\tilde{G}^{ZZ}\partial_{Z}\tilde{A}_{x_2}\right)
-e^{-\phi_{IIA}}\sqrt{-\tilde{G}}~\tilde{G}^{x_2x_2}\left(q^2\tilde{G}^{x_1x_1}+w^2\tilde{G}^{tt}\right)\tilde{A}_{x_2}=0
\end{split}
\end{equation}
}
Now, we define the following gauge invariant variables: $E_{x_1}=q\tilde{A}_{t}+w\tilde{A}_{x_1}$ and $E_{x_2}=E_{x_3}=E_{T}=w\tilde{A}_{x_2/x_3}$. With these new variable the first three equations in (\ref{EOMtildeA1}) for $\beta=\{Z,t,x_1\}$ can be cast into a single second order equation involving $E_{x_1}$. Even more obviously, the fourth one for $\beta=x_2/x_3$, can be rewritten in terms of the new variable $E_T$. Moreover, in the zero momentum limit $(q\rightarrow 0)$, it can be shown that the equation involving $E_{x_1}$ is the same as the one involving $E_{T}$, given as:
\begin{equation}\label{EOMET}
\partial_{Z}\left(e^{-\phi_{IIA}}\sqrt{-\tilde{G}}~\tilde{G}^{x_2x_2}\tilde{G}^{ZZ}\partial_{Z}E_{T}\right)
-e^{-\phi_{IIA}}\sqrt{-\tilde{G}}~\tilde{G}^{x_2x_2}\left(w^2\tilde{G}^{tt}\right)E_{T}=0
\end{equation}
Equation (\ref{EOMET}) can be rewritten as:
\begin{equation}
\partial_{Z}\Biggl[P(Z)\partial_{Z}E_{T}(Z)\Biggr]+Q(Z)w^2E_{T}(Z)=0,
\end{equation}
where $P(Z)=e^{-\phi_{IIA}}\sqrt{-\tilde{G}}\tilde{G}^{ZZ}\tilde{G}^{x_2x_2}$, $Q(Z)=-e^{-\phi_{IIA}}\sqrt{-\tilde{G}}\tilde{G}^{x_2x_2}\tilde{G}^{tt}$.
One can show:
{\footnotesize
\begin{eqnarray}
\label{P+Q-up-to-Obeta}
& & \hskip -0.8in P(Z) = \frac{\sqrt[10]{N} {r_h}^3 e^{- Z} \left(e^{4 Z}-1\right) \left(-{g_s} {N_f} \log \left({r_h}^6 e^{4 Z}
   \left(e^{2 Z}+3\right)\right)+2 {g_s} \log N  {N_f}+8 \pi
   \right)}{8 \sqrt{2} \sqrt[3]{3} \pi ^{3/2} {g_s}^2 \alpha _{\theta _2}^4 \sqrt{\frac{ \left({r_h}^2 e^{2 Z} \left(6 b^2+e^{2 Z}\right)-\left(9 b^2+e^{2 Z}\right) {A_t}'(Z)^2\right)}{{g_s}
   \alpha _{\theta _2}^4 \left(9 b^2+e^{2 Z}\right)}}}\nonumber\\
& & \hskip -0.8in-\frac{1}{32 \sqrt{2} \sqrt[3]{3} \pi ^{5/2} \left(1-3 b^2\right)^5
   \left(6 b^2+1\right)^4 N^{23/20} {N_f} \alpha _{\theta _2}^3 \log ^4(N) \left({r_h}^3 e^{2 Z} \left(6
   b^2+e^{2 Z}\right)-{r_h} \left(9 b^2+e^{2 Z}\right) {A_t}'(Z)^2\right)^2}\nonumber\\
& & \hskip -0.8in\times\Biggl\{{r_h}^3 e^Z \left(1-e^{4 Z}\right) \left(9 b^2+e^{2 Z}\right) \sqrt{\frac{{r_h}^2 e^{2 Z} \left(6
   b^2+e^{2 Z}\right)-\left(9 b^2+e^{2 Z}\right) {A_t}'(Z)^2}{{g_s} \alpha _{\theta _2}^4 \left(9 b^2+e^{2
   Z}\right)}} \Biggl[-{N_f} \Biggl\{\log \left(9 e^{-2 Z} \left(\frac{{g_s} M^2 ({c_1}+{c_2} \log
   ({r_h}))}{N}+\frac{1}{\sqrt{3}}\right)^2+1\right)\nonumber\\
& & \hskip -0.8in +6 (\log ({r_h})+Z)\Biggr\}+\frac{8 \pi }{{g_s}}+4
   {N_f} \log \left(\frac{4 \sqrt{N}}{\alpha _{\theta _1} \alpha _{\theta _2}}\right)\Biggr] \Biggl[45 \left(9
   b^2+1\right)^4 b^{10} M e^{-Z} \left(e^Z-2\right) \left(19683 \sqrt{6} \alpha _{\theta _1}^6+6642 \alpha _{\theta
   _2}^2 \alpha _{\theta _1}^3-40 \sqrt{6} \alpha _{\theta _2}^4\right) \log ^3({r_h})\nonumber\\
& & \hskip -0.8in\times \left({r_h}^2 e^{2 Z}
   \left(6 b^2+e^{2 Z}\right)-\left(9 b^2+e^{2 Z}\right) {A_t}'(Z)^2\right)+2 \pi  \left(3 b^2-1\right)^5 \left(6
   b^2+1\right)^4 N^{5/4} {N_f} {r_h}^4 \alpha _{\theta _2}^3 \left(6 b^2+e^{2 Z}\right) \log ^4(N)\nonumber\\
& & \hskip -0.8in\times
    \left({\cal C}_{zz} - 2 {\cal C}_{\theta_1z} + 2 
  {\cal C}_{\theta_1x}\right)\Biggr]\Biggr\}\beta = P_{\beta^0}(Z) + \beta P_\beta(Z);\nonumber\\
   & & \hskip -0.8in Q(Z)= \frac{N^{11/10} {r_h}^4
   e^{5 Z} \left(6 b^2+e^{2 Z}\right) \left(-{g_s} {N_f} \left(\log \left(\frac{9 a^2 e^{-2
   Z}}{{r_h}^2}+1\right)+6 (\log ({r_h})+Z)\right)-4 {g_s} {N_f} \log \left(\frac{4 \sqrt{N}}{\alpha
   _{\theta _1} \alpha _{\theta _2}}\right)+8 \pi \right)}{2 \sqrt[3]{3} \sqrt{2 \pi } \sqrt{{g_s}} \left(e^{4
   Z}-1\right) \alpha _{\theta _2}^2 \left(9 b^2+e^{2 Z}\right) \sqrt{\frac{{r_h}^2 e^{2 Z} \left(6 b^2+e^{2
   Z}\right)-\left(9 b^2+e^{2 Z}\right) {A_t}'(Z)^2}{9 b^2+e^{2 Z}}}}\nonumber\\
& & \hskip -0.8in + \beta  \Biggl(\frac{1}{4 \sqrt[3]{3} \sqrt{2 \pi } \sqrt{{g_s}} \left(e^{4 Z}-1\right) \alpha _{\theta _2}^2
   \left(9 b^2+e^{2 Z}\right)^2 \left(\frac{{r_h}^2 e^{2 Z} \left(6 b^2+e^{2 Z}\right)-\left(9 b^2+e^{2 Z}\right)
   {A_t}'(Z)^2}{9 b^2+e^{2 Z}}\right)^{3/2}}\Biggl\{N^{11/10} {r_h} e^{5 Z} \left(6 b^2+e^{2 Z}\right)  \left({\cal C}_{zz} - 2 {\cal C}_{\theta_1z} + 2 
  {\cal C}_{\theta_1x}\right)\nonumber\\
& & \hskip -0.8in \times \left({r_h}^2 e^{2 Z} \left(6 b^2+e^{2 Z}\right)-2 \left(9 b^2+e^{2 Z}\right)
   {A_t}'(Z)^2\right) \left(-{g_s} {N_f} \left[\log \left(\frac{9 a^2 e^{-2 Z}}{{r_h}^2}+1\right)+6
   (\log ({r_h})+Z)\right]-4 {g_s} {N_f} \log \left(\frac{4 \sqrt{N}}{\alpha _{\theta _1} \alpha _{\theta
   _2}}\right)+8 \pi \right)\Biggr\}\nonumber\\
& & \hskip -0.8in -\frac{1}{8 \sqrt{2} \pi ^{3/2} \left(3
   b^2-1\right)^5 \left(6 b^2+1\right)^4 \sqrt{{g_s}} {N_f} {r_h} \left(e^{4 Z}-1\right) \alpha _{\theta
   _2}^5 \left(9 a^2+{r_h}^2\right) \left(9 b^2+e^{2 Z}\right) \log ^4(N) \sqrt{\frac{{r_h}^2 e^{2 Z} \left(6
   b^2+e^{2 Z}\right)-\left(9 b^2+e^{2 Z}\right) {A_t}'(Z)^2}{9 b^2+e^{2 Z}}}}\nonumber\\
& & \hskip -0.8in \Biggl\{9\ 3^{2/3} b^{10} \left(9 b^2+1\right)^4 M
   \left(\frac{1}{N}\right)^{3/20} e^{6 Z} \left(19683 \sqrt{6} \alpha _{\theta _1}^6+6642 \alpha _{\theta _2}^2 \alpha
   _{\theta _1}^3-40 \sqrt{6} \alpha _{\theta _2}^4\right) \left(6 a^2+{r_h}^2\right) \left(6 b^2+e^{2 Z}\right)
   \left(-6 e^Z+3 e^{2 Z}+2 e^Z \left(e^Z-2\right)\right) \nonumber\\
& & \hskip -0.8in\log ^3({r_h}) \left(-{g_s} {N_f} \left[\log
   \left(\frac{9 a^2 e^{-2 Z}}{{r_h}^2}+1\right)+6 (\log ({r_h})+Z)\right]-4 {g_s} {N_f} \log
   \left(\frac{4 \sqrt{N}}{\alpha _{\theta _1} \alpha _{\theta _2}}\right)+8 \pi \right)\Biggr\}\Biggr) = Q_{\beta^0} + \beta Q_\beta,
\end{eqnarray}
}
Defining a new variable $\tilde{E_{T}}(Z)=\sqrt{P(Z)}E_{T}(Z)$, one can write the above equation as the following Schr\"{o}dinger like form:
\begin{equation}
\label{Schroedinger-like-ET}
\left(\partial^{2}_{Z}+V_{E_{T}}\right)\tilde{E_{T}}=0,
\end{equation}
where the potential term $V_{E_{T}}$ is given as:
\begin{equation}
V_{E_{T}}=\frac{1}{4}\left(\frac{1}{P}\frac{\partial P}{\partial Z}\right)^2-\frac{1}{2}\frac{1}{P}\frac{\partial^2 P}{\partial Z^2}+w^2\frac{Q}{P}.
\end{equation}

Also, using the equation of motion we get the following on-shell DBI action for the $D6$ brane:
\begin{equation}\begin{split}\label{D6DBIonshell}
S^{on-shell}_{D6}&=\frac{T_{D6}}{4}\Omega_{2}\int d^{4}x e^{-\phi_{IIA}}\sqrt{-\tilde{G}}\left\{(\tilde{G}^{\mu Z}\tilde{G}^{\beta\gamma}
-\tilde{G}^{\mu\beta}\tilde{G}^{Z\gamma})\partial_{[\gamma}\tilde{A}_{\mu]}
-\frac{1}{2}\tilde{G}^{[Z\beta]}\tilde{G}^{\mu\nu}\partial_{[\mu}\tilde{A}_{\nu]}
\right\}\tilde{A}_{\beta}\\ &
=\frac{T_{D6}}{2}\Omega_{2}\int d^{4}x\Biggl[ e^{-\phi_{IIA}}\sqrt{-\tilde{G}}
\Biggl\{(\tilde{G}^{tZ})^2\tilde{A}_{t}(\partial_{Z}\tilde{A}_{t})-\tilde{G}^{tt}\tilde{G}^{ZZ}\tilde{A}_{t}(\partial_{Z}\tilde{A}_{t})
\\& -\tilde{G}^{xx}\tilde{G}^{ZZ}\tilde{A}_{x}(\partial_{Z}\tilde{A}_{x})-\tilde{G}^{yy}\tilde{G}^{ZZ}\tilde{A}_{y}(\partial_{Z}\tilde{A}_{y})
-\tilde{G}^{zz}\tilde{G}^{ZZ}\tilde{A}_{z}(\partial_{Z}\tilde{A}_{Z})\Biggr\}\Biggr]\Biggr|^{Z_{\infty}}_{Z_{h}}.
\end{split}
\end{equation}
As the above on-shell action has to be evaluated on the boundary $(Z=Z_{\infty})$, where $F_{tZ}=-F_{Zt}=0$, we must set $\tilde{G}^{tZ}=0$ and $\sqrt{-\tilde{G}}$ has to be replaced by $\sqrt{-G}$. So the boundary value of the on-shell action is given as:
\begin{equation}\begin{split}\label{D6DBIonshell2}
S^{on-shell}_{D6}&=-\frac{T_{D6}}{2}\Omega_{2}\int d^{4}x \Biggl[e^{-\phi_{IIA}}\sqrt{-\tilde{G}}\tilde{G}^{ZZ}
\Biggl\{\tilde{G}^{tt}\tilde{A}_{t}(Z,-k)\partial_{Z}\tilde{A}_{t}(Z,k)
 +\tilde{G}^{xx}\tilde{A}_{x}(Z,-k)\partial_{Z}\tilde{A}_{x}(Z,k)\\ &+\tilde{G}^{yy}\tilde{A}_{y}(Z,-k)\partial_{Z}\tilde{A}_{y}(Z,k)
+\tilde{G}^{zz}\tilde{A}_{z}(Z,-k)\partial_{Z}\tilde{A}_{Z}(Z,k)\Biggr\}\Biggr]\Biggr|_{Z_{\infty}}.
\end{split}
\end{equation}
Using the first equation in (\ref{EOMtildeA1}), the above action can be rewritten in terms of the gauge invariant variables $E_{x_1}$, $E_{x_2}$ and $E_{x_3}$ as:
\begin{equation}\begin{split}\label{D6DBIonshell3}
S^{on-shell}_{D6}&=-\frac{T_{D6}}{2}\Omega_{2}\int d^{4}x \Biggl[e^{-\phi_{IIA}}\sqrt{-\tilde{G}}\tilde{G}^{ZZ}\tilde{G}^{x_1x_1}
\Biggl\{\frac{E_{x_1}(Z,-k)\partial_{Z}E_{x_1}(Z,k)}{w^2-\frac{q^2}{1-e^{-4Z}}}\\ &+\frac{1}{w^2}E_{x_2}(Z,-k)\partial_{Z}E_{x_2}(Z,k)
+\frac{1}{w^2}E_{x_3}(Z,-k)\partial_{Z}E_{x_3}(Z,k)\Biggr\}\Biggr]\Biggr|_{Z_{\infty}}.
\end{split}
\end{equation}
Defining $E_{x_1}(Z,k)=E_{0}(k)\frac{E_{k}(Z)}{E_{k}(Z_{\infty})}$, we get in the zero momentum limit for the $x_1$ component of the gauge field:
\begin{equation}
\mathcal{F}=-\frac{T_{D6}}{2}\frac{\Omega_{2}}{w^2}\Biggl[e^{-\phi_{IIA}}\sqrt{-\tilde{G}}~\tilde{G}^{ZZ}\tilde{G}^{x_1x_1}\left(\frac{\partial_{Z}E_{k}(Z)}{E_{k}(Z)}\right)\Biggr]\Biggr|_{Z\rightarrow{\infty}}
\end{equation}

The $x_1$ component of the retarded Green's function in the zero momentum limit is defined as:
\begin{equation}
G^{R}_{x_1x_1}(w,q=0)=-2w^2\mathcal{F}.
\end{equation}
Finally the spectral function is given as:
\begin{equation}
\mathcal{R}_{x_1x_1}=\rho(T\neq0,w)=-2 \Im m G^{R}_{x_1x_1}(w,q=0).
\end{equation}

{\footnotesize
\begin{eqnarray}
\label{P+Q-large-N-large-Z}
& & \hskip -0.8in P(Z) = \frac{\sqrt[5]{N} {r_h}^6 e^{2 Z} \left(e^{4 Z}-1\right) \left(-{g_s} {N_f} \left[\log
   \left(\frac{9 a^2 e^{-2 Z}}{{r_h}^2}+1\right)+6 (\log ({r_h})+Z)\right]-4 {g_s} {N_f} \log
   \left(\frac{4 \sqrt{N}}{\alpha _{\theta _1} \alpha _{\theta _2}}\right)+8 \pi \right)}{8 \sqrt{2}
   \sqrt[3]{3} \pi ^{3/2} {g_s}^2 \alpha _{\theta _2}^4 \sqrt{\frac{\sqrt[5]{N} {r_h}^6
   e^{6 Z} \left({r_h}^2 e^{2 Z} \left(6 b^2+e^{2 Z}\right)-\left(9 b^2+e^{2 Z}\right)
   {At}'(Z)^2\right)}{{g_s} \alpha _{\theta _2}^4 \left(9 b^2+e^{2
   Z}\right)}}}\nonumber\\
& & \hskip -0.8in -\frac{\beta  N^{2/5} {r_h}^{14} e^{10 Z} \left(e^{4 Z}-1\right) \left(6
   b^2+e^{2 Z}\right)  \left({\cal C}_{zz} - 2 {\cal C}_{\theta_1z} + 2 
  {\cal C}_{\theta_1x}\right) \left(-{g_s} {N_f} \left[\log
   \left(\frac{9 a^2 e^{-2 Z}}{{r_h}^2}+1\right)+6 (\log ({r_h})+Z)\right]-4 {g_s} {N_f} \log
   \left(\frac{4 \sqrt{N}}{\alpha _{\theta _1} \alpha _{\theta _2}}\right)+8 \pi \right)}{16 \sqrt{2} \sqrt[3]{3} \pi ^{3/2} {g_s}^3 \alpha _{\theta _2}^8 \left(9
   b^2+e^{2 Z}\right) \left(\frac{\sqrt[5]{N} {r_h}^6 e^{6 Z} \left({r_h}^2 e^{2 Z}
   \left(6 b^2+e^{2 Z}\right)-\left(9 b^2+e^{2 Z}\right) {At}'(Z)^2\right)}{{g_s}
   \alpha _{\theta _2}^4 \left(9 b^2+e^{2 Z}\right)}\right){}^{3/2}}\nonumber\\
& & \hskip -0.8 in = \frac{\sqrt[10]{N} {r_h}^2 e^{2 Z} \left(-{g_s} {N_f} \log
   \left({r_h}^6 e^{4 Z} \left(e^{2 Z}+3\right)\right)+2 {g_s} {N_f} \log (N)-4
   {g_s} {N_f} \log \left(\alpha _{\theta _1}\right)-4 {g_s} {N_f} \log
   \left(\alpha _{\theta _2}\right)+4 {g_s} {N_f} \log (4)+8 \pi \right)}{8 \sqrt{2}
   \sqrt[3]{3} \pi ^{3/2} {g_s}^{3/2} \alpha _{\theta _2}^2}\nonumber\\ 
& & \hskip -0.8in + \frac{3^{2/3} \beta  \sqrt[10]{N} {N_f} {r_h}^2 e^{2 Z} \log ({r_h})
    \left({\cal C}_{zz} - 2 {\cal C}_{\theta_1z} + 2 
  {\cal C}_{\theta_1x}\right)}{8 \sqrt{2} \pi ^{3/2} \sqrt{{g_s}}
   \alpha _{\theta _2}^2};\nonumber\\
& & \hskip -0.8in Q(Z) = \frac{N^{6/5} {r_h}^4 e^{8 Z} \left(6 b^2+e^{2 Z}\right)
    \left(-{g_s} {N_f} \left[\log
   \left(\frac{9 a^2 e^{-2 Z}}{{r_h}^2}+1\right)+6 (\log ({r_h})+Z)\right]-4 {g_s} {N_f} \log
   \left(\frac{4 \sqrt{N}}{\alpha _{\theta _1} \alpha _{\theta _2}}\right)+8 \pi \right)}{2 \sqrt[3]{3} \sqrt{2 \pi } {g_s} \left(e^{4 Z}-1\right) \alpha _{\theta
   _2}^4 \left(9 b^2+e^{2 Z}\right) \sqrt{\frac{\sqrt[5]{N} {r_h}^6 e^{6 Z}
   \left({r_h}^2 e^{2 Z} \left(6 b^2+e^{2 Z}\right)-\left(9 b^2+e^{2 Z}\right)
   {At}'(Z)^2\right)}{{g_s} \alpha _{\theta _2}^4 \left(9 b^2+e^{2 Z}\right)}}}\nonumber\\
& & \hskip -0.8in+\frac{\beta  N^{11/10}r_h^3 e^{5 Z} \left(6 b^2+e^{2 Z}\right)  \left({\cal C}_{zz} - 2 {\cal C}_{\theta_1z} + 2 
  {\cal C}_{\theta_1x}\right) \left({r_h}^2 e^{2 Z} \left(6 b^2+e^{2 Z}\right)-2 \left(9 b^2+e^{2
   Z}\right) {At}'(Z)^2\right) \sqrt{\frac{ 
   \left({r_h}^2 e^{2 Z} \left(6 b^2+e^{2 Z}\right)-\left(9 b^2+e^{2 Z}\right)
   {At}'(Z)^2\right)}{{g_s} \alpha _{\theta _2}^4 \left(9 b^2+e^{2 Z}\right)}}
  }{4 \sqrt[3]{3} \sqrt{2 \pi } \left(e^{4 Z}-1\right) \left({r_h}^3 e^{2 Z}
   \left(6 b^2+e^{2 Z}\right)-{r_h} \left(9 b^2+e^{2 Z}\right)
   {At}'(Z)^2\right)^2}\nonumber\\
& & \hskip -0.8in \times  \left(-{g_s} {N_f} \left[\log
   \left(\frac{9 a^2 e^{-2 Z}}{{r_h}^2}+1\right)+6 (\log ({r_h})+Z)\right]-4 {g_s} {N_f} \log
   \left(\frac{4 \sqrt{N}}{\alpha _{\theta _1} \alpha _{\theta _2}}\right)+8 \pi \right)\nonumber\\
& & \hskip -0.8in =\frac{N^{11/10} \left(-{g_s} {N_f} \left[\log
   \left(\frac{9 a^2 e^{-2 Z}}{{r_h}^2}+1\right)+6 (\log ({r_h})+Z)\right]-4 {g_s} {N_f} \log
   \left(\frac{4 \sqrt{N}}{\alpha _{\theta _1} \alpha _{\theta _2}}\right)+8 \pi \right)}{2 \sqrt[3]{3}
   \sqrt{2 \pi } \sqrt{{g_s}} \alpha _{\theta _2}^2}\nonumber\\
& & \hskip -0.8in -\frac{\beta  N^{11/10}  \left({\cal C}_{zz} - 2 {\cal C}_{\theta_1z} + 2 
  {\cal C}_{\theta_1x}\right)\left(-{g_s} {N_f} \left[\log
   \left(\frac{9 a^2 e^{-2 Z}}{{r_h}^2}+1\right)+6 (\log ({r_h})+Z)\right]-4 {g_s} {N_f} \log
   \left(\frac{4 \sqrt{N}}{\alpha _{\theta _1} \alpha _{\theta _2}}\right)+8 \pi \right)}{4 \sqrt[3]{3} \sqrt{2 \pi
   } \sqrt{{g_s}} \alpha _{\theta _2}^2}.\nonumber\\
& & 
\end{eqnarray}
}
One can show:
{\footnotesize
\begin{eqnarray}
\label{R+S-beta0}
& & \frac{1}{4}\left(\frac{P'(Z)}{P(Z)}\right)^2-\frac{1}{2}\frac{P''(Z)}{P(Z)}
\nonumber\\
& & \hskip -0.8in = \frac{\left(-4 {g_s} {N_f} \log
   \left(\frac{4 \sqrt{N}}{\alpha _{\theta _1} \alpha _{\theta _2}}\right)+6 {g_s} {N_f}
   (\log ({r_h})+Z)+3 {g_s} {N_f}-4 {g_s} {N_f} \log (4)-8 \pi
   \right){}^2}{\left(-4 {g_s} {N_f} \log
   \left(\frac{4 \sqrt{N}}{\alpha _{\theta _1} \alpha _{\theta _2}}\right)-6 {g_s}
   {N_f} (\log ({r_h})+Z)+4 {g_s} {N_f} \log (4)+8 \pi \right){}^2}\nonumber\\
& & \hskip -0.8in +\frac{2
   \left(-4 {g_s} {N_f} \log
   \left(\frac{4 \sqrt{N}}{\alpha _{\theta _1} \alpha _{\theta _2}}\right)+6 {g_s} {N_f}
   (\log ({r_h})+Z)+6 {g_s} {N_f}-4 {g_s} {N_f} \log (4)-8 \pi \right)}{2 {g_s} {N_f} \log
   \left(\frac{4 \sqrt{N}}{\alpha _{\theta _1} \alpha _{\theta _2}}\right)-6 {g_s} {N_f} (\log
   ({r_h})+Z)+4 {g_s} {N_f} \log (4)+8 \pi }\nonumber\\
& & \hskip -0.8in + \frac{9 {g_s}^2 {N_f}^2 \log ({r_h})  \left({\cal C}_{zz} - 2 {\cal C}_{\theta_1z} + 2 
  {\cal C}_{\theta_1x}\right) \left(-4 {g_s} {N_f} \log
   \left(\frac{4 \sqrt{N}}{\alpha _{\theta _1} \alpha _{\theta _2}}\right)+6 {g_s} {N_f} \log ({r_h})+6 {g_s} {N_f} Z-3 {g_s}
   {N_f}-4 {g_s} {N_f} \log (4)-8 \pi \right)}{4 \left({g_s} \log N 
   {N_f}-2 {g_s} {N_f} \log \left(\alpha _{\theta _1}\right)-2 {g_s} {N_f}
   \log \left(\alpha _{\theta _2}\right)-3 {g_s} {N_f} \log ({r_h})-3 {g_s}
   {N_f} Z+{g_s} {N_f} \log (16)+4 \pi \right){}^3}\beta.\nonumber\\
& & 
\end{eqnarray}
Similarly,
\begin{eqnarray}
\label{T}
& & \hskip -0.8in w^2\frac{Q(Z)}{P(Z)} = \frac{4 \pi  {g_s} N w^2 e^{-2Z}}{ {r_h}^2}
\nonumber\\
& & \hskip -0.8in + \frac{4 \pi  {g_s} N w^2 e^{-2 Z}  \left({\cal C}_{zz} - 2 {\cal C}_{\theta_1z} + 2 
  {\cal C}_{\theta_1x}\right)
   \left(2 {g_s} {N_f} \log
   \left(\frac{4 \sqrt{N}}{\alpha _{\theta _1} \alpha _{\theta _2}}\right)-6 {g_s} {N_f}
   \log ({r_h})-3 {g_s} {N_f} Z+{g_s} {N_f} \log (16)+4 \pi
   \right)}{{r_h}^2 \left(-6{g_s} {N_f} (Z + \log r_h)+4 {g_s} {N_f} \log
   \left(\frac{4 \sqrt{N}}{\alpha _{\theta _1} \alpha _{\theta _2}}\right)+4 {g_s} {N_f} \log (4)+8
   \pi \right)}\beta.
\end{eqnarray}
}
Performing now a small-$N^f_{\rm UV}$-expansion, one obtains:
{\footnotesize
\begin{eqnarray}
\label{VET-UV}
& & V_{\tilde{E}_T} = -\frac{9 \beta  {g_s}^2 {N_f}^2 \log ({r_h})
   \left({\cal C}_{zz} - 2 {\cal C}_{\theta_1z} + 2 
  {\cal C}_{\theta_1x}\right)}{32 \pi ^2}+\frac{9 {g_s}^2
   {N_f}^2 \log ({r_h})}{16 \pi ^2}+\frac{3 {g_s} {N_f}}{4 \pi }-1\nonumber\\
& & + e^{-2 Z} \left(\frac{\beta  {g_s} N w^2  \left({\cal C}_{zz} - 2 {\cal C}_{\theta_1z} + 2 
  {\cal C}_{\theta_1x}\right)
   \left(-9 {g_s}^2 {N_f}^2 \log ^2({r_h})-12 \pi  {g_s} {N_f} \log
   ({r_h})+16 \pi ^2\right)}{8 \pi  {r_h}^2}+\frac{4 \pi  {g_s} N
   w^2}{{r_h}^2}\right)
\end{eqnarray}
}

The Schr\"{o}dinger-like EOM:
\begin{equation}
\label{ETtilde-1}
\tilde{E}''_T(Z) + V_{\tilde{E}_T}(Z)\tilde{E}_T(Z) = 0,
\end{equation}
with $V_{\tilde{E}_T}(Z) = A + B e^{-2Z}$, is given by:
\begin{equation}
\label{ETtilde-2}
\tilde{E}_T(Z) = c_1 J_{-i \sqrt{A}}\left(\sqrt{B} e^{-Z}\right) + c_2 J_{i \sqrt{A}}\left(\sqrt{B} e^{-Z}\right), c_{1,2}\in\mathbb{C}.
\end{equation}
Writing $A=A^{\beta^0}, B = B^{\beta^0} + \beta B^\beta$, expanding up to ${\cal O}(\beta)$, and then performing a small-$N_f$ and then a large-$Z$ expansion, the LO terms in $\tilde{E}_T(Z)$ are:
{\footnotesize
\begin{eqnarray}
\label{ETtilde-4}
& & \hskip -0.8in \tilde{E}_T(Z) =\frac{3 \sqrt[4]{\frac{1}{{B^{\beta^0}}}} {g_s} {N^f_{\rm UV}} e^{Z/2} \left(3+8
   \pi  (c_1+c_2) \sin \left(\sqrt{{B^{\beta^0}}} e^{-Z}+\frac{\pi }{4}\right)\right)}{64 \sqrt{2} \pi
   ^{3/2}}+\sqrt{\frac{2}{\pi }} \sqrt[4]{\frac{1}{{B^{\beta^0}}}} (c_1-c_2) e^{Z/2} \sin \left(\sqrt{{B^{\beta^0}}}
   e^{-Z}-\frac{\pi }{4}\right)\nonumber\\
& & \hskip -0.8in +  \beta  \left(\frac{\left(\frac{1}{256}+\frac{i}{256}\right) {B^{\beta}} \left(\frac{1}{{B^{\beta^0}}}\right)^{7/4}
   e^{-Z/2} \mathbb{N}_1}{\sqrt{2 \pi }}-\frac{{B^{\beta}}
   \left(\frac{1}{{B^{\beta^0}}}\right)^{3/4} (c_1-c_2) e^{-Z/2} \sin \left(\sqrt{{B^{\beta^0}}} e^{-Z}+\frac{\pi
   }{4}\right)}{\sqrt{2 \pi }}\right),  
\end{eqnarray}
}
where
{\footnotesize
\begin{eqnarray}
\label{N1}
& & \hskip -0.8in \mathbb{N}_1 \equiv 24 (-1)^{3/4} {B^{\beta^0}} (c_1+c_2) {g_s} {N^f_{\rm UV}} \left(\cos \left(\sqrt{{B^{\beta^0}}}
   e^{-Z}\right)-\sin \left(\sqrt{{B^{\beta^0}}} e^{-Z}\right)\right)-(128-128 i) {B^{\beta^0}} (c_1-c_2) \sin
   \left(\sqrt{{B^{\beta^0}}} e^{-Z}+\frac{\pi }{4}\right).
\end{eqnarray}
}
One thus obtains:
{\footnotesize
\begin{eqnarray}
\label{ImdEToverET-i}
& & \hskip -0.8in \Im m\left[\frac{\frac{d}{dZ}\left(\frac{\tilde{E}_T(Z)}{\sqrt{P}(Z)}\right)}{\tilde{E}_T(Z)/\sqrt{P}(Z)}\right] \nonumber\\
& & \hskip -0.8in = \frac{3 \Im m\left(\frac{\beta  {B^{\beta}} \sqrt{\frac{1}{{B^{\beta^0}}}} {g_s} {N^f_{\rm UV}} e^{-2 Z} \csc
   ^2\left(\sqrt{{B^{\beta^0}}} e^{-Z}-\frac{\pi }{4}\right) \mathbb{N}_2}{(c_1-c_2){}^2}\right)}{256 \pi }+\frac{3
   \Im m\left(\frac{\sqrt{{B^{\beta^0}}} {g_s} {N^f_{\rm UV}} e^{-Z} \csc ^2\left(\sqrt{{B^{\beta^0}}} e^{-Z}-\frac{\pi
   }{4}\right) \left(3 \sin \left(\sqrt{{B^{\beta^0}}} e^{-Z}+\frac{\pi }{4}\right)+8 \pi 
   (c_1+c_2)\right)}{c_1-c_2}\right)}{128 \pi },
\end{eqnarray}
}
where
{\footnotesize
\begin{eqnarray}
\label{N2}
& &  \mathbb{N}_2 \equiv 3 \sqrt{2} (c_1-c_2)
   \left(\sqrt{{B^{\beta^0}}}+e^Z\right) \cos \left(\sqrt{{B^{\beta^0}}} e^{-Z}\right)+3 \sqrt{2} (c_1-c_2)
   \left(e^Z-\sqrt{{B^{\beta^0}}}\right) \sin \left(\sqrt{{B^{\beta^0}}} e^{-Z}\right)\nonumber\\
& &   +4 \pi 
   \left(c_1{}^2-c_2{}^2\right) e^Z \left(3 \sin \left(2 \sqrt{{B^{\beta^0}}} e^{-Z}\right)+1\right)+2 (c_1-c_2)
   \left(e^Z \cos \left(2 \sqrt{{B^{\beta^0}}} e^{-Z}\right)+2 \sqrt{{B^{\beta^0}}}\right) \csc
   \left(\sqrt{{B^{\beta^0}}} e^{-Z}-\frac{\pi }{4}\right)\nonumber\\
   & & \times \left(3+8 \pi  (c_1+c_2) \sin \left(\sqrt{{B^{\beta^0}}}
   e^{-Z}+\frac{\pi }{4}\right)\right).
\end{eqnarray}
}
In the $w\rightarrow0$-limit and retaining terms only linear in $w$,
{\footnotesize
\begin{eqnarray}
\label{ImdEToverET-ii}
& & \hskip -0.8in \lim_{w\rightarrow0}\frac{3
   \Im m\left(\frac{\sqrt{{B^{\beta^0}}} {g_s} {N^f_{\rm UV}} e^{-Z} \csc ^2\left(\sqrt{{B^{\beta^0}}} e^{-Z}-\frac{\pi
   }{4}\right) \left(3 \sin \left(\sqrt{{B^{\beta^0}}} e^{-Z}+\frac{\pi }{4}\right)+8 \pi 
   (c_1+c_2)\right)}{c_1-c_2}\right)}{128 \pi }  =\frac{3 \left(\frac{3}{\sqrt{2}}+8 \pi  (c_1+c_2)\right) {g_s} \sqrt{N} {N^f_{\rm UV}} w e^{-{ZUV}}}{64 \sqrt{\pi
   } (c_1-c_2) {r_h}};\nonumber\\
& & \hskip -0.8in \lim_{w\rightarrow0}\frac{3 \Im m\left(\frac{\beta  {B^{\beta}} \sqrt{\frac{1}{{B^{\beta^0}}}} {g_s} {N^f_{\rm UV}} e^{-2 Z} \csc
   ^2\left(\sqrt{{B^{\beta^0}}} e^{-Z}-\frac{\pi }{4}\right) \mathbb{N}_2}{(c_1-c_2){}^2}\right)}{256 \pi }
 \nonumber\\          
& & \hskip -0.8in = \Im m \left[\frac{9 \beta  \left(\sqrt{2}+4 \pi  (c_1+c_2)\right) {g_s} {N^f_{\rm UV}} w e^{-Z} \left({\cal C}_{zz} - 2 {\cal C}_{\theta_1z} + 2 
  {\cal C}_{\theta_1x}\right) \left(9 {g_s}^2 {N^f_{\rm UV}}^2 \log ^2({r_h})+12 \pi  {g_s} {N^f_{\rm UV}} \log ({r_h})-16 \pi
   ^2\right)}{2048 \pi ^{5/2} (c_1-c_2) {r_h} \sqrt{\frac{1}{{g_s} N}}}\right].   
\end{eqnarray}
}
Hence,
{\footnotesize
\begin{eqnarray}
\label{spectral-function-final-result-1}
& & \hskip-0.8in \lim_{w\rightarrow0}\left(\frac{\rho(T\neq0)}{\omega}-\frac{\rho(T=0,w)}{w}\right) \nonumber\\
& & \hskip -0.8in = \tilde{\kappa}_1 \frac{e^{Z_{\rm UV}} N^{3/5}N^f_{\rm UV} r_h\left(8\pi - g_s N^f_{\rm UV}\log r_h\right)}{\sqrt{g_s}} + \tilde{\kappa}_2 e^{Z_{\rm UV}}N^f_{\rm UV}N^{3/5}\left({\cal C}_{zz} - 2 {\cal C}_{\theta_1z} + 2 
  {\cal C}_{\theta_1x}\right)\left(64\pi^3-16\sqrt{g_s}N^f_{\rm UV}\pi^2\log r_h + 27 g_s^3 N^f_{\rm UV}(\log r_h)^3\right).\nonumber\\
  & & 
\end{eqnarray}
}
As summarized in (\ref{Neff}) - (\ref{rh-estimate}),
\begin{equation}
\label{r_h-est}
r_h\sim e^{-\frac{\sqrt[3]{N}}{3 \sqrt[3]{6 \pi } {g_s} M^{2/3} {N_f}^{2/3}}}
\equiv e^{-\kappa_{r_h}N^{1/3}},
\end{equation}
for $N_f=M=3, g_s=0.1,$ one obtains $\kappa_{r_h}\approx 0.3$. Also, assuming $N^f_{\rm UV}\sim N^{-\alpha_{N^f}}, \alpha_{N^f}>1/3$, (\ref{spectral-function-final-result-1}) may be rewritten as:
\begin{eqnarray}
\label{spectral-function-final-result-2}
& & \hskip-0.8in \lim_{w\rightarrow0}\left(\frac{\rho(T\neq0)}{\omega}-\frac{\rho(T=0,w)}{w}\right) \nonumber\\
& & \hskip -0.8in = \kappa_1 \frac{e^{Z_{\rm UV}} N^{3/5}N^f_{\rm UV} r_h\left(\log r_h\right)^3}{\sqrt{g_s}N} + \kappa_2 \frac{e^{Z_{\rm UV}}N^f_{\rm UV}N^{3/5}r_h\left({\cal C}_{zz} - 2 {\cal C}_{\theta_1z} + 2 
  {\cal C}_{\theta_1x}\right)\left(\log r_h\right)^3}{\kappa_{r_h}^3N}\beta.\nonumber\\
  & & 
\end{eqnarray}
As $e^{Z_{\rm UV}}\sim N^{1/\mathbb{N}_{>4}}, \mathbb{N}_{>4}$ being an integer greater than four, choose: $N^f_{\rm UV}\sim N^{-3/5-1/\mathbb{N}_{>4}}$. Further, ${\cal C}_{zz} - 2 {\cal C}_{\theta_1z}=0, |{\cal C}_{\theta_1x}|\ll1$ if one considers the self-consistent truncation wherein the ${\cal O}(R^4)$ corrections to $C_{MNP}$ in ${\cal M}$-theory dual are set to zero \cite{OR4}. We hence $| {\cal C}_{\theta_1x}|\sim\kappa_{r_h}^3$. Therefore, using (\ref{sq-1over3-cssq}),
\begin{eqnarray}
\label{spectral-function-final-result-3}
& &  \lim_{w\rightarrow0}\left(\frac{\rho(T\neq0)}{\omega}-\frac{\rho(T=0,w)}{w}\right) = 
\hat{\kappa}_1 \frac{r_h\left(\log r_h\right)^3}{N} + \hat{\kappa}_2 \frac{r_h\left(\log r_h\right)^3}{N}\beta.
\end{eqnarray}
Now, one can show that for QCD-inspired values of Table \tcb{1} and $N=100$, for $<0\epsilon<1$ and hence 
$T\in[T_c, 1.45T_c)$, $\alpha = 5.6{\cal O}(1)(g_s N_f)|\log r_h|^3$. As $\frac{1}{3} - c_s^2 = -\frac{2g_sM^2\alpha}{\sqrt{3}N}$, the above discussion using (\ref{TT-JJ}), suggests :
\begin{eqnarray}
\label{spectral-function-final-result-4}
& & \zeta = {\cal C}_1(r_h; g_s, M_{\rm UV}, N_f^{\rm UV} N)\left(\frac{1}{3} - c_s^2\right) + {\cal C}_2(r_h; g_s, M_{\rm UV}, N_f^{\rm UV} N)\left(\frac{1}{3} - c_s^2\right)^2.
\end{eqnarray}
Note, (\ref{spectral-function-final-result-4}) will be put on a much more rigorous footing in Section \ref{Eling+Oz}. From (\ref{1over3-cssq}), e.g., near the QCD-inspired values of $(g_s, M, N_f)$ of Table \tcb{1} and $N=100$, one can show that if ${\rm Sgn}({\cal C}_1)=-1$ (by appropriate choice of the two constants of integration appearing in (\ref{ETtilde-2})), then for $\epsilon\in(0,1), \frac{\zeta}{\eta}(T)>0, \frac{\zeta}{\eta}'(T)<0$ (see \cite{dzetaovereta-negative}; also see \cite{Trace anomaly_AM+CG} for a discussion on QCD trace anomaly's temperature variation from the ${\cal M}$-theory dual of this paper) $\forall T\in[T_c, (1.2 - 2.3)T_c)$.  Hence using the Kubo's formula for the bulk viscosity for computing the spectral function $\mathcal{R}_{x_1x_1}$, we obtained the result that the speed-of-sound dependence of the bulk viscosity for large-but-finite-$N$ thermal QCD-like theories at intermediate coupling dual to $\cal M$ theory with higher derivative (quartic in curvature), is a linear combination of the dependences on the speed of sound in the strong and weak coupling limits.

\section{$\frac{\zeta}{\eta}$ using Eling-Oz Formula \cite{EO}}
\label{Eling+Oz}

In the last stage of this journey of hydrodynamical studies of large-but-finite-$N$ thermal QCD-like theories dual to $\cal M$-theory, we consider the  Eling and Oz's results \cite{EO} to compute the bulk-to-shear viscosity ratio. Using the formula (\ref{EO-1}), with the ${\cal M}$-theory three-form potential $C_3$ given in terms of $B_{\rm NS\ NS}^{\rm IIA}$ (and $dx^{10}$ corresponding to the ${\cal M}$-theory circle), by considering the most dominant and sub-dominant components of $C_3/B_2^{\rm IIA}$ in the near horizon limit/IR, after an appropriate angular regularization, we will show that one obtains the bulk-to-shear-viscosity ratio upto terms quartic in curvature in the ${\cal M}$-theory action. Then, using (\ref{EO-2}) or its eleven-dimensional generalization (\ref{EO-D=11-i}), retaining only the most dominant aforementioned $C_3/B_2^{\rm IIA}$ component, we again work out the shear-to-bulk-viscosity ratio up to ${\cal O}(R^4)$.  In all computations, we will show that one obtains the speed-of-sound dependence of the abovementioned ratio to be given as linear combination of weak  and strong coupling results.

Now \cite{EO}, 
\begin{equation}
\label{EO-1}
\frac{\zeta}{\eta} = c_s^4 T^2\sum_i \left.\left(\frac{d\Phi^i}{dT}\right)^2\right|_{r=r_h}.
\end{equation}
Further, from \cite{MQGP}, near (\ref{Ouyang-definition}) effected via (\ref{alpha_theta_12}), $B_{\rm NS-SN}^{\rm IIA}$ components are listed in (\ref{BNSNSIIA}). 
Near (\ref{r_h-est}), one hence sees that $B_{\theta_1\theta_2}^{\rm IIA}$ and $B_{\theta_1x}^{\rm IIA}$ are the most dominant components. The corresponding kinetic term will be given as: 
\begin{eqnarray}
\label{kinetic-term-Ctheta1theta2x10}
& & \int \sqrt{-g_5}\sqrt{g_6}\left( G^{r\theta_1\theta_2x^{10}} G_ {r\theta_1\theta_2x^{10}} + 
  G^{r\theta_1xx^{10}} G_ {r\theta_1xx^{10}}\right)\nonumber\\
& &  = \int  \sqrt{-g_5}\sqrt{g_6}\Biggl[
 g^{rr} g^{x^{10}x^{10}} \left(\left[
     g^{\theta_1\theta_1} g^{\theta_2\theta_2} - (g^{\theta_1\theta_2})^2\right] 
G_{r\theta_1\theta_2x^{10}} + \left[
     g^{\theta_1x} g^{\theta_2\theta_1} - 
      g^{\theta_1\theta_1} g^{\theta_2x}\right] G_ {rx\theta_1x^{10}}\right) 
G_{r\theta_1\theta_2x^{10}}\nonumber\\
& & + g^{rr} g^{x^{10}x^{10}} \left( \left[
    g^{\theta_1\theta_1} g^{xx} - (g^{\theta_1x})^2\right] G_ {r\theta_1xx^{10}} - \left[
    g^{\theta_1\theta_2} g^{x\theta_1} - 
     g^{\theta_1\theta_1} g^{x\theta_2}\right] G_ {r\theta_1\theta_2x^{10}}\right) 
G_{r\theta_1xx^{10}}\Biggr]\nonumber\\
& & \approx \int \sqrt{-g_5}\sqrt{g_6}g^{rr} g^{x^{10}x^{10}} \left[
     g^{\theta_1\theta_1} g^{\theta_2\theta_2} - (g^{\theta_1\theta_2})^2\right] 
\left(\partial_r B^{\rm IIA}_{\theta_1\theta_2}\right)^2. 
\end{eqnarray}
From (\ref{sqrtdetg6}) - (\ref{regularization}), one can show that the principal part of the $\theta_{1,2}$ double integral, is given by:
\begin{equation}
\label{Principal-Obeta0}
\frac{5987513439\ 3^{5/6}}{5600 \sqrt{2}}\approx10^6. 
\end{equation}

One hence finally obtains:
{\footnotesize
\begin{eqnarray}
\label{PVbeta0}
& & \hskip -0.8in {\cal P}\left.\int_{S_{\rm squashed}^2\times_w S_{\rm squashed}^3}\sqrt{g_6}
 \left(g^{rr} g^{x^{10}x^{10}}\left[
     g^{\theta_1\theta_1} g^{\theta_2\theta_2} - (g^{\theta_1\theta_2})^2\right]\left(\partial_rC_{\theta_1\theta_2x^{10}}\right)^2\right)^{\beta^0}\right|_{r=r_h}\nonumber\\
& & \hskip -0.8in = \frac{0.5\times10^{-2} c_2 N^{5/4} {N_f}^{10/3} g_s^{21/4} }{(c_1+c_2 \log ({r_h}))^2 (\log (N)-3 \log ({r_h}))^{2/3}r_h^2}.
\end{eqnarray} 
}

Now, at ${\cal O}(\beta)$, one needs to evaluate:
{\footnotesize
\begin{eqnarray}
\label{intbetatheta1theta2LON-main-text}
& & {\cal P}\left.\int_{S_{\rm squashed}^2\times_w S_{\rm squashed}^3}\sqrt{g_6}
 g^{rr} g^{x^{10}x^{10}}\left[
     g^{\theta_1\theta_1} g^{\theta_2\theta_2} - (g^{\theta_1\theta_2})^2\right]\left(\partial_rC_{\theta_1\theta_2x^{10}}\right)^2\right|_{r=r_h}^{\beta}.
\end{eqnarray}
}
From (\ref{intbetatheta1theta2LON}) - (\ref{delta2intermsofdelta1}), one sees that:
\begin{eqnarray}
\label{PVbeta}
& & \hskip -0.8in {\cal P}\left.\int_{S_{\rm squashed}^2\times_w S_{\rm squashed}^3}\sqrt{g_6}
 \left(g^{rr} g^{x^{10}x^{10}}\left[
     g^{\theta_1\theta_1} g^{\theta_2\theta_2} - (g^{\theta_1\theta_2})^2\right]\left(\partial_rC_{\theta_1\theta_2x^{10}}\right)^2\right)^{\beta}\right|_{r=r_h}
\nonumber\\
& & = \frac{2\times 10^{-8}{\cal C}_2 g_s^{21/4}MN^{5/4}N_f^{4/3}\left(\log N - 3 \log r_h\right)^{1/3}}{r_h^2(\log r_h)^2(c_1 + c_2 \log r_h)}.
\end{eqnarray}

As $c_s^4 = \left(\frac{1}{3} - c_s^2\right)^2 -\frac{2}{3}\left(\frac{1}{3} - c_s^2\right)  +\frac{1}{9}$, one hence obtains:
\begin{eqnarray}
\label{EO-2}
& & \frac{\zeta}{\eta} = M N^{5/4}N_f^{4/3}g_s^{21/4}\Biggl[-\frac{10^{-4}\left(c_2\left\{-\frac{1}{6} + \left(\frac{1}{3} - c_s^2\right)^{\beta^0}\right\} N_f^{2}\right)}{3(\log N - 3 \log r_h)^{2/3}(c_1 + c_2\log r_h)^2} \nonumber\\
& & + \Biggl(-\frac{2\times10^{-7}{\cal C}_2\left(\frac{1}{9} - \frac{2\left(\frac{1}{3} - c_s^2\right)^{\beta^0}}{3}\right)(\log N - 3 \log r_h)^{1/3}}{(\log r_h)^2(c_1 + c_2\log r_h)}\nonumber\\
& & +\frac{5\times10^{-5}c_2N_f^2\left(-\frac{2}{3} \left(\frac{1}{3} - c_s^2\right)^{\beta}+ \left(\frac{1}{3} - c_s^2\right)^{2,\ \beta}\right)}{(\log N - 3 \log r_h)^{1/3}}\Biggr)\beta\Biggr].
\end{eqnarray} 
Now, $\beta\sim l_p^6=\left(\frac{\sqrt{G_{x^{10}x^{10}}^{\cal M}}}{g_s^{2/3}}\right)^6$ where $\sqrt{G_{x^{10}x^{10}}^{\cal M}}$ is the size of the ${\cal M}$-theory circle near the $\psi=2n\pi, n=0, 1, 2$-patches. Now, near (\ref{alpha_theta_12}),
{\footnotesize
\begin{eqnarray}
\label{M-theory-circle-metric}
& & G_{x^{10}x^{10}}^{\cal M} = \frac{16\ 2^{2/3} \pi ^{7/3} \left(\frac{24 a^2 {g_s} M^2 N_f
   \left(c_1+c_2 \log \left(r_h\right)\right)}{9 a^2+r^2}+\frac{3 N N_f (\log (N)-3 \log (r))}{4 \pi
   }\right)}{9 \sqrt[3]{3} N (N_f (\log (N)-3 \log (r)))^{7/3}}\nonumber\\
& & + \frac{12\ 6^{2/3} \sqrt[3]{\pi } \left(9 b^2+1\right)^4   b^{10} M r \left(19683 \sqrt{6} \alpha _{\theta
   _1}^6+6642 \alpha _{\theta _2}^2 \alpha _{\theta _1}^3-40 \sqrt{6} \alpha _{\theta _2}^4\right) \left(r-2 r_h\right)
   \left(6 a^2+r_h^2\right) \log ^3\left(r_h\right)}{\left(3 b^2-1\right)^5 \left(6 b^2+1\right)^4 N^{5/4} r_h^4
   \alpha _{\theta _2}^3 N_f \log ^4(N) \left(9 a^2+r_h^2\right) \left(N_f (\log (N)-3 \log
   (r))\right){}^{4/3}}\beta. 
\end{eqnarray}
}
As $G_{x^{10}x^{10}}^{{\cal M},\ \beta}\sim \frac{\beta}{N^{5/4}}$, one can set $G_{x^{10}x^{10}}^{{\cal M},\ \beta} = 0$. 

From (\ref{Neff}) - (\ref{rh-estimate}),  one ses that 
\begin{equation}
\label{interpolating-i}
G_{x^{10}x^{10}}^{{\cal M},\ \beta^0}(r\in{\rm IR})\sim \frac{1}{|\log r_h|^{4/3}}\sim N^{-4/9}.
\end{equation}
Thus, the ${\cal O}(l_p^6)$ contribution to $\frac{\zeta}{\eta}$ in (\ref{summary-zetaovereta}) will be $\frac{|\log r_h|^{8/3}}{g_s^4 |\log r_h|^{13/3}}\sim \frac{1}{g_s^3 |\log r_h|^{5/3}}\sim\frac{1}{g_s^{22/9} L^{20/9}}$-suppressed ($L\equiv (4\pi g_s N)^{1/4}$) in the $N\gg1$ limit as compared to the leading order/$\beta^0$ contribution. 

One hence sees that: 
\begin{enumerate}
\item in the weak-coupling limit - $L\ll1$ - the ${\cal O}(R^4)/{\cal O}(\beta)$ contributions to $\frac{\zeta}{\eta}$ encode the weak coupling result $\frac{\zeta}{\eta} \sim \left(\frac{1}{3} - c_s^2 \right)^2$; the novelty here is that from (\ref{1over3-cssq}) we see that at weak coupling, $\frac{\zeta}{\eta}$ receives contribution from the ${\cal O}(R^4)/{\cal O}(\beta)$ component of $\frac{1}{3} - c_s^2$  as well; 
\item in the strong-coupling limit - $L\gg1$ - the ${\cal O}(R^4)/{\cal O}(\beta)$ contributions become negligible and one hence obtains the strong-coupling result
$\frac{\zeta}{\eta}\sim\left(\frac{1}{3} - c_s^2\right)^{\beta^0}$;
\item in the {\bf intermediate coupling limit} $L\sim{\cal O}(1)$, one obtains a linear combination of the strong-coupling and weak-coupling results, i.e.,  the ${\cal O}(R^4)/{\cal O}(\beta)$ contributions interpolate between the strong and weak coupling results of $\frac{\zeta}{\eta}$.
\end{enumerate}

One also notes from (\ref{EO-2}) that, as expected, $\frac{\zeta}{\eta}\rightarrow0$ as one turns off the non-conformality by  setting $M$ and $N_f$ to zero. 

Now \cite{Gopal-Tc-Vorticity}
\begin{equation}
\label{T-rh}
T = \frac{r_h}{\sqrt{3}\pi^{3/2}\sqrt{g_s N}}\left(1 +\frac{\beta}{2}\left[- {\cal C}_{zz} + 2 {\cal C}_{\theta_1 z} - 3 {\cal C}_{\theta_1 x}\right]\right), 
\end{equation}
and $T_c = \frac{r_h}{2\pi^{3/2}\sqrt{g_s N}}$ (as well as (\ref{rh-r0-beta0})) with $r_h$ related to the IR cut-off $r_0$ of the thermal ${\cal M}$-theory background dual to thermal QCD-like theories for low temperatures (i.e., $T<T_c$) via (\ref{r_h-r_0-relation}).  In the consistent truncation of the ${\cal M}$-theory uplift wherein the ${\cal O}(R^4)$- or equivalently ${\cal O}(\beta)$-corrections to the three-form potential $C_{MNP}$ can be set to zero, it was shown that ${\cal C}_{zz} = 2 {\cal C}_{\theta_1z}$ and $|{\cal C}_{\theta_1x}|\ll1$ implying the non-renormalization of $T$ up to ${\cal O}(R^4)$ (in the zero-instanton sector). Further, $r$ (and hence $r_h$) is in fact $\frac{r}{{\cal R}_{D5/\overline{D5}}^{\rm bh}},$ where 
\begin{equation}
\label{RD5antiD5}
{\cal R}_{D5/\overline{D5}}=\sqrt{3}r_h\left(\frac{1}{\sqrt{3}} + \varepsilon \right), 
\end{equation}
with \cite{OR4}
\begin{equation}
\label{varepsilon-value}
\varepsilon  = -\kappa_b r_h^2 (|\log r_h|)^{9/2}N^{-9/10 - \tilde{\kappa}_{b}} + \frac{g_s M^2(c_1 + c_2\log r_h)}{N}>0, \tilde{\kappa}_b>0.
\end{equation}
Hence,
\begin{equation}
\label{rh-Tc}
r_h = \frac{\sqrt{3}T}{2(1 + \sqrt{3}\varepsilon )T_c}.
\end{equation}

\begin{tcolorbox}[enhanced,width=6.8in,center upper,size=fbox,
    drop shadow southwest,sharp corners]
    \begin{flushleft}
As $r_h<1$, hence our ${\cal M}$-theory computations are valid for the following range of temperatures above $T_c$:
\begin{equation}
\label{range-T}
T\in[T_c, \frac{2(1 + \sqrt{3}\varepsilon )}{\sqrt{3}}T_c].
\end{equation}
\end{flushleft}
\end{tcolorbox}

 Near the QCD-inspired values of $g_s, M, N_f$ of Table \tcb{1}, (\ref{EO-2}) yields, 
\begin{eqnarray}
\label{EO-4}
& & \hskip -0.8in \left(\frac{\zeta}{\eta}\right)^{\beta^0}(g_s=0.1, M=N_f=3, N=100; T = T_c(1 + \delta): 0<\delta<\frac{2 - \sqrt{3} + 2\sqrt{3}\varepsilon }{\sqrt{3}}) \nonumber \\
& & \hskip -0.8in= \frac{1.5\times10^{-7} (c_1 - 5.2 c_2)c_2\delta}{(c_1 - 0.14 c_2)^3} + {\cal O}(\delta\varepsilon ).
\end{eqnarray}
As $c_{1,2}<0$ \cite{IITR-McGill-bulk-viscosity}, $\left(\frac{\zeta}{\eta}\right)^{\beta^0\ \prime}\left(T\in[T_c, \frac{2(1 + \sqrt{3}\varepsilon )}{\sqrt{3}}T_c)\right)<0$ provided $c_2>2c_1$.

One can similarly show that:
\begin{eqnarray}
\label{EO-5}
& & \hskip -0.8in \left(\frac{\zeta}{\eta}\right)^{\beta}\ '(g_s=0.1, M=N_f=3, N=100; T = T_c(1 + \delta): 0<\delta<\frac{2 - \sqrt{3} + 2\sqrt{3}\varepsilon }{\sqrt{3}})<0 \nonumber \\
\end{eqnarray}
provided:
\begin{eqnarray}
\label{EO-6}
& & {\cal C}_2>\left|\frac{0.89 c_1^2 c_2\kappa_{\alpha_\beta} - 3.82 c_1 c_2^2\kappa_{\alpha_\beta} + 14 c_2^3 \kappa_{\alpha_\beta}}{c_1^3 - 0.53 c_1^2c_2 + 0.1 c_1 c_2^2 - 0.01 c_2^3}\right|.
\end{eqnarray}

Following another approach as spelled out in \cite{EO}, assuming $C_{\theta_1\theta_2 x^{10}} = B_{\theta_1\theta_2}^{\rm IIA}$ to be the only scalar in the ${\cal M}$-theory uplift and then working in the gauge $r = C_{\theta_1\theta_2 x^{10}}$, from a five-dimensional perspective (delocalized along the six-dimensional compact space $M_6(\theta_{1,2}, \phi_{1,2}, \psi, x^{10})$) 
\begin{equation}
\label{EO-7}
\frac{\zeta}{\eta} = \left(\frac{1}{3 A'(C_{\theta_1\theta_2 x^{10}})}\right)^2,
\end{equation}
where \cite{IITR-McGill-bulk-viscosity},
\begin{equation}
\label{A-def}
A\equiv\frac{e^{-\frac{2\phi^{\rm IIA}(C_{\theta_1\theta_2 x^{10}})}{3}}}{\sqrt{h(C_{\theta_1\theta_2 x^{10}})}}.
\end{equation}

From (\ref{A+h-defs}) - (\ref{Sigma12-defs}), one obtains:
{
\begin{eqnarray}
\label{A}
& & \hskip -0.8in A(C_{\theta_1\theta_2 x^{10}}) = \frac{1}{2} \log \left(\frac{\left(\frac{3}{\pi
   }\right)^{2/3} \Sigma_2^{2/3}}{4 \sqrt{\frac{64 \pi  b^2 \left(b-\frac{1}{\sqrt{3}}\right)
   {g_s} {N_f} {r_h}^2}{\left(9 b^2 {r_h}^2+C_{\theta_1\theta_2 x^{10}}^2\right)
   \left(g_s\Sigma_2\right)}+1}}\right)-\frac{3 {g_s} M^2
   \log (C_{\theta_1\theta_2 x^{10}}) \Sigma_2}{64 \pi ^2 N}\nonumber\\
& & \hskip -0.8in +\frac{1}{2} \log \left(\frac{C_{\theta_1\theta_2 x^{10}}^2}{2 \sqrt{\pi }
   \sqrt{{g_s}} \sqrt{N}}\right) -\frac{3\ 3^{2/3} \left(9 b^2+1\right)^3 \beta  b^{10} M
   \left(\frac{1}{N}\right)^{5/4} r \Sigma_1 (C_{\theta_1\theta_2 x^{10}}-2 {r_h}) \log ^3({r_h}) }{8 \pi ^{5/3} \left(3
   b^2-1\right)^5 \left(6 b^2+1\right)^3 {N_f} {r_h}^4 \alpha _{\theta
   _2}^3 \log ^4(N) \left(9 b^2 {r_h}^2+r^2\right)^2 \left(g_s \Sigma_2 \right)^2}\nonumber\\
& & \hskip -0.8in \times \Biggl(6 {g_s}
   {N_f} \log (N) \left(9 b^2 {r_h}^2+C_{\theta_1\theta_2 x^{10}}^2\right)+8 \pi  \left[b^2
   {r_h}^2 \left(27-8 \left(\sqrt{3}-3 b\right) {g_s} {N_f}\right)+3
   r^2\right]\nonumber\\
& & \hskip -0.8in   -3 {g_s} {N_f} \left(9 b^2 {r_h}^2+C_{\theta_1\theta_2 x^{10}}^2\right) \log
   \left(9 b^2 C_{\theta_1\theta_2 x^{10}}^4 {r_h}^2+C_{\theta_1\theta_2 x^{10}}^6\right)\Biggr)^2.
\end{eqnarray}
}
Writing $\frac{1}{\left(A_{\beta^0}'(C_{\theta_1\theta_2 x^{10}}) + \beta A_{\beta}'(r)\right)^2} = \frac{1}{\left(A_{\beta^0}'(C_{\theta_1\theta_2 x^{10}})\right)^2}-2\frac{A_{\beta}'(C_{\theta_1\theta_2 x^{10}})}{\left(A_{\beta^0}(C_{\theta_1\theta_2 x^{10}})\right)^3}\beta$, we see that up to LO in $N$,
{\footnotesize
\begin{eqnarray}
\label{1overdAbeta0sq}
& & \hskip -0.8in \left(\frac{1}{\left(A_{\beta^0}'(C_{\theta_1\theta_2 x^{10}}=r_h)\right)^2}\right)^{\rm LO} \nonumber\\
& & \hskip -0.8in = \frac{{r_h}^2}{\left(\frac{32 \pi  \left(3 b-\sqrt{3}\right) b^2 {g_s}
   {N_f} \left({g_s} {N_f} \left(-18 b^2+2
   \log N -3\right)-{g_s} {N_f} \log \left(\left(9 b^2+1\right)
   {r_h}^6\right)+8 \pi \right)}{\left(9 b^2+1\right) \left({g_s}\Sigma_1 \right) \left(6 \left(9 b^2+1\right) {g_s}
   \log N  {N_f}-3 \left(9 b^2+1\right) {g_s} {N_f} \log
   \left(\left(9 b^2+1\right) {r_h}^6\right)+64 \pi  \left(3
   b-\sqrt{3}\right) b^2 {g_s} {N_f}+24 \left(9 \pi  b^2+\pi
   \right)\right)}-\frac{2 \left(6 b^2+1\right) {g_s} {N_f}}{\left(9
   b^2+1\right) \left(g_s\Sigma_1\right)}+1\right)^2}\nonumber\\
& & \hskip -0.8in +\frac{3 {g_s} M^2 {r_h}^2 (-{g_s}
   \log N  {N_f}+24 {g_s} {N_f} \log ({r_h})+8 \pi )}{32 \pi
   ^2 N \left(\frac{32 \pi  \left(3 b-\sqrt{3}\right) b^2 {g_s} {N_f}
   \left({g_s} {N_f} \left(-18 b^2+2 \log N -3\right)-{g_s}
   {N_f} \log \left(\left(9 b^2+1\right) {r_h}^6\right)+8 \pi
   \right)}{\left(9 b^2+1\right) \left(g_s\Sigma_1
   \right) \left(6 \left(9 b^2+1\right) {g_s} \log N  {N_f}-3
   \left(9 b^2+1\right) {g_s} {N_f} \log \left(\left(9 b^2+1\right)
   {r_h}^6\right)+64 \pi  \left(3 b-\sqrt{3}\right) b^2 {g_s}
   {N_f}+24 \left(9 \pi  b^2+\pi \right)\right)}-\frac{2 \left(6
   b^2+1\right) {g_s} {N_f}}{\left(9 b^2+1\right) \left(g_s\Sigma_1
   \right)}+1\right)^3}\nonumber\\
& & \hskip -0.8in   + {\cal O}\left(\frac{1}{N^2}\right).
\end{eqnarray}
}
Performing a large-$|\log r_h|$ expansion, $\left(\frac{1}{\left(A_{\beta^0}'(C_{\theta_1\theta_2 x^{10}}=r_h)\right)^2}\right)^{\rm LO} = \sum_{n=1}\frac{{\cal C}_n(r_h, N, N_f, g_s)}{(\log r_h)^n}$, near $N=100, g_s=0.1, N_f=3, r_h\sim e^{-0.3 N^{1/3}}$, one sees that by truncating the sum at $\left({\cal O}(1)\right)^2$ number of terms,  $\left.\left(\frac{1}{\left(A_{\beta^0}'(C_{\theta_1\theta_2 x^{10}}=r_h)\right)^2}\right)^{\rm LO}\right|_{b\sim\frac{1}{{\cal O}(1)}}=0$ up to LO in $N$. Up to NLO in $N$, one sees that
{\footnotesize
\begin{eqnarray}
\label{1overdAbeta0NLON}
& &\hskip -0.8in \left(\frac{1}{\left(A_{\beta^0}'(C_{\theta_1\theta_2 x^{10}}=r_h)\right)^2}\right)^{\rm NLO} \nonumber\\
& & \hskip -0.8in = \frac{3 {g_s} M^2 {r_h}^2 (-{g_s} \log N  {N_f}+24 {g_s} {N_f} \log
   ({r_h})+8 \pi )}{32 \pi ^2 N \left(\frac{8 \pi  \left(3 b-\sqrt{3}\right) b^2 {g_s} {N_f}
   \left({g_s} {N_f} \left(-18 b^2+2 \log N -3\right)-6 {g_s} {N_f} \log ({r_h})+8
   \pi \right)}{\left(9 b^2+1\right)  \left(\frac{g_s\Sigma_1}{2}\right) \left({g_s} {N_f} \left(3 \left(9 b^2+1\right) \log N +32 \pi  \left(3
   b-\sqrt{3}\right) b^2\right)-9 \left(9 b^2+1\right) {g_s} {N_f} \log ({r_h})+12 \left(9 \pi 
   b^2+\pi \right)\right)}-\frac{\left(6 b^2+1\right) {g_s} {N_f}}{\left(9 b^2+1\right)  \left(\frac{g_s\Sigma_1}{2}\right)}+1\right)^3}.
\end{eqnarray}
}
One can show that
{\footnotesize
\begin{eqnarray}
& & \hskip -0.8in 1 + \frac{8 \pi  \left(3 b-\sqrt{3}\right) b^2 {g_s} {N_f} \left({g_s} {N_f} \left(-18 b^2+2
   \log N -3\right)-6 {g_s} {N_f} \log ({r_h})+8 \pi \right)}{\left(9 b^2+1\right)
   \left(\frac{g_s\Sigma_1}{2}\right) \left({g_s} {N_f}
   \left(3 \left(9 b^2+1\right) \log N +32 \pi  \left(3 b-\sqrt{3}\right) b^2\right)-9 \left(9
   b^2+1\right) {g_s} {N_f} \log ({r_h})+12 \left(9 \pi  b^2+\pi \right)\right)}\nonumber\\
& & \hskip -0.8in-\frac{\left(6
   b^2+1\right) {g_s} {N_f}}{\left(9 b^2+1\right)  \left(\frac{g_s\Sigma_1}{2}\right)}=\frac{\xi}{\left(\log r_h\right)^{2/3}}, \xi\equiv {\cal O}(1).
\end{eqnarray}
}

The above is hence suggestive of:
\begin{equation}
\label{zetaoveretabeta0}
\left(\frac{\zeta}{\eta}\right)^{\beta^0}\sim\frac{(g_s M^2)(g_s N_f) r_h^2\left(\log r_h\right)^3}{N}\sim r_h^2\left(\frac{1}{3} - c_s^2\right)^{\beta^0}.
\end{equation}

One similarly sees:
\begin{eqnarray}
\label{zetaoveretabetaLON}
& & \hskip -0.8in \left(\frac{\zeta}{\eta}\right)^\beta \sim -\frac{{\cal O}(10)^{-2}) b^{12} \beta {g_s}^2 M^2
   \left(\frac{1}{N}\right)^{7/4} \sqrt{{N_f}}
  \Sigma_1 \sqrt{-\log ({r_h})}
   \left[{g_s} {N_f} \left(-18 b^2+2 \log
   (N)-3\right)-6{g_s}{N_f} \log ({r_h})+8 \pi
   \right]}{\left(b+\frac{1}{\sqrt{3}}\right)^5
   \left(6 b^2+1\right)^3 \kappa_{A_{\beta^0}'}^3 \epsilon
   ^4 \alpha _{\theta _2}^3 \log ^4(N) ({g_s} {N_f}
   \log (N)-3 {g_s} {N_f} \log ({r_h})+4 \pi )^3}\nonumber\\
& &\hskip -0.8in \times \left[ \left\{-{g_s} {N_f} \left(3 \left(9
   b^2+1\right) \log (N)+32 \pi  \left(3 b-\sqrt{3}\right)
   b^2\right)+9 \left(9 b^2+1\right) {g_s}{N_f} \log
   {r_h}\right\}-12 \left(9 \pi  b^2+\pi
   \right)\right]\left(\frac{1}{3} - c_s^2\right)^2\nonumber\\
& & \downarrow {\rm Table} \tcb{1}\nonumber\\
& &\hskip -0.8in \frac{{{\cal O}(10^{-13})} \beta 
   \left(48213 \alpha _{\theta _1}^6+6642 \alpha _{\theta_2}^2 \alpha _{\theta _1}^3-98 \alpha _{\theta_2}^4\right)}{\kappa_{A_{\beta^0}'}^3 \epsilon ^4 
\alpha_{\theta _2}^3}\left(\frac{1}{3} - c_s^2\right)^2
\end{eqnarray}
where $ b = \frac{1}{\sqrt{3}} + \epsilon, \Sigma_1 \equiv  \left(19683 \sqrt{6} \alpha _{\theta _1}^6+6642 \alpha
   _{\theta _2}^2 \alpha _{\theta _1}^3-40 \sqrt{6} \alpha
   _{\theta _2}^4\right)$. Using (\ref{Sigma1overalphatheta2cubed}), and $\left.\frac{r_h^2}{L^2}\right|_{\rm Table\ \tcb{1}}\sim\frac{e^{-0.3 N^{1/3}}}{\sqrt{4\pi g_s N}}\stackrel{N=100, g_s=0.1}{\longrightarrow}10^{-3}$, one hence sees that (\ref{zetaoveretabetaLON}) is equal to:
{\footnotesize
\begin{eqnarray}
\label{PVSigma1overalphatheta2cubed-2}
& & \hskip -0.8in \left.\frac{{\cal O}(10^{-13}){\cal P}\int_{S^2_{\rm squashed}\times_w S^3_{\rm squashed}}\sin\theta_1 \sin\theta_2 \Biggl\{ 48213.3 N^{21/10} \left(-\frac{\sin \left(\theta
   _2\right)}{500}+\sin ^6\left(\theta _1\right) \csc
   ^3\left(\theta _2\right)+0.14 \sin ^3\left(\theta
   _1\right) \csc \left(\theta _2\right)\right)\Biggr\}}{\kappa_{A_{\beta^0}'}^3 \epsilon ^4 }\right|_{{\rm Table}\ \tcb{1}}\nonumber\\
   & & \hskip -0.8in \sim\frac{10^{-4}}{\kappa_{A_{\beta^0}'}^3 \epsilon ^4 \epsilon_\beta}\frac{r_h^2}{L^2},
\end{eqnarray}
}
where one regularizes (\ref{PVSigma1overalphatheta2cubed-2}) by choosing $\epsilon_\beta, \epsilon: \frac{10^{-4}}{\epsilon_\beta\xi^{3/2}\epsilon^4}\sim{\cal O}(1)$.

If one considers the full D=11 metric then the expression for the entropy becomes: 
$\frac{e^{3A(r_h)}\sqrt{g_6}(r_h)}{4},$ and hence generalizing (\ref{EO-7}) to: 
\begin{equation}
\label{EO-D=11-i}
\frac{\zeta}{\eta} = \frac{1}{\left(3 A'(r_h) + \frac{g_6'(r_h)}{g_6(r_h)}\right)^2}.
\end{equation}
 For 
\begin{equation}
\label{c_2}
c_2\sim
\frac{g_s\sqrt{N N_f}\log N}{\xi_2 M\sqrt{|\log r_h|}}\sim\frac{g_s\sqrt{N_f}|\log r_h|}{\xi_2M},
\end{equation}
 where $\xi_2$ is defined via: 
\begin{eqnarray}
\label{xi2-def}
& & \hskip -0.8in \frac{N (-2 \log (N)+6 \log ({r_h})+9)}{2 \sqrt{3} {g_s} M^2
   (\log (N)-3 \log ({r_h})) ({c_1}+{c_2} \log
   ({r_h}))}+\frac{-6 \log (N) \log
   ({r_h})+8 \log (N)+18 \log ^2({r_h})-9 \log ({r_h})}{4
   \log (N) \log ({r_h})-12 \log ^2({r_h})}\nonumber\\
   & & \hskip - 0.8in \sim \frac{\sqrt{N} {\xi_2}}{(-\log ({r_h}))^{3/2}
   \sqrt{{g_s}^2 M^2 {N_f}}},
\end{eqnarray}  
on obtains: 
\begin{equation}
\label{dg6overg6}
\frac{g_6'(r_h)}{g_6(r_h)}\sim\frac{\xi_2\sqrt{N}}{r_h\sqrt{(g_sM^2)(g_sN_f)|\log r_h|}}>A'(r_h).
\end{equation}
 Now, 
\begin{equation}
\label{EO-D=11-ii}
A'(r_h) + \frac{g_6'(r_h)}{2g_6(r_h)} \approx \left(\frac{g_6'(r_h)}{2g_6(r_h)}\right)^{\beta^0} + \beta\left(A'(r_h)\right)^\beta + {\cal O}(\beta^2),
\end{equation}
 implying
\begin{eqnarray}
\label{EO-D=11-iii}    
& & \frac{1}{\left(A'(r_h) + \frac{g_6'(r_h)}{2g_6(r_h)}\right)^2} = \frac{1}{\left[\left(\frac{g_6'(r_h)}{2g_6(r_h)}\right)^{\beta^0}\right]^2}  - \beta\frac{2\left(A'(r_h)\right)}{\left[\left(\frac{g_6'(r_h)}{2g_6(r_h)}\right)^{\beta^0}\right]^3} + {\cal O}(\beta^2)\nonumber\\
& & = \xi_3\frac{(g_sM^2)(g_sN_f)|\log r_h|^3}{N} + \frac{\xi_2 |\log r_h|^{5/2}\Sigma_1}{3^{11/6}\pi^{5/3}N^{3/4}N_f^{1/2}(\log N)^8\alpha_{\theta_2}^3}\beta.
\end{eqnarray}
Now, from (\ref{Sigma1overalphatheta2cubed}), $\frac{\Sigma_1}{\alpha_{\theta_2}^3}\rightarrow \frac{10^4 N^{3/10}}{\epsilon_\beta^2}$. Choosing $\xi_2\sim \frac{g_s^6}{\kappa_{\xi_2}},\ \kappa_{\xi_2}\sim{\cal O}(1)$ and regularizing via
 \footnote{Near QCD-inspired values of $g_s, M, N_f$ of Table \tcb{1} (implying $\kappa_{r_h}\approx0.3$ (\ref{r_h})) and $N=100$ (\ref{EO-D=11-regularization}) implies $\epsilon_\beta\sim10^{-5}$.} : 
\begin{eqnarray}
\label{EO-D=11-regularization}
M^2N_f^{3/2}\frac{\sqrt{\kappa_{r_h}N^{1/3}}}{N^{1/4+3/10+1/2}}e^{-2\kappa_{r_h}N^{1/3}}(\log N)^8\epsilon_\beta^2\sim{\cal O}(1),
\end{eqnarray}
with $\frac{10^4}{\kappa_{\xi_2}3^{11/6}\pi^{11/3}}\sim{\cal O}(1)$, one obtains:
\begin{eqnarray}
\label{EO-D=11-ii}
& & \frac{\zeta}{\eta} = \frac{(g_sM^2)(g_sN_f)|\log r_h|^3}{N}\frac{r_h^2}{L^2}\left(\xi_3 + \beta \xi_4\right).
\end{eqnarray} 

Hence,  (\ref{zetaoveretabeta0}), (\ref{zetaoveretabetaLON}) or (\ref{EO-D=11-ii}) are suggestive of:
\begin{eqnarray}
\label{zetaoveretaEO}
& & \frac{\zeta}{\eta} = \frac{r_h^2}{L^2}\left[ \chi_1 \left(\frac{1}{3} - c_s^2\right)^{\beta^0} + \chi_2\left(\frac{1}{3} - c_s^2\right)^2 \right],
\end{eqnarray}
($\left(\frac{1}{3} - c_s^2\right)^2\equiv{\cal O}(\beta)$)
in conformity with (\ref{spectral-function-final-result-3}). 
Analogous to the result (\ref{spectral-function-final-result-3}), (\ref{zetaoveretabeta0}) is put on a more rigorous footing in (\ref{EO-2}) - (\ref{M-theory-circle-metric}) resulting in (\ref{summary-zetaovereta}) and Fig. \tcb{2}.

In short via results of \cite{EO}, we found out that turning off the non-conformality, i.e. choosing $M$ and $N_f$ to zero, results in $\frac{\zeta}{\eta}\rightarrow 0$ as expected. Working in the gauge $r = C_{\theta_1\theta_2 x^{10}} = B_{\theta_1\theta_2}^{IIA}$ and in the intermediate-$N$ MQGP limit (\ref{MQGP_limit}), we establish the relation of the ratio $\frac{\zeta}{\eta}$ and the deviation of the square of the speed of sound from its conformal value - $\Biggl(\frac{1}{3}-c_s^2\Biggr)$ - both from a five-dimensional perspective with delocalized angular values, as well as  the full eleven-dimensional perspective. We find  that the ratio $\frac{\zeta}{\eta}$ is the linear combination of  $\Biggl(\frac{1}{3}-c_s^2\Biggr)$ (the strong coupling result) and $\Biggl(\frac{1}{3}-c_s^2\Biggr)^2$ (the weak coupling result). From a Math perspective, using the results of \cite{ACMS}, the aforementioned behavior is tied up to the existence of Contact 3-Structures in the intermediate-$N$ limit (\ref{MQGP_limit}).

\section{Summary and Significance of Results}
\label{Summary}
In an endeavor to study the hydrodynamical properties of strongly interacting Quark Gluon Plasma via top-down holographic QCD, recent pieces of literature have explored several hydrodynamical features of strongly coupled plasmas like shear-viscosity-to-entropy-density ratio, bulk-viscosity-to-shear-viscosity ratio, etc., in different coupling limits. In this paper, we answer the question of what the bulk-to-shear-viscosity ratio of thermal QCD-like theories will be at intermediate coupling effected via finite-$N\ {\cal M}$-theory dual having incorporated higher derivative corrections (quartic in the curvature) in the latter. In \ref{summary}, we provide a summary of the important results, and in \ref{significance}, we discuss the Physics- and Math-related significance of the results obtained.

\subsection{Summary of Results}
\label{summary}

\begin{itemize}
\item The linearized equation of motion of  the gauge invariant combination $Z_s(r)$ (\ref{Zs}) up to  $\cal O(\eta)$, resulted in the differential equation (\ref{Zs-EOM}),
\begin{equation}
Z_s''(r) = l(r) Z_s'(r) + m(r) Z_s(r).
\end{equation}
Utilizing the dispersion relation (\ref{wuptoqsquared}) of \cite{IITR-McGill-bulk-viscosity} it turns out that $r=r_h$ is an irregular singular point due to the presence of $\frac{1}{(r-r_h)^3}$ in the $m(r)$. Hence we have chosen $\alpha$ for which the residue of the $\frac{\lambda_1}{r_h^3(r-r_h)}$-term vanishes. From solutions of the  equations of motion of $h_{\mu\nu}^{\beta^{0}}$ in the limit $q\rightarrow 0$, we worked the deviation of the square of the speed of sound from its conformal value, $\Biggl(\frac{1}{3}-c_s^2\Biggr)$  in terms of $g_s, M , N, N_f$, and log$r_h$, i.e.
\begin{eqnarray}
& & \frac{1}{3} - c_s^2 =  \frac{517\left(g_s M^2\right)(g_s N_f)(16 N\pi^2\log r_h - 4653 g_s^2 M^2 N_f (\log r_h)^4)}{113,400N\pi^2(2 + \log r_h)} - \kappa_{\alpha_\beta}\beta,\nonumber\\
& &  |\kappa_{\alpha_\beta}|\equiv{\cal O}(1),
\end{eqnarray}
(dropping terms of ${\cal O}\left(\frac{1}{N^2}\right),\ {\cal O}(\beta^2)$), and
\begin{eqnarray}
& & \hskip -0.8in \left(\frac{1}{3} - c_s^2\right)^2\sim  -\frac{517\left(g_s M^2\right)(g_s N_f)(16 N\pi^2\log r_h - 4653 g_s^2 M^2 N_f (\log r_h)^4)}{113,400N\pi^2(2 + \log r_h)} \kappa_{\alpha_\beta}\beta.
\end{eqnarray}

\item Utilizing the Schr\"odinger-like equation of motion (\ref{Schroedinger-like-ET}) obtained from the DBI action of $D6$-brane via introduction of the perturbation in the $U(1)$ gauge field, we formed the on-shell DBI action. Using the fact that the gauge field strength vanishes at the boundary, i.e. $F_{Zt}=-F_{tZ}=0$, under the zero-momentum limit we obtained the retarded Green function, the imaginary part of which provides the spectral function for the Kubo's formula. Using Kubo's formula, 
\begin{eqnarray}
& & \hskip-0.8in \lim_{w\rightarrow0}\left(\frac{\rho(T\neq0)}{\omega}-\frac{\rho(T=0,w)}{w}\right) \nonumber\\
& & \hskip -0.8in = \kappa_1 \frac{e^{Z_{\rm UV}} N^{3/5}N^f_{\rm UV} r_h\left(\log r_h\right)^3}{\sqrt{g_s}N} + \kappa_2 \frac{e^{Z_{\rm UV}}N^f_{\rm UV}N^{3/5}r_h\left({\cal C}_{zz} - 2 {\cal C}_{\theta_1z} + 2 
  {\cal C}_{\theta_1x}\right)\left(\log r_h\right)^3}{\kappa_{r_h}^3N}\beta.\nonumber\\
  & & 
\end{eqnarray}
Since $e^{Z_{\rm UV}}\sim N^{1/\mathbb{N}_{>4}}$, where $\mathbb{N}_{>4}$ is an integer greater than four, choose: $N^f_{\rm UV}\sim N^{-3/5-1/\mathbb{N}_{>4}}$. Taking, $C_{zz}-2C_{\theta_1x}=0, |C_{\theta_1x}|<<1$ under a self-consistent truncation of $C_{MNP}$ within the $\mathcal{O}(R^{4})$ corrections to $\cal M$-theory dual \cite{OR4},  the above expression does not receive $\mathcal{O}(R^{4})$ corrections. So via  (\ref{1over3-cssq}) and (\ref{sq-1over3-cssq}), we have proved that the bulk viscosity at intermediate 't-Hooft coupling for our $\cal M$-theoretic background is the linear combination of the factor appearing in the strong $\left(\frac{1}{3} - c_s^2\right)$, and weak coupling limits $\left(\frac{1}{3} - c_s^2\right)^2$, i.e.
\begin{equation}
 \zeta = {\cal C}_1(r_h; g_s, M_{\rm UV}, N_f^{\rm UV} N)\left(\frac{1}{3} - c_s^2\right) + {\cal C}_2(r_h; g_s, M_{\rm UV}, N_f^{\rm UV} N)\left(\frac{1}{3} - c_s^2\right)^2.
\end{equation}

\item We used the Eling and Oz's proposal \cite{EO} for computing the $\frac{\zeta}{\eta}$ ratio using the scalar fields. For simplicity, we have chosen the most dominant components of $C_{MNP} (C_3 = B^{\rm IIA}\wedge dx^{10})$ in eleven dimensions near $r_h$ (\ref{r_h-est}) which are $B_{\theta_1\theta_2}^{IIA}$ and $B_{\theta_1x}^{IIA}$; $B_{\theta_1\theta_2}^{IIA}$ contributes most dominantly to the kinetic terms. Choosing the proper angular regularization (\ref{regularization}), we obtained the expected result in (\ref{EO-2}) that if one switches off the non-conformality by taking $M$, and $N_f$ to zero, we get $\frac{\zeta}{\eta}\rightarrow 0$. Analysing close to QCD-inspired values from Table (\ref{Parameters-real-QCD}),
{\footnotesize
\begin{eqnarray}
\label{summary-zetaovereta}
& & \hskip -0.8in \frac{\zeta}{\eta} = \left.c_s^4 T^2\int_{S^2_{\rm squashed}\times_w S^3_{\rm squashed}} \sqrt{g_6}g^{rr} g^{x^{10}x^{10}} \left[
     g^{\theta_1\theta_1} g^{\theta_2\theta_2} - (g^{\theta_1\theta_2})^2\right]\left(\frac{d C_{\theta_1\theta_2x^{10}}}{dT}\right)^2\right|_{r=r_h}\nonumber\\
 & &\hskip 1in  \downarrow\ {\rm near}\ (\ref{alpha_theta_12})\nonumber\\    
& & \hskip -0.8in  M N^{5/4}N_f^{4/3}g_s^{21/4}\Biggl[-\frac{10^{-4}\left(c_2\left\{-\frac{1}{6} + \left(\frac{1}{3} - c_s^2\right)^{\beta^0}\right\} N_f^{2}\right)}{3\left(\log N - 3 \log \left(\frac{\sqrt{3}T}{2(1 + \sqrt{3}\varepsilon )T_c}\right)\right)^{2/3}\left(c_1 + c_2\log  \left(\frac{\sqrt{3}T}{2(1 + \sqrt{3}\varepsilon )T_c}\right)\right)^2} \nonumber\\
& & \hskip -0.8in  + \Biggl(-\frac{2\times10^{-7}{\cal C}_2\left(\frac{1}{9} - \frac{2\left(\frac{1}{3} - c_s^2\right)^{\beta^0}}{3}\right)\left(\log N - 3 \log  \left(\frac{\sqrt{3}T}{2(1 + \sqrt{3}\varepsilon )T_c}\right)\right)^{1/3}}{\left(\log \left(\frac{\sqrt{3}T}{2(1 + \sqrt{3}\varepsilon )T_c}\right)\right)^2(c_1 + c_2\log  \left(\frac{\sqrt{3}T}{2(1 + \sqrt{3}\varepsilon )T_c}\right))}\nonumber\\
& & \hskip -0.8in +\frac{5\times10^{-5}c_2N_f^2\left(-\frac{2}{3} \left(\frac{1}{3} - c_s^2\right)^{\beta}+ \left(\frac{1}{3} - c_s^2\right)^{2,\ \beta}\right)}{\left(\log N - 3 \log  \left(\frac{\sqrt{3}T}{2(1 + \sqrt{3}\varepsilon )T_c}\right)\right)^{1/3}}\Biggr) \frac{\kappa_\beta}{(g_s N_f)^4}\frac{ \left(\frac{24 a^2(T/T_c) {g_s} M^2 
   \left(c_1+c_2 \log \left(\frac{\sqrt{3}T}{2(1 + \sqrt{3}\varepsilon )T_c}\right)\right)}{N\left[9 a^2(T/T_c)+\left(\frac{\sqrt{3}T}{2(1 + \sqrt{3}\varepsilon )T_c}\right)^2\right]}+\frac{3  \left[\log (N)-3 \log \left(\frac{\sqrt{3}T}{2(1 + \sqrt{3}\varepsilon )T_c}\right)\right]}{4 \pi
   }\right)^3}{\left[\log (N)-3 \log \left(\frac{\sqrt{3}T}{2(1 + \sqrt{3}\varepsilon )T_c}\right)\right]^{7}}\Biggr],
\nonumber\\
& & \hskip -0.8in \forall T\in[T_c,\frac{2(1 + \sqrt{3}\varepsilon )}{\sqrt{3}}T_c],
\end{eqnarray}
}
$\kappa_\beta$ being a numerical factor, $a(T/T_c) = \frac{\sqrt{3}T}{2(1 + \sqrt{3}\varepsilon )T_c}\left(\frac{1}{\sqrt{3}} + \frac{g_sM^2\left[c_1 + c_2 \log\left(\frac{\sqrt{3}T}{2(1 + \sqrt{3}\varepsilon )T_c}\right)\right]}{N}\right)$ and $\varepsilon $ given by (\ref{varepsilon-value}). Now, $r_h$ was estimated in \cite{IITR-McGill-bulk-viscosity} by estimating the deep IR-valued $r={\cal R}_0$ where the {\it effective} number of type IIB three-brane charge:
\begin{equation}
\label{Neff}
N_{\rm eff}(r) = \int_{\mathbb{M}_5} F_5^{\rm IIB} + \int_{\mathbb{M}_5} B_2^{\rm IIB} \wedge F_3^{\rm IIB}, 
\end{equation}
(with $B_2^{\rm IIB}, F_3^{\rm IIB}$, and $F_5^{\rm IIB}$ are given in \cite{metrics}, and  $\mathbb{M}_5$ is the base of the resolved warped-defomed conifold), vanishes after the completion of a thermal Seiberg-like duality cascade similar to the the Klebanov-Strassler model. The deep IR-valued ${\cal R}_0$ is estimated by solving $N_{\rm eff}({\cal R}_0)=0$ yielding
\begin{equation}
\label{rh-estimate}
{\cal R}_0\sim r_h\sim e^{-\kappa_{r_h}(M, N_f g_s)N^{\frac{1}{3}}}, \kappa_{r_h} = 
\frac{1}{3(6\pi)^{1/3}(g_sN_f)^{2/3}(g_sM^2)^{1/3}}.
\end{equation}
Hence, assuming $|\log r_h| = |\log\left(\frac{\sqrt{3}T}{2(1 + \sqrt{3}\varepsilon )T_c}\right)|>\log N$, we had observed beneath (\ref{M-theory-circle-metric}) the following. The coefficient of $\left(1/3 - c_s^2\right)^2$ is accompanied by $\frac{1}{L^{20/9}}$-suppression relative to the coefficient of $1/3 - c_s^2$, with $L = \left(4\pi g_s N\right)^{1/4}$ playing the role of `t-Hooft coupling. 

One may wish to rewrite (\ref{summary-zetaovereta}) in terms of the running/temperature-dependent YM coupling constant $g(T)$. Using standard KS(Klebanov-Strassler)-like RG-flow equations \cite{Klebanov:2000hb}, \cite{IITR-McGill-bulk-viscosity,Czajka:2018egm}, one sees that (similar to \cite{effec_kin_model_zetaovereta}):
{\footnotesize
\begin{equation}
\label{g(T)}
g^2(T/T_c) \sim \left.\left(e^{-\phi^{\rm IIA}}\int_{S^2}B^{\rm IIA}\right)^{-2}\right|_{r=r_h}
\sim\frac{1}{(g_s M) (g_s N_f)}\left[\left|\log \left(\frac{\sqrt{3}T}{2(1 + \sqrt{3}\varepsilon )T_c}\right)\right| \left(\log N - 3 \log \left(\frac{\sqrt{3}T}{2(1 + \sqrt{3}\varepsilon )T_c}\right)\right)\right]^{-1}, 
\end{equation}
}
and consequently, setting ${\cal C}_2=0$, (\ref{summary-zetaovereta}) yields:
{\footnotesize
\begin{eqnarray}
\label{zetaovereta-g}
& & \hskip -0.8in \frac{\zeta}{\eta} =  M N^{5/4}N_f^{4/3}g_s^{21/4}\Biggl[-{\cal O}(10^{-4})c_2\left\{-\frac{1}{6} + \left(\frac{1}{3} - c_s^2\right)^{\beta^0}\right\} N_f^2(g_s^2 M N_f)^{2/3}\left( g^2(T/T_c)\right)^{2/3} \frac{\left(\left|\log \left(\frac{\sqrt{3}T}{2(1 + \sqrt{3}\varepsilon )T_c}\right)\right|\right)^{2/3}}{\left(c_1 + c_2\log  \left(\frac{\sqrt{3}T}{2(1 + \sqrt{3}\varepsilon )T_c}\right)\right)^2} \nonumber\\
& & \hskip -0.8in  + {\cal O}(10^{-5})c_2N_f^2\left\{-\frac{2}{3} \left(\frac{1}{3} - c_s^2\right)^{\beta}+ \left(\frac{1}{3} - c_s^2\right)^{2,\ \beta}\right\}(g_s^2 M N_f)^{1/3}\left( g^2(T/T_c)\right)^{1/3}\left(\left|\log \left(\frac{\sqrt{3}T}{2(1 + \sqrt{3}\varepsilon )T_c}\right)\right|\right)^{1/3}\nonumber\\
& & \hskip -0.8in \times \frac{\kappa_\beta}{(g_s N_f)^4}\left(\left|\log \left(\frac{\sqrt{3}T}{2(1 + \sqrt{3}\varepsilon )T_c}\right)\right|\right)^8  \left(g_s^2 M N_f T \frac{d g^2(T/T_c)}{d T}\right)^4\Biggr].
\nonumber\\
& & \hskip -0.8in \forall T\in[T_c,\frac{2(1 + \sqrt{3}\varepsilon )}{\sqrt{3}}T_c].
\end{eqnarray}
}
Two plots for two sets of choices of $(c_1, c_2, {\cal C}_2, \kappa_{\alpha_\beta}, \kappa_\beta, N)$
are given in Figures \tcb{2} and \tcb{3}, for QCD-inspired values of $(g_s, M, N_f)$ of Table \tcb{1}.  In \cite{lattice-SU3_Glue}, there are lattice-calculations results available for $\frac{\zeta}{s}$ (i.e., bulk-viscosity-to-entropy-density ratio) in $SU(3)$ Gluodynamics at only two temperatures below $1.45T_c$ (which sets the limit of validity of our ${\cal M}$-theory computations - see (\ref{range-T}) for $\varepsilon=\frac{0.3}{\sqrt{3}}$): $T = 1.02 T_c, 1.24 T_c$ for which the lattice results yield $\frac{\zeta}{s} = 0.73(3), 0.065(17)$ with the statistical errors given by the numbers in the parentheses. As shown in \cite{IITR-McGill-bulk-viscosity}, $\frac{\eta}{s} = \frac{1}{4\pi} + {\cal O}\left(\frac{g_s M^2}{N}, \frac{(g_s M^2)(g_s N_f)}{N}\right)$. Now disregarding the non-conformal corrections as the same are $\frac{1}{N}$-suppressed, the lattice calculations of \cite{lattice-SU3_Glue} hence imply:
\begin{equation}
\label{perfect-match-Meyer-lattice-SU3_glue}
\left.\frac{\frac{\zeta}{\eta}(T=1.02T_c)}{\frac{\zeta}{\eta}(T=1.24T_c)}\right|_{\rm Lattice}\leq15.
\end{equation}
 This is precisely the ratio that one obtains from Fig. \tcb{2}, but not Fig. \tcb{3}. Further, \\ $\frac{\left(\frac{\zeta}{\eta}\right)^\beta}{\left(\frac{\zeta}{\eta}\right)^{\beta^0}}\biggl(T>T_c\biggr)>1,\ \frac{\left(\frac{\zeta}{\eta}\right)^\beta}{\left(\frac{\zeta}{\eta}\right)^{\beta^0}}'\biggl(T>T_c\biggr)>0$ (see Fig. \tcb{4}) as expected. This hence rules out the (curvature of the) plot as given in Fig. \tcb{3}. This is expected to be a useful guide for lattice practitioners looking for more points on the plot.

\begin{figure}
\begin{center}
\includegraphics[width=0.60\textwidth]{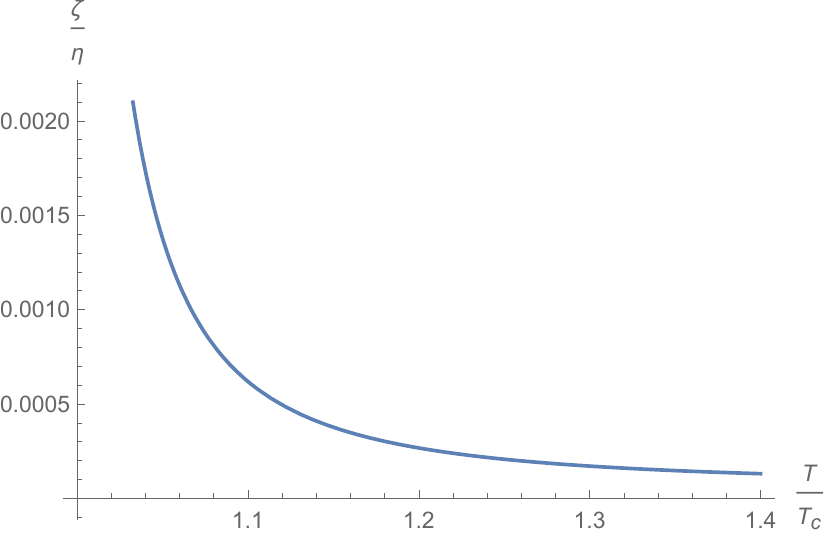}
\end{center}
\caption{$\frac{\zeta}{\eta}$-vs-$\frac{T}{T_c}$ for $(g_s, M, N_f)$ of Table \tcb{1}, $N=100, c_1 = -109, c_2 = - 10, \varepsilon = 0.3/\sqrt{3}, \kappa_{\alpha_\beta} = -1, \kappa_\beta = 1, {\cal C}_2 = 0$; The graph is meaningful $\forall T\in[T_c, 1.45 T_c)$, which shows $\frac{\zeta}{\eta}'(T)<0$, consistent with results of \cite{dzetaovereta-negative} effective kinetic theory of quasiparticle excitations with medium-modified dispersion relation, and most importantly matching beautifully with the $SU(3)$ Gluodynamics lattice results of  \cite{lattice-SU3_Glue} within the statistical error in the latter.}
\label{S-cs-l}
\end{figure}

\begin{figure}
\begin{center}
\includegraphics[width=0.60\textwidth]{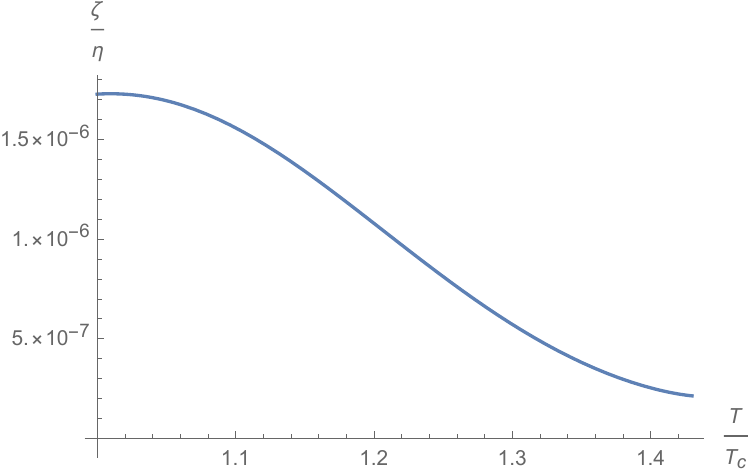}
\end{center}
\caption{$\frac{\zeta}{\eta}$-vs-$\frac{T}{T_c}$ for $(g_s, M, N_f)$ of Table \tcb{1}, $N=80, c_1 = -76, c_2 = - 60, \varepsilon = 0.3/\sqrt{3}, \kappa_{\alpha_\beta} = -1, \kappa_\beta = 1, {\cal C}_2 = (-200,200)$; The graph is meaningful $\forall T\in[T_c, 1.45 T_c)$, which shows $\frac{\zeta}{\eta}'(T)<0$ but with different curvature; the plot however is inconsistent with the lattice results of \cite{lattice-SU3_Glue}.}
\label{S-cs-l}
\end{figure}

\begin{figure}
\begin{center}
\includegraphics[width=0.60\textwidth]{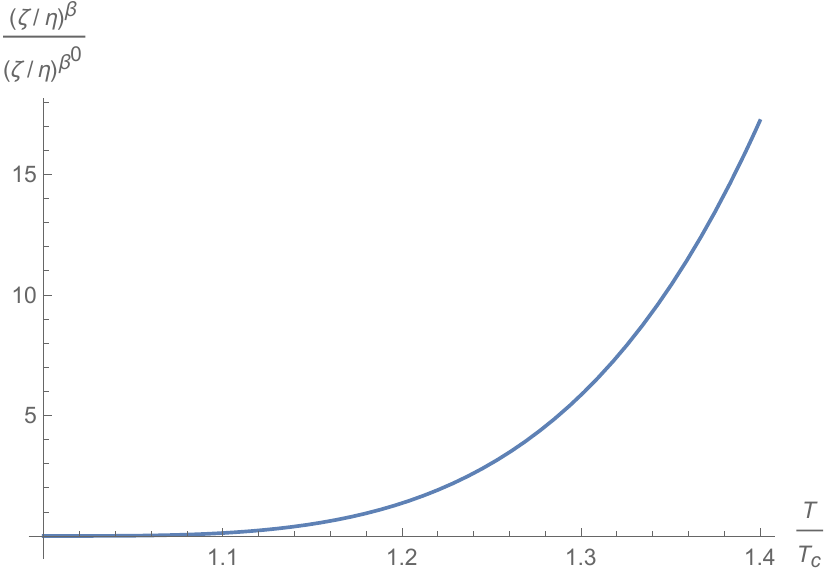}
\end{center}
\caption{$\frac{\left(\frac{\zeta}{\eta}\right)^{\beta}}{\left(\frac{\zeta}{\eta}\right)^{\beta^0}}$-vs-$\frac{T}{T_c}$ for $(g_s, M, N_f, c_1, c_2, {\cal C}_2, N, \varepsilon, \kappa_{\alpha_\beta}, \kappa_\beta)$ of Fig. \tcb{2}.}
\label{S-cs-l}
\end{figure}

Assuming $C_{\theta_1\theta_2 x^{10}} = B_{\theta_1\theta_2}^{\rm IIA}$ to be the only scalar in the ${\cal M}$-theory uplift, since $B_{\theta_1\theta_2}^{\rm IIA}$ component of the NS-NS B-field is the most dominating term, and hence  working in the gauge $r = C_{\theta_1\theta_2 x^{10}}$,
\begin{eqnarray}
& & \frac{\zeta}{\eta} = \left(1/\left(3 A'(C_{\theta_1\theta_2 x^{10}})\right)\right)^2,\ {\rm from\ a\ D=5\ perspective},\nonumber\\
& & {\rm or}\ \frac{1}{\left(3 A'(r_h) + \frac{g_6'(r_h)}{g_6(r_h)}\right)^2}\ {\rm from\ D=11\ perspective} \nonumber\\
& & = \frac{r_h^2}{L^2}\left[\chi_1^{(5)\ {\rm or}\ (11)} \left(\frac{1}{3} - c_s^2\right) + \chi_2^{(5)\ {\rm or}\ (11)}\left(\frac{1}{3} - c_s^2\right)^2\right],
\end{eqnarray}
where $\chi_{1,2}^{(5)\ {\rm or}\ (7)} = \chi_{1,2}^{(5)\ {\rm or}\ (7)}\left(g_s, N, M, N_f\right)$. Therefore, the bulk-to-shear viscosity ratio for the $\cal M$-theoretic dual thermal QCD-like theories which are UV conformal, IR confining, and having bi-fundamental quarks follows the linear combination of the bulk-to-shear viscosity ratio at strong and weak 't-Hooft coupling limits.
\end{itemize}

\subsection{Significance of Results}
\label{significance}

\begin{itemize}

\item Physics
\begin{itemize}
\item
\begin{tcolorbox}[enhanced,width=6.5in,center upper,size=fbox,
    drop shadow southwest,sharp corners]
    \begin{flushleft}
The higher derivative (${\cal O}(R^4)$) corrections of ${\cal M}$-theory \cite{OR4} to the ${\cal M}$-theory uplift (\cite{MQGP}, \cite{NPB}) of the type IIB string dual \cite{metrics} of thermal QCD-like theories, are shown to interpolate between the weak and strong coupling results for the bulk-to-shear-viscosity ratio $\frac{\zeta}{\eta}$. 
\end{flushleft}
\end{tcolorbox}
We hence see that at intermediate coupling, both $1/3 - c_s^2$ and $\left(1/3 - c_s^2\right)^2$ contribute to $\frac{\zeta}{\eta}$. 

\item
\begin{tcolorbox}[enhanced,width=6.5in,center upper,size=fbox,
    drop shadow southwest,sharp corners]
    \begin{flushleft}
The dependence of the coefficients of $\frac{1}{3} - c_s^2$ and $\left(\frac{1}{3} - c_s^2\right)^2$
in $\frac{\zeta}{\eta}$ on the temperature-dependent/running coupling constant as obtained in (\ref{zetaovereta-g}) is similar to the one  in the quasiparticle description of the gluon plasma as obtained in \cite{dzetaovereta-negative} with the difference that apart from the similar dependence on $\left(T \frac{d g^2(T/T_c)}{d T}\right)^{m\in\mathbb{Z}^+}$, $\frac{\zeta}{\eta}$ also depends on fractional powers ($\frac{1}{3}, \frac{2}{3}$) of $g^2(T/T_c)$. 
\end{flushleft}
\end{tcolorbox}

The fractional-power dependence on the running coupling could be understood from QCD at large temperatures as follows. 
\begin{itemize}
\item
The renormalized coupling constant of QCD at temperature $T$ is \cite{Fujimoto+Yamada}:
\begin{equation}
\label{gR_T}
g^2_R(T) = \frac{g_R^2(T_0)}{1+\frac{2N}{(4\pi)^2}g_R^2(T_0)\left(\Omega_{\rm glue}\left(\frac{p}{T}\right) - \Omega_{\rm glue}\left(\frac{p}{T_0}\right)\right)},
\end{equation}
where ($a\equiv\frac{p}{T}$)
\begin{equation}
\label{Omega}
\Omega_{\rm glue}\left(a\right) = \frac{2\pi^2}{a^2} - \frac{23}{6} -\frac{3}{2}F_2 - \frac{7}{3}G_0 + 16 G_2 + \frac{3}{8}J_1^{(2)} + 2 J_3^{(2)} + \frac{1}{4}J_1^{(3)} + \frac{4}{3}J_3^{(3)},
\end{equation}
with
\begin{eqnarray}
\label{integrals-def-Omega}
& & J_n^{(m)}(a\ll1) \equiv (a/\sqrt{3})^{1-n}\int_0^\infty dx x^{n-1}(x^2 + a^2/3)^{-(1+m)/2}\approx a^{-m};\nonumber\\
& & F_n \equiv \int_0^\infty dx \frac{x^n}{e^{p x/(2 T)} - 1}\log\left|\frac{1+x}{1-x}\right|;\nonumber\\
& & G_n \equiv \int_0^1 dy \int_0^\infty dx \frac{x^{n+1}}{e^{p x/T} - 1}\frac{1}{x^2(a^2/3 + 3) - 1}. 
\end{eqnarray}
At large temperatures, $\Omega(a\ll1)\sim a^{-3}$ implying that (from the three-gluon vertex) \cite{g2T3}: 
\begin{equation}
\label{g_T-large-T}
g_R(T)\sim T^{-3/2}.
\end{equation}

\item We can map the QCD to Ising model and on the basis of universality arguments we can workout the temperature dependence of bulk viscosity near critical temperature by comparing the results of liquid-gas phase transition as explored by Moore et al \cite{Moore:2008ws}. The mapping between the QCD variables, e.g., temperature(T) and baryon chemical potential $(\mu)$, and Ising model variables, e.g., reduced temperature $(t)$ and reduced magnetic field ($h$), is linear \cite{Moore:2008ws}, \cite{Martinez:2019bsn}. Based on universality arguments, the temperature dependence of bulk viscosity reads as $\zeta\sim t^{-z/\nu+\alpha}$ \cite{Moore:2008ws}, where $t=\frac{T-T_{c}}{T_{c}}$, and the critical coefficients are $\nu=0.630$, $z\approx 3 $ (for $T>T_c, \mu=\mu_c$ where the $\mu_c$ corresponds to the critical value of the chemical potential at which the smooth cross-over becomes a first-order phase transition, and small values of the chemical potential (See appendix A in \cite{Trace anomaly_AM+CG} for the string-theoretical explanation dependent on the small Ouyang-embedding modulus chosen in the type IIB holographic dual, to see that the chemical potential is indeed very small), which can be written as $ \zeta \sim t^{\Upsilon} $, $\Upsilon$ being an appropriate fraction. Hence for high temperatures, one may write $\zeta\sim T^{\Upsilon^{\prime}} $, where $\Upsilon^{\prime} $ is a fraction.
\item
For high temperatures utilizing the equation(\ref{g_T-large-T}), we can conclude that at large temperatures, bulk viscosity $\zeta$ in QCD, has some fractional power dependence on coupling constant, where the fractional power depends on the kind of vertex chosen (e.g.,  three gluon vertex (\ref{g_T-large-T}).
\end{itemize}

\item
From (\ref{summary-zetaovereta})/(\ref{zetaovereta-g}), one can show that for QCD-inspired values of $(g_s, M, N_f)$ and, e.g., for $(N, c_1, c_2, {\cal C}_2, \kappa_\beta, \kappa_{\alpha_\beta})$ of Fig. \tcb{2}:
\begin{table}[h]
\begin{center}
\begin{tcolorbox}[enhanced,width=5in,center upper,size=fbox,
    drop shadow southwest,sharp corners]
\begin{tabular}{|c|c|}\hline
Lattice $SU(3)$ Gluodynamics & Our Result  \\ \hline
$\frac{\frac{\zeta}{\eta}(T = 1.02 T_c)}{\frac{\zeta}{\eta}(T = 1.24 T_c)}\leq15$  & $\frac{\frac{\zeta}{\eta}(T = 1.02 T_c)}{\frac{\zeta}{\eta}(T = 1.24 T_c)}\sim 15$ \\ \hline
\end{tabular}
\end{tcolorbox}
\end{center}
\caption{Match, within statistical errors, with lattice $SU(3)$ Gluodynamics computations for $T\in[T_c, 1.45 T_c]$}
\end{table}
\\
and
\begin{tcolorbox}[enhanced,width=6.5in,center upper,size=fbox,
    drop shadow southwest,sharp corners]
\begin{flushleft}
\begin{eqnarray}
\label{weakoverstronghightemp}
& &  \left.\frac{\left(\frac{\zeta}{\eta}\right)^{\beta}}{\left(\frac{\zeta}{\eta}\right)^{\beta^0}}\right|_{\rm Table \tcb{1},\ Fig. \tcb{2}}>1, \ \left(\frac{\left(\frac{\zeta}{\eta}\right)^{\beta}}{\left(\frac{\zeta}{\eta}\right)^{\beta^0}}\right)'\bigl(T\bigr)_{\rm Table \tcb{1},\ Fig. \tcb{2}}>0\ \forall T\in[T_c, 1.45 T_c],
\end{eqnarray}
\end{flushleft}
\end{tcolorbox}
implying the following: 
\begin{tcolorbox}[enhanced,width=6.5in,center upper,size=fbox,
    drop shadow southwest,sharp corners]
    \begin{flushleft}
At large temperatures (but within the domain of applicability of our ${\cal M}$-theory computation),  $\frac{\zeta}{\eta}$ receives the dominant contribution from the weak-coupling result, i.e., from $\left(\frac{1}{3} - c_s^2\right)^2$ (which is entirely ${\cal O}(\beta)$, i.e., dependent on the ${\cal O}(R^4)$ corrections), as expected, and the ${\cal O}(\beta)$ contribution to $\frac{1}{3} - c_s^2$.
\end{flushleft}
\end{tcolorbox}

\end{itemize}

\item Math
\begin{itemize}
\item
The aforementioned results depend crucially on taking the intermediate-$N$ MQGP limit of (\ref{MQGP_limit}) and it was precisely in this limit that the closed seven-fold $M_7$ (which supports a $G_2$ structure, appearing in the ${\cal M}$-theory dual of thermal QCD-like theories at intermediate coupling) which is a warped product of the ${\cal M}$-theory circle and a non-K\"{a}hler six-fold with the latter itself being a warped product of the thermal circle with a non-Einsteinian deformation of $T^{1,1}$, was shown in \cite{ACMS} to support Contact 3-Structures (C3S). The deep connection between the existence of C3S and the speed-of-sound dependence of $\frac{\zeta}{\eta}$ at intermediate coupling can be understood as follows. 

Noting that,
\begin{eqnarray}
\label{fluxes}
& & \hskip -0.8in {\rm near}\ (\ref{alpha_theta_12}): G^2(r\in{\rm IR})\sim \frac{\left(\frac{1}{N}\right)^{19/10}}{\sqrt{{g_s}} {N_f}^{2/3} |\log r|^{\frac{14}{3}}} - \frac{\left(\frac{1}{N}\right)^{11/10}}{\sqrt{{g_s}} ({N_f} |\log r|)^{2/3}};\nonumber\\
& & \hskip -0.8in {\rm globally}: G^2(r\in{\rm IR}) \sim \frac{{\sin}^4\left(\theta _1\right)  \left(2187 \sin ^6\left(\theta _1\right)+270 \sqrt{6} \sin ^2\left(\theta _2\right) \sin^3\left(\theta _1\right)+50 \sin ^4\left(\theta _2\right)\right) \csc ^2\left(\theta _2\right)}{\sqrt{{g_s} N}{N_f}^{2/3} |\log r|^{14/3}} \nonumber\\
& & \hskip -0.8in -\frac{ {\sin}^6\left(\theta _1\right) }{\sqrt{{g_s N}} \sin^2\left(\theta _2\right)  ({N_f} |\log r|)^{2/3}},
\end{eqnarray}
and,
\begin{equation}
\label{R_IR_BH}
R(r\in{\rm IR})\sim \frac{1}{(r - r_h)r_h N_f^{8/3}(|\log r_h|)^{11/3}},
\end{equation}
one sees that:
\begin{equation}
\label{EH-more-than-flux}
R > G^2.
\end{equation}
Now, Ricci scalar (essentially the Einstein-Hilbert term in the $D=11$ supergravity action) can be written in terms of the four  $G_2$-structure torsion classes \cite{Bryant-G2} as follows:
\begin{eqnarray}
\label{Ricci-scalar-G2}
& & \hskip -0.3in R(M_7(r,\theta_1,\theta_2,\phi_1,\phi_2,\psi,x^{10})) = 12 d^\dagger W_7 + \frac{21}{8}W_1^2 + 30 |W_7|^2 - \frac{1}{2}|W_{14}|^2 - \frac{1}{2}|W_{27}|^2.\nonumber\\
& &
\end{eqnarray}
Apart from (\ref{EH-more-than-flux}), using also, e.g.,
\begin{eqnarray*}
\label{flux-beta0-subdominant}
& &  \left.G_{rMNP}G_r^{\ \ MNP}\right|_{(\ref{MQGP_limit})} <\left.R_{rr}^{\beta^0}\right|_{(\ref{MQGP_limit})},
\end{eqnarray*}
one drops flux-dependent contributions at ${\cal O}(\beta^0)$, and using 
$J_0>E_8>t_8^2G^2R^3$, one drops the flux-dependent contributions at
${\cal O}(\beta)$ in the $D=11$ supergravity action. 

\begin{tcolorbox}[enhanced,width=6.5in,center upper,size=fbox,
    drop shadow southwest,sharp corners]
    \begin{flushleft}
One hence sees that the dynamics of metric fluctations in the intermediate-$N$ limit (\ref{MQGP_limit}) will be given predominantly by the $G_2$-structure torsion classes ($W_{7, 14, 27}$) and hence the Contact structure arising from the same, and not the four-form fluxes.
\end{flushleft}
\end{tcolorbox}

\item

\begin{tcolorbox}[enhanced,width=6.5in,center upper,size=fbox,
    drop shadow southwest,sharp corners]
    \begin{flushleft}
One can hence show that a rich variety of phenomena such as post Page-time black hole entanglement entropy \cite{MQGP-Page}, the "Lyapunov exponent" and "butterfly velocity" pertaining to a chaotic behavior \cite{MChaos}, and dependence of "bulk viscosity" on the speed of sound (this paper), all evaluated in the aforementioned ${\cal M}$-theoretic dual, match expectations precisely in the aforementioned intermediate-$N$ MQGP limit (\ref{MQGP_limit})  supporting Contact 3-Structures.
\end{flushleft}
\end{tcolorbox}

\item
As \\
\noindent (i) the $N\gg1$ MQGP limit of footnote \tcb{5}, which corresponds to the strong coupling result
$\frac{\zeta}{\eta}\sim\frac{1}{3} - c_s^2$, supports Almost Contact 3-Structures \cite{ACMS} and the intermediate-$N$($N>1$) MQGP limit (\ref{MQGP_limit}), which corresponds to the intermediate-coupling result  (schematically) $\frac{\zeta}{\eta}$ given by a superposition of the weak coupling ($\frac{\zeta}{\eta}\sim\left(\frac{1}{3} - c_s^2\right)^2$) and strong coupling ($\frac{\zeta}{\eta}\sim\frac{1}{3} - c_s^2$) results, supports Contact 3-Structures \cite{ACMS},\\
and\\
\noindent (ii) the parameter space (determined by the brane content and string coupling of the parent type IIB dual\cite{metrics}) is not ``$N$-path connected'' with reference to Contact Structures in the IR \cite{ACMS}, i.e., the $N\gg1$ Almost Contact Metric 3-Structures arising from the $G_2$ structure, do not connect to  Contact 3-Structures (in the IR) which are shown to exist only for the aforementioned intermediate $N>1$-limit (\ref{MQGP_limit}),\\
we conjecture that:
\begin{tcolorbox}[enhanced,width=6.5in,center upper,size=fbox,
    drop shadow southwest,sharp corners]
    \begin{flushleft}
The $N$-path disconnectedness of the parameter space with reference to Contact Structures in the IR arising from the $G_2$ structure, is mapped to the appearance of fractional powers of the coupling constants (which if complexified, would have corresponding to branch-point singularities in the complex coupling constant space) in $\frac{\zeta}{\eta}$ as seen in (\ref{zetaovereta-g}).
\end{flushleft}
\end{tcolorbox}

\end{itemize}

\end{itemize}

\section*{Acknowledgements}

SSK is supported by a Junior
Research Fellowship (JRF) from the Ministry of Human Resource and Development (MHRD), Govt. of India. AM is partly supported by a Core Research Grant number SER-1829-PHY from the Science and Engineering Research Board, Govt. of India. We thank G.~Yadav for useful discussions and collaboration on \cite{MChaos} which was the precursor of this work.

\appendix

\section{Review of the Basics of (Almost) Contact Structures}
\label{review-(A)C3S}
\setcounter{equation}{0}
\seceqaa

In this section we briefly review what (Almost) Contact 3-Structures are and how they are obtained from $G_2$ structures. We also summarize the relevant results of \cite{ACMS}.

Let $M_7$ be an odd-dimensional  Riemannian manifold equipped with a metric $g$, assumed to admit an endomorphism $J$ of the tangent bundle $T M_7$, a unit vector field $R$ (with respect to the metric $g$) called the Reeb vector field, and a one form $\sigma$ which satisfy
    $$J^2=-\mathbf{1}+R\otimes\sigma,\quad\sigma(R)=1.$$
    $M_7$ is then said to admit an almost contact structure $(J,R,\sigma)$ (ACS) and $\sigma$ is called the contact potential \cite{Sasaki}. The structure group of the tangent space reduces to $U(n)\times{\bf 1}$ where $2n+1$ is the dimension of $M_7$. 

The ACS is said to be a contact structure if
  $$\sigma\wedge d{\sigma}....\wedge d{\sigma}\neq 0\quad \forall{\rm points }\in M_7.$$
 A Riemannian manifold $M_7$ with an ACS $(J,R,\sigma)$ has an almost contact metric structure $(J,R,\sigma,g)$ (ACMS) if,
    $$g(Ju,Jv)=g(u,v)-\sigma(u)\sigma(v),\quad\forall u,v\in\Gamma(T M_7).$$ The fundamental two-form $\omega$ of the almost contact manifold is then defined as,
    $$\omega(u,v)=g(Ju,v),\quad\forall u,v\in\Gamma(T M_7)$$
    and satisfies
    $$\sigma\wedge\omega\wedge....\wedge\omega\neq 0.$$

Almost contact 3-structures (AC3S) on a manifold $M_7$ are defined by three distinct ACSs $(J^{\alpha},R^{\alpha},\sigma^{\alpha}),\alpha=1,2,3$ on $M_7$ which satisfy \cite{Kuo-AC3S},
    \begin{eqnarray*}
    \label{AC3S-conditions}
        J^{\gamma}&=&J^{\alpha}J^{\beta}-R^{\alpha}\otimes\sigma^{\beta}=J^{\beta}J^{\alpha}-R^{\beta}\otimes\sigma^{\alpha}\nonumber\\
        R^{\gamma}&=&J^{\alpha}(R^{\beta}=-J^{\beta}(R^{\alpha}\nonumber\\
        \sigma^{\gamma}&=&\sigma^{\alpha}\circ J^{\beta}=-\sigma^{\beta}\circ J^{\alpha}\\
        \sigma^{\alpha}(R^{\beta})&=&\sigma^{\beta}(R^{\alpha})=0
    \end{eqnarray*}
where $\{\alpha,\beta,\gamma\}$ are cyclic permutation of $\{1,2,3\}$. An AC3S consisting of three contact structures satisfying (\ref{AC3S-conditions}), defines a 3-Sasakian geometry \cite{3-Sasakian-geometry}.
    
$M_7$ admitting  AC3S must have dimensionality $4n+3, n\in\mathbb{Z}^+$, and the structure group reduces to $Sp(n)\times\mathbf{1}_3$. An almost contact metric 3-structure (ACM3S) on a Riemannian manifold $M_7$ with metric $g$ possesses AC3S satisfying,
    $$g(J^{\alpha}u,J^{\alpha}v)=g(u,v)-\sigma^{\alpha}(u)\sigma^{\alpha}(u),\quad\forall u,v\in\Gamma(T M_7)$$ for $\alpha\in{1,2,3}$. With $J^\alpha(u) = R^\alpha\times_\Phi u$,  $\Phi$ being the positive 3-form defining a $G_2$ structure (see footnote \tcb{7}), one can see that with $R^\alpha.R^\beta=\delta^{\alpha\beta}$,  $R^1, R^2, \left(R^1\times R^2\right)$ provide  AC3S.

\section{Computation of $T_c$ from ${\cal M}$ Theory at Intermediate Coupling}
\label{Tc-M-theory}
\setcounter{equation}{0}
\seceqbb

In this appendix we summarize the pair of computations carried out in \cite{Gopal+Vikas+Aalok} pertaining to evaluation of the deconfinement temperature from ${\cal M}$ theory - (i) a semi-classical computation involving a Hawking-Page phase transition between the thermal background dual to low-temperature QCD-like theories at $T<T_c$, and black-hole background dual to high-temperature QCD-like theories at $T>T_c$, and (ii) from the entanglement entropy between and a spatial interval and its complement.

\subsection{$T_c$ From Semi-Classical Computation as a Hawking-Page Phase Transition}

\begin{itemize}

\item Let $\beta_{\rm th}$ and $\beta_{\rm BH}$ are periodicities for thermal circle in thermal (\ref{TypeIIA-from-M-theory-Witten-prescription-T<Tc}) and black hole background (\ref{TypeIIA-from-M-theory-Witten-prescription-T>Tc}) and compute action densities for both backgrounds.Then \cite{Witten:1998qj}
\begin{equation}
\label{equality-actions-Tc}
\left.\beta_{\rm BH}\tilde{S}_{\rm BH} = \beta_{\rm Th}\tilde{S}_{\rm th}\right|_{\beta_{\rm BH}\sqrt{ G^{\rm BH}_{tt}} = \beta_{\rm Th} \sqrt{G^{\rm Th}_{tt}}},
\end{equation}
where $\tilde{S}_{\rm BH/th}$ excludes the coordinate integral of $x^{0}$.
\item From the previous step we obtain relation between black hole horizon radius $r_h$ and IR cut off for thermal background $r_0$.
\item Deconfinement temperature on gauge theory side is given by the following expression \cite{NPB}:
\begin{equation}
\label{Tc-definition}
T_c = \frac{r_h}{\pi L^2}.
\end{equation}
\item
 One notes that $t\rightarrow x^3,\ x^3\rightarrow t$ in the BH metric following by a Double Wick rotation in the new $x^3, t$ coordinates obtains the thermal metric, i.e.,  $-g_{tt}^{\rm BH}(r_h\rightarrow r_0) = g_{x^3x^3}\ ^{\rm Thermal}(r_0),$ $ g_{x^3x^3}^{\rm BH}(r_h\rightarrow r_0) = -g_{tt}\ ^{\rm Themal}(r_0)$ (see \cite{Kruczenski et al-2003}) in the context of Euclidean/black $D4$-branes in type IIA).

\item
  In the thermal metric, we will assume the spatial part of the solitonic $D3$ brane world volume to be given by $\mathbb{R}^2(x^{1,2})\times S^1(x^3)$ where the period of $S^1(x^3)$ is given by a very large: $\frac{2\pi}{M_{\rm KK}}$, where the very small $M_{\rm KK} = \frac{2r_0}{ L^2}\left[1 + {\cal O}\left(\frac{g_sM^2}{N}\right)\right]$, $r_0$ being the very small IR cut-off in the thermal background \cite{Armoni et al-2020} and $L = \left( 4\pi g_s N\right)^{\frac{1}{4}}$. So, $\lim_{M_{\rm KK}\rightarrow0}\mathbb{R}^2(x^{1,2})\times S^1(x^3) = \mathbb{R}^3(x^{1,2,3})$.

\end{itemize}

\begin{itemize}

\item
$g_{MN} = g_{MN}^{\rm MQGP}\left(1 + \beta f_{MN}\right)$, $g_{MN}^{\rm MQGP}$ being the MQGP metric worked out at ${\cal O}(\beta^0)$ in \cite{MQGP}, \cite{NPB}, and $f_{MN}$ are the ${\cal O}(\beta)$-corrections; $f_{MN}\approx0$ in the UV \cite{Gopal+Vikas+Aalok}.
\item
Partition $r$ into the IR ($r\in[r_h,{\cal R}_{D5/\overline{D5}}^{\rm bh} = \sqrt{3}a^{\rm bh}]$) and the UV ($r\in[{\cal R}_{D5/\overline{D5}}^{\rm bh},{\cal R}_{\rm UV}^{\rm bh}]$), utilizing the results of \cite{OR4} as regards ${\cal O}(R^4)$ corrections to the ${\cal M}$-theory uplift of large-$N$ thermal QCD-like cousins as worked out in \cite{MQGP}, \cite{NPB}, and realizing the dominant contributions to the EH/GHY/{\footnotesize $\sqrt{-G}J_0$}/{\footnotesize $\sqrt{-G}G^{MN}\frac{\delta J_0}{\delta G^{MN}}$} arise from the small-$\theta_{1,2}$ values, introduce polar angular cut-offs $\epsilon_{1,2}$: 
$\theta_1\in\left[\epsilon_1,\pi-\epsilon_1\right]$ and 
$\theta_2\in\left[\epsilon_2,\pi-\epsilon_2\right]$.
\end{itemize}

\begin{itemize}
\item
The LO result for $T_c$ also holds even after inclusion of the ${\cal O}(R^4)$ corrections. The dominant contribution from the ${\cal O}(R^4)$ terms in the large-$N$ limit arises from the $t_8t_8R^4$ terms, which from a type IIB perspective in the zero-instanton sector, correspond to the tree-level contribution at ${\cal O}\left((\alpha^\prime)^3\right)$ as well as one-loop contribution to four-graviton scattering amplitude. As from the type IIB perspective, the $SL(2,\mathbb{Z})$ completion of these $R^4$ terms \cite{Green and Gutperle} suggests that they are not renormalized  perturbatively beyond one loop in the zero-instanton sector, this {\it therefore suggests the non-renormalization of $T_c$ at all loops in ${\cal M}$ theory at ${\cal O}(R^4)$}.
\end{itemize}

\begin{itemize}
\item
Upon equating the ${\cal O}(\beta^0)$ terms for black-hole and thermal backgrounds, one obtains: 
\begin{equation}
\label{r_h-r_0-relation}
r_h = \frac{\sqrt[4]{\frac{\kappa_{\rm GHY}^{{\rm th},\ \beta^0}}{\kappa_{\rm GHY}^{{\rm bh},\ \beta^0}}} {r_0} {{\cal R}_{D5/\overline{D5}}^{\rm bh}} \sqrt[4]{\frac{\log
   \left(\frac{{{\cal R}_{\rm UV}}}{{{\cal R}_{D5/\overline{D5}}^{\rm th}}}\right)}{\log
   \left(\frac{{{\cal R}_{\rm UV}}}{{{\cal R}_{D5/\overline{D5}}^{\rm bh}}}\right)}}}{{{\cal R}_{D5/\overline{D5}}^{\rm th}}}\approx \frac{\sqrt[4]{\frac{\kappa_{\rm GHY}^{{\rm th},\ \beta^0}}{\kappa_{\rm GHY}^{{\rm bh},\ \beta^0}}} {r_0} {{\cal R}_{D5/\overline{D5}}^{\rm bh}}}{{{\cal R}_{D5/\overline{D5}}^{\rm th}}},
\end{equation} 
(as ${\cal R}_{\rm UV}\gg{\cal R}_{D5/\overline{D5}}^{\rm bh}$ with ${\cal R}_{D5/\overline{D5}}^{\rm thermal/bh}$ being the $D5/\overline{D5}$-brane separation for the parent thermal/black hole type IIB dual of thermal QCD-like theories) where $\kappa_{\rm GHY}^{{\rm th/bh},\ \beta^0}$ are numerical pre-factors appearing in the evaluation of the GHY surface terms for the thermal/black hole backgrounds at ${\cal O}(\beta^0)$.
\item  Identifying $\frac{r_0}{L^2}$ with $\frac{m^{0^{++}}}{4}$ \cite{Glueball-Roorkee}, where $m^{0^{++}}$ is the mass of $0^{++}$ glueball and using $T_c = \frac{r_h}{\pi L^2}$, the deconfinement temperature is given by the following expression:
\begin{equation}
\label{rh-r0-beta0}
T_c = \frac{\sqrt[4]{\frac{\kappa_{\rm GHY}^{{\rm th},\ \beta^0}}{\kappa_{\rm GHY}^{{\rm bh},\ \beta^0}}} {m^{0^{++}}} {{\cal R}_{D5/\overline{D5}}^{\rm bh}}}{4 \pi {{\cal R}_{D5/\overline{D5}}^{\rm th}}}.
\end{equation}
\end{itemize}


\begin{itemize}

\item Now equating $O(\beta)$ term for blackhole and thermal background obtains:
\begin{eqnarray}
\label{rh-r0-beta-3}
& & \hskip -0.5in f_{x^{10}x^{10}}({r_0})
\nonumber\\
& & \hskip -0.7in \sim -\frac{b^2 {g_s}^3 M^2 {N_f}^{14/3}  \left(\frac{r_h}{{\cal R}_{D5/\overline{D5}}^{\rm bh}}\right)^4  \log ^{{3}}(N) \log
   \left(\frac{{r_0}}{{\cal R}_{D5/\overline{D5}}^{\rm th}}\right) \log \left(\frac{{r_h}}{{\cal R}_{D5/\overline{D5}}^{\rm bh}}\right) \log
   \left(1-\frac{{r_h}}{{\cal R}_{D5/\overline{D5}}^{\rm bh}}\right)}{  {\kappa_{\rm EH, th}^{\beta,\ \rm IR}} {N} \left(\frac{r_0}{{\cal R}_{D5/\overline{D5}}^{\rm th}}\right)^3}\nonumber\\
& & \hskip -0.5in \times \left(-11  {\cal C}_{\theta_1x} {\kappa_{\sqrt{G^{(1)}}R^{(0)}}^{\rm IR}}
    \log ^3(N)-10 {\kappa_{\rm EH,bh}^{\beta,\ \rm IR}} \left(-{\cal C}_{zz}^{\rm bh} + 2 {\cal C}_{\theta_1z}^{\rm bh} - 3 {\cal C}_{\theta_1x}^{\rm bh}\right)\right),
\end{eqnarray}
with the understanding that one substitutes $r_h$ in terms of $r_0$ as obtained in (\ref{r_h-r_0-relation}). Above equation relates ${\cal O}(R^4)$ corrections to the thermal background along M theory circle and combination of integrations constant appearing in the ${\cal O}(R^4)$ corrections to the black hole background along the compact part of non-compact four cycle in type IIB setup around which flavor branes are wrapping. This relation is valid in the IR. Which is a version of {\it ``UV-IR mixing"}. 
 \item The aforementioned combination of integrations constant appearing in the ${\cal O}(R^4)$ corrections to the black hole background encodes information about the flavor branes in parent type IIB dual. We refer this as {\it `` Flavor Memory"} effect in the context of M theory dual.

\end{itemize}


\subsection{$T_c$ From Entanglement Entropy}

\begin{itemize}

\item In this section we will discuss confinement-deconfinement phase transition in QCD$_{2+1}$-like theory from entanglement entropy point of view based on \cite{Tc-EE}. We will calculate entanglement entropy for "connected" and "disconnected" regions - suitably defined. We will show that at a critical value of ${\it l}$ of a spatial interval which we are denoting here by ${\it l_{crit}}$, phase transition will occur from confined phase to the deconfined phase.  Contact with results of the previous section is made by looking at the $M_{KK}\rightarrow0$- or equivalently $r_0\rightarrow0$-limit, i.e., the 4D-limit of results of this section.
\par
Authors in \cite{Tc-EE} have investigated entanglement entropy between an interval and its complement in the gravity dual of large $N_c$ gauge theories and they found that there are two RT surfaces - disconnected and connected. Below a critical value of ${\it  l}$ connected surface dominates while above that critical value of ${\it  l}$ disconnected surface dominates. This is analogous to finite temperature deconfinement transition in dual theories. \par
For $AdS_{d+2}/CFT_{d+1}$ correspondence, quantum entanglement entropy between the regions $A \equiv \mathbb{R}^{d-1}\times l$ and $B \equiv \mathbb{R}^{d-1}\times(\mathbb{R}-l)$, $l$ being an interval of length $l$, is given by the following expression,
\begin{equation}
\label{EE-def}
S_A=\frac{1}{4 G_N^{(d+2)}} \int_{\gamma} d^d\sigma\sqrt{G_{ind}^{(d)}} .
\end{equation}
where, $G_N^{(d+2)}$ is Newton constant in $(d+2)$ dimensions and $G_{ind}^{(d)}$ is the determinant of the induced string frame metric on co-dimension 2 minimal surface $\gamma$. Equation $(\ref{EE-def})$ can be generalised to non-conformal theories as below \cite{Tc-EE}:
\begin{equation}
\label{EE-def-NCT}
S_A=\frac{1}{4 G_N^{(d+2)}} \int_{\gamma} d^d\sigma e^{-2\phi}\sqrt{G_{ind}^{(d)}},
\end{equation}
where $\phi$ is the dilaton profile. In an ${\mathscr {M}}$-theory dual since there is no dilaton, therefore equation (\ref{EE-def-NCT}) will be assumed to be replaced by the following expression,
\begin{equation}
\label{EE-def-M Theory}
S_A=\frac{1}{4 G_N^{(11)}} \int_{\gamma} d^9\sigma \sqrt{G_{ind}^{(9)}} .
\end{equation}  
 Let us define two regions $A$ and $B$ in two dimensions as below,
\begin{eqnarray}
&&
A={\rm I\!R}\times l , \\
&& B={\rm I\!R} \times ({\rm I\!R}-l).
\end{eqnarray}
Now, we are going to calculate enganglement entropy between  $A$ and $B$ for the metric $(\ref{metric})$ using equation $(\ref{EE-def-M Theory})$.

\item Entanglement entropies for the connected and disconnected \cite{Ryu-Takayanagi (RT) surfaces} are ($r^*$ is the turning point of the RT surface):
\begin{eqnarray*}
\label{final-S-connected+S-disconnected}
& & \frac{S_{\rm disconnected}^{\rm UV-finite}}{{\cal V}_1} \sim  -{{g_s}^{2} M \sqrt[20]{\frac{1}{N}} {N_f}^{8/3} r_0^4 \log ^2(r_0) ({\log N}-3 \log
   (r_0))^{5/3}};\nonumber\\
& & \frac{S_{\rm connected}^{\rm UV-finite}}{{\cal V}_1}  \approx -{{g_s}^{2/3}{M_{\rm UV}} {N_f^{\rm UV}}^{4/3} {r^*}^4 \log ^{\frac{4}{3}}(N) \log ({r^*})\sqrt[20]{\frac{1}{N}}},\nonumber
\end{eqnarray*}

\end{itemize}

\begin{itemize}
\item
\begin{eqnarray*}
\frac{S_{\rm (dis)connected}^{{\rm UV-finite},\ \beta}}{{\cal V}_1}\propto {\cal C}_{zz}^{\rm th} - 2 {\cal C}_{\theta_1z}^{\rm th};
\end{eqnarray*}
\item
Now, ${\cal C}_{zz}^{\rm th} - 2 {\cal C}_{\theta_1z}^{\rm th}=0$ \cite{Vikas+Gopal+Aalok}.

\item
Therefore, no contribution to the entanglement entropy for (dis)connected surface at ${\cal O}(R^4)$ supporting the previously conjectured non-renormalization of $T_c$ at ${\cal O}(R^4)$.
\end{itemize}

\begin{itemize}

\item At $r^* = r_{\rm criticial}$, $S_{\rm connected}^{\rm UV-Finite} = S_{\rm disconnected}^{\rm UV-Finite}$ and $r_{\rm critical}$ is given by:
\begin{equation*}
\label{r-critical}
r_{\rm critical} \sim \frac{\sqrt[3]{{g_s}} M^{1/4} {N_f}^{2/3} \log ^{\frac{7}{6}}(N)}{N^{3/20}} r_0. \nonumber
\end{equation*}
\item For $N=100, M=N_f=3, g_s=0.1-1$, $r_{\rm criticial} \sim 3.8 r_0$ and $l(r=r_{\rm criticial}) \sim 5.4$.

\item For $N=100, M_{\rm UV} = N_f^{\rm UV} = 0.01, M=N_f=3$, and $r_0=N^{-\frac{f_{r_0}}{3}}, f_{r_0}\approx 1$ (\cite{Vikas+Gopal+Aalok}), we obtained the following plot for the entanglement entropies for the connected and disconnected surface versus $l$. This graph depicts that $l<l_{\rm crit}$ and $l>l_{\rm crit}$ correspond to confined and deconfined phases of thermal QCD-like theories:
\begin{figure}
\begin{center}
\includegraphics[width=0.60\textwidth]{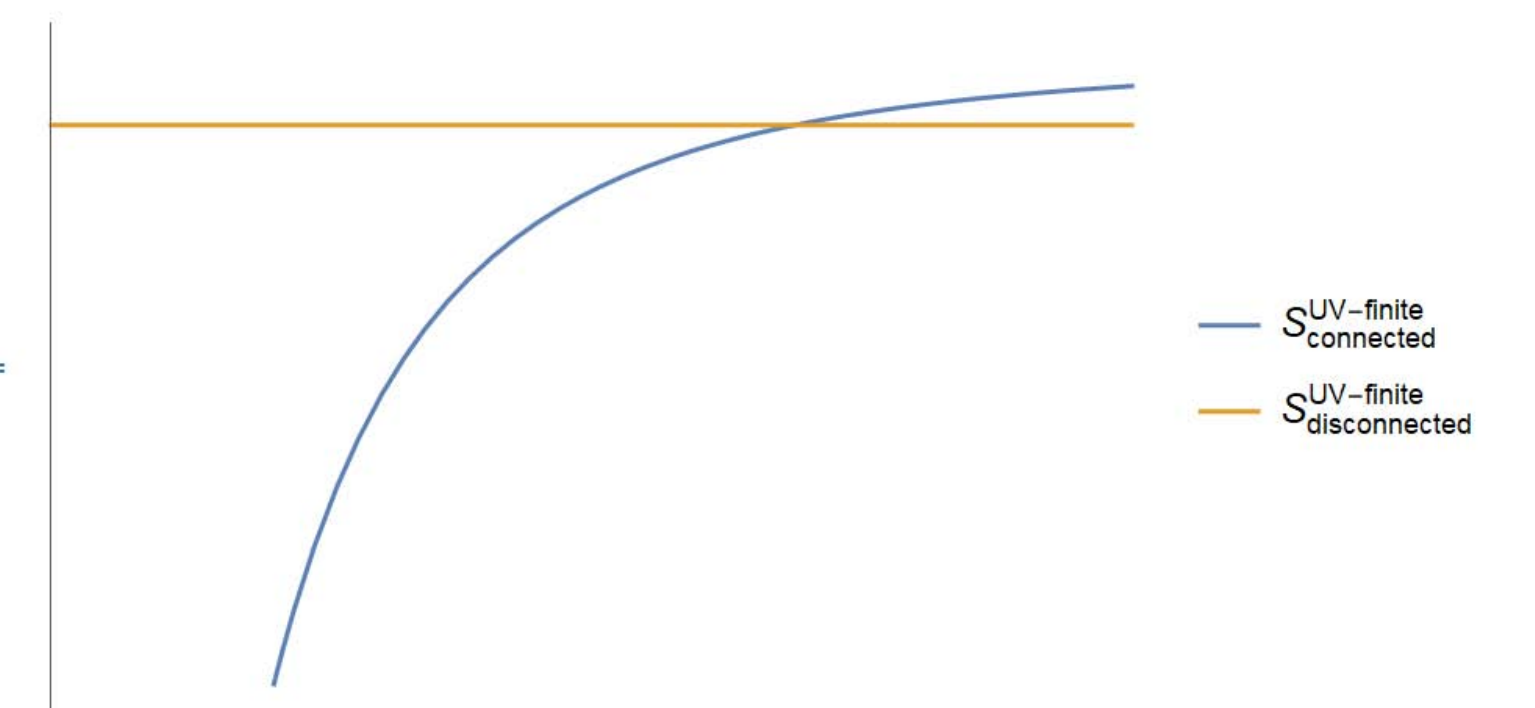}
\end{center}
\caption{\scriptsize{Plot $S_{\rm connected}^{\rm UV-finite}$ (blue) and $S_{\rm disconnected}^{\rm UV-finite}$(orange) versus $l(r^*)$}}
\label{S-cs-l}
\end{figure}

\end{itemize}

\section{Details of Sec. \ref{Eling+Oz} - Regularization and Principal Values of Angular ($\theta_{1,2}$) Integrals}
\label{details of Sec 5}
\setcounter{equation}{0}
\seceqcc

\begin{itemize}
\item
The $B_{\rm NS-NS}^{\rm IIA}$ components close to (\ref{alpha_theta_12}), are given below:
\begin{eqnarray}
\label{BNSNSIIA}
& & B^{\rm IIA}_{\theta_1\theta_2} = -\frac{M \sqrt{N} {N_f} g_s^2 \log (r) \left(36 a^2 \log (r)+r\right)}{4 \pi  r \alpha _{\theta _1} \alpha _{\theta
   _2}};\ B^{\rm IIA}_{\theta_1x} = -\frac{3\ 3^{2/3} M N^{7/20} {N_f} g_s^{7/4} \log (r)}{\sqrt{2} \pi ^{5/4} \alpha _{\theta _1}^3};\nonumber\\
& & B^{\rm IIA}_{\theta_2x} = -\frac{\sqrt{3} \sqrt[4]{\pi } \mu_{\rm def}^2 \sqrt[4]{N} \sqrt[4]{g_s}}{r^3};\  B^{\rm IIA}_{\theta_1y} = \frac{\sqrt[4]{\pi } N^{3/20} \alpha _{\theta _2} \sqrt[4]{g_s}}{\sqrt{3} \alpha _{\theta _1}};\nonumber\\
& & B^{\rm IIA}_{xy} = -\frac{2 \sqrt{\frac{2}{3}} \left(3^{2/3}-3\right) \sqrt[10]{N} \alpha _{\theta _2}}{\alpha _{\theta _1}^2};\  B^{\rm IIA}_{\theta_1z} = \frac{9\ 3^{2/3} M \sqrt[20]{\frac{1}{N}} {N_f} g_s^{7/4} \log (r)}{4 \sqrt{2} \pi ^{5/4} \alpha _{\theta _1}};\nonumber\\
& & B^{\rm IIA}_{\theta_2z} = -\frac{9\ 3^{2/3} {g_s}^{7/4} M \left(\frac{1}{N}\right)^{7/20} {N_f} \alpha _{\theta _1}^2 \log (r)}{128
   \sqrt{2} \pi ^{5/4} \alpha _{\theta _2}};\  B^{\rm IIA}_{xz} = -\frac{2 N^{2/5}}{3 \alpha _{\theta _1}^2};\nonumber\\
& & B^{\rm IIA}_{yz} = -\frac{3 \sqrt[6]{3} \left(\frac{1}{N}\right)^{3/10} \alpha _{\theta _2}}{\sqrt{2}};\  B^{\rm IIA}_{xy} = -\frac{2 \sqrt{\frac{2}{3}} \left(3^{2/3}-3\right) \sqrt[10]{N} \alpha _{\theta _2}}{\alpha _{\theta _1}^2}.     
\end{eqnarray}

\item
Noting that:
\begin{itemize}
\item
\begin{eqnarray}
\label{sqrtdetg6}
& & \sqrt{g_6} = -\frac{M \left(r^2-3 a^2\right)
   N_f^2 g_s^2 \csc ^2\left(\theta _2\right) \csc ^3\left(\theta _1\right) \log
   ^2(r) (3 \log (r)-\log (N))}{6 \sqrt{6} \pi ^2 r^2}
\nonumber\\
& &  -\frac{\beta  M \left(r^2-3 a^2\right) ({C_{zz}}-2 {C_{\theta_1z}}) N_f^2
   g_s^2 \csc ^2\left(\theta _2\right) \csc ^3\left(\theta _1\right) \log ^2(r)
   (\log (N)-3 \log (r))}{12 \sqrt{6} \pi ^2 r^2},\nonumber\\
   & & 
\end{eqnarray}
\item the LO-in-$N$ terms in the product of the inverse metric components as given below:
{\footnotesize
\begin{eqnarray}
\label{metric-inverses-drBtheta1theta2}
& & \hskip -1.5in g^{rr} g^{x^{10}x^{10}} \left[
     g^{\theta_1\theta_1} g^{\theta_2\theta_2} - (g^{\theta_1\theta_2})^2\right] 
 \sim \frac{\sqrt[6]{\pi } N^{3/2} g_s^{3/2} \csc ^2\left(\theta
   _1\right) \left(49 \sin ^2\left(\theta _2\right) \csc
   ^2\left(\theta _1\right)-81\right) }{3188646\ 6^{2/3} M^4 \log
   ^4\left(r_h\right) \left(c_2 \log
   \left(r_h\right)+c_1\right){}^2 \left(N_f \left(\log
   (N)-3 \log \left(r_h\right)\right)\right){}^{2/3}} \nonumber\\
& & \hskip -1.5in \times \Biggl(81 \sqrt[3]{3}
   \sin ^4\left(\theta _2\right) \csc ^4\left(\theta
   _1\right) \left(-729 \sqrt{6} \sin \left(\theta
   _1\right)+243 \sqrt{6} \sin \left(3 \theta _1\right)+162
   \cos \left(2 \theta _1\right)+82 \cos \left(2 \theta
   _2\right)-244\right)\nonumber\\
& & \hskip -1.5in -2 \left(49 \sin ^4\left(\theta
   _2\right) \csc ^4\left(\theta _1\right)-49 \sin
   ^2\left(\theta _2\right) \csc ^2\left(\theta
   _1\right)+81\right)\Biggr)+ \frac{\sqrt[3]{2} \sqrt[6]{\pi } \beta  N^{3/2} g_s^{7/2}
   \sin ^4\left(\theta _2\right)
  }{19683\ 3^{2/3}
   {g_s}^{13/2} M^4 \log ^4({r_h})
   (c_1+c_2 \log ({r_h}))^2 ({N_f}
   (\log N -3 \log ({r_h})))^{2/3}}\beta\nonumber\\
& & \hskip -1.5in \times  \left(({\cal C}_{\theta_1z}-{\cal C}_{yz}) \sqrt{g_s} \csc
   ^2\left(\theta _2\right) \left(49 \csc ^4\left(\theta
   _1\right)-130 \csc ^2\left(\theta _2\right) \csc
   ^2\left(\theta _1\right)+81 \csc ^4\left(\theta
   _2\right)\right)-81 \sqrt[3]{3} {\cal C}_{\theta_1z}(1)
   \sqrt{{g_s}} \left(81 \cos \left(2 \theta
   _1\right)-49 \cos \left(2 \theta _2\right)-32\right)
   \csc ^6\left(\theta _1\right)\right),\nonumber\\
& & 
\end{eqnarray}
}
\end{itemize}
and using (\ref{xyz-defs}), one obtains:
{\footnotesize
\begin{eqnarray}
\label{PVtheta1}
& & \hskip -0.5in {\cal P}\left.\int_{S_{\rm squashed}^2\times_w S_{\rm squashed}^3}\sqrt{g_6}
 g^{rr} g^{x^{10}x^{10}}\left[
     g^{\theta_1\theta_1} g^{\theta_2\theta_2} - (g^{\theta_1\theta_2})^2\right]\left(\partial_rC_{\theta_1\theta_2x^{10}}\right)^2\right|_{r=r_h}^{\beta^0}
\nonumber\\
& & \hskip -0.5in \sim 2(2\pi)^3\frac{(g_s^{3/4} N^{3/4}M^2N_f^2 g_s^4}{16\pi^2r_h^2}\lim_{\epsilon_{1,2}\rightarrow0^+}\int_{\epsilon_2}^{\pi-\epsilon_2} d\theta_2\int_{\epsilon_1}^{\pi-\epsilon_1}d\theta_1\frac{1}{\sin\theta_1 \sin\theta_2}\sqrt{g_6}
 g^{rr} g^{x^{10}x^{10}}\left[
     g^{\theta_1\theta_1} g^{\theta_2\theta_2} - (g^{\theta_1\theta_2})^2\right]\nonumber\\
& & \hskip -0.5in =  - (2\pi)^3\frac{g_s^{3/4} N^{3/4}M^2N_f^2 g_s^4}{16\pi^2r_h^2}\int_{\epsilon_2}^{\pi-\epsilon_2} d\theta_2\int_{\epsilon_1}^{\pi-\epsilon_1}d\theta_1 \frac{c_2 \sqrt{N} {N_f}^2 g_s^4 \csc
   ^6\left(\theta _1\right) \csc ^3\left(\theta _2\right)
   \left(49 \sin ^2\left(\theta _2\right) \csc
   ^2\left(\theta _1\right)-81\right) \Omega_1}{6377292 \sqrt[6]{2}
   3^{2/3} \pi ^{11/6} {g_s}^{7/2} M
   (c_1+c_2 \log ({r_h}))^2 ({N_f}
   (\log (N)-3 \log ({r_h})))^{2/3}},\nonumber\\
& & 
\end{eqnarray}
where 
\begin{eqnarray}
\label{Omega1-def}
& & \hskip -0.8in \Omega_1 \equiv \Biggl(-98 \sin
   ^4\left(\theta _2\right) \csc ^4\left(\theta
   _1\right)+98 \sin ^2\left(\theta _2\right) \csc
   ^2\left(\theta _1\right)\nonumber\\
& & \hskip -0.8in +81 \sqrt[3]{3} \sin
   ^4\left(\theta _2\right) \csc ^4\left(\theta _1\right)
   \left(-729 \sqrt{6} \sin \left(\theta _1\right)+243
   \sqrt{6} \sin \left(3 \theta _1\right)+162 \cos \left(2
   \theta _1\right)+82 \cos \left(2 \theta
   _2\right)-244\right)-162\Biggr).\nonumber\\
& & 
\end{eqnarray}
}
One chooses the following regularization:
\begin{equation}
\label{regularization}
\epsilon_2 = \frac{5\times 10^{-15}}{\epsilon_1^4} - 2\times10^{-14},
\end{equation}
and one hence obtains the principal value of the double angular ($\theta_{1,2}$) integral (\ref{Principal-Obeta0}).

\item
Now, at ${\cal O}(\beta)$,
{\footnotesize
\begin{eqnarray}
\label{intbetatheta1theta2LON}
& & {\cal P}\left.\int_{S_{\rm squashed}^2\times_w S_{\rm squashed}^3}\sqrt{g_6}
 g^{rr} g^{x^{10}x^{10}}\left[
     g^{\theta_1\theta_1} g^{\theta_2\theta_2} - (g^{\theta_1\theta_2})^2\right]\left(\partial_rC_{\theta_1\theta_2x^{10}}\right)^2\right|_{r=r_h}^{\beta}\nonumber\\
& & \hskip -0.8in = - 2(2\pi)^3\frac{(g_s^{3/4} N^{3/4}M^2N_f^2 g_s^4}{16\pi^2r_h^2}\int_{\delta_2}^{\pi-\delta_2}d\theta_2\int_{\delta_1}^{\pi-\delta_1}d\theta_1
\Biggl[\frac{\beta  \sqrt{N} \sqrt{g_s} \sin\left(\theta
   _2\right) \csc ^4\left(\theta _1\right)
   \sqrt[3]{{N_f} (\log N -3 \log ({r_h}))}
   }{59049 \sqrt[6]{2} 3^{2/3} \pi ^{11/6}
   M {N_f} \log ^2({r_h}) ({c_1}+{c_2} \log
   ({r_h}))}\nonumber\\
& & \hskip -0.8in \times \Biggl(({\cal C}_{\theta_1z}-{\cal C}_{yz}) \csc
   ^2\left(\theta _2\right) \left(49 \csc ^4\left(\theta
   _1\right)-130 \csc ^2\left(\theta _2\right) \csc
   ^2\left(\theta _1\right)+81 \csc ^4\left(\theta
   _2\right)\right)\nonumber\\
& &\hskip -0.8in -81 \sqrt[3]{3} {\cal C}_{\theta_1z} \left(81
   \cos \left(2 \theta _1\right)-49 \cos \left(2 \theta
   _2\right)-32\right) \csc ^6\left(\theta
   _1\right)\Biggr)\left(\partial_rC_{\theta_1\theta_2x^{10}}(r)\right)^2\Biggr]^\beta\nonumber\\
& & \hskip -0.8in = -(2\pi)^310^{-10} \frac{\beta  {g_s}^{21/4} M N^{5/4} {N_f} \sqrt[3]{{N_f}
   ({\log N}-3 \log ({r_h}))} }{2.5 {r_h}^2
   \log^2({r_h}) ({c_1}+{c_2} \log ({r_h}))}\nonumber\\
& & \hskip -0.8in \times \frac{\left(-3640 {\delta_1}^4 {\delta_2}^2
   ({\cal C}_{\theta_1z}-{\cal C}_{yz})-1960 {\delta_1}^2 {\delta_2}^4
   ({\cal C}_{\theta_1z}-{\cal C}_{yz}) \log ({\delta_2})+9 {\delta_1}^6 (272
   {\cal C}_{\theta_1z}-210 {\cal C}_{yz})-237452 {\cal C}_{\theta_1z}
   {\delta_2}^4\right)}{ {\delta_1}^9 {\delta_2}^4}
\end{eqnarray}
}
The regularization of (\ref{intbetatheta1theta2LON}) is:
{\footnotesize
\begin{eqnarray}
\label{regularizationbeta}
& & \hskip -0.8in\frac{\left(-3640 {\delta_1}^4 {\delta_2}^2
   ({\cal C}_{\theta_1z}-{\cal C}_{yz})-1960 {\delta_1}^2 {\delta_2}^4
   ({\cal C}_{\theta_1z}-{\cal C}_{yz}) \log ({\delta_2})+9 {\delta_1}^6 (272
   {\cal C}_{\theta_1z}-210 {\cal C}_{yz})-237452 {\cal C}_{\theta_1z}
   {\delta_2}^4\right)}{ {\delta_1}^9 {\delta_2}^4} ={\cal C}_2,
\end{eqnarray}
}
which assuming ${\cal C}_{\theta_1z} = {\cal C}_{yz}$, implies:
\begin{eqnarray}
\label{delta2intermsofdelta1}
& & \hskip -0.8in \delta_2 = \frac{99.2\delta_1^{3/2}{\cal C}_{yz}^{1/4}}{\left(173559 {\cal C}_2\delta_1^9 + 4\times10^{10}{\cal C}_{yz}\right)^{1/4}
  } = 0.2\delta_1^{3/2} - \frac{2.3\times10^{-7}}{{\cal C}_{yz}}\delta_1^{21/2} \approx 0.2 \delta_1^{3/2}.
\end{eqnarray}

\item
Using
{\footnotesize
\begin{eqnarray}
\label{A+h-defs}
& & \hskip -0.8in e^{-\frac{2\phi^{\rm IIA}}{3}} = \frac{\left(\frac{3}{\pi }\right)^{2/3} \Sigma_2^{2/3}}{4
   \sqrt{\frac{64 \pi  a^2 {g_s} {N_f}
   \left(\frac{a}{{r_h}}-\frac{1}{\sqrt{3}}\right)}{\left(9 a^2+C_{\theta_1\theta_2 x^{10}}^2\right)
   \left( g_s\Sigma_2 \right)}+1}} -\frac{1}{32 \pi ^{7/3} \left(3 b^2-1\right)^5 \left(6
   b^2+1\right)^3 {g_s^2} {N_f} {r_h}^4 \alpha _{\theta _2}^3 \log
   ^4(N) \left(9 b^2 {r_h}^2+C_{\theta_1\theta_2 x^{10}}^2\right)^2 {\Sigma_2}^{4/3}}\nonumber\\
& & \hskip -0.8in \times\Biggl\{9 \sqrt[3]{3} b^{10} \left(9
   b^2+1\right)^3 \beta  M \left(\frac{1}{N}\right)^{5/4} r \Sigma_1 (C_{\theta_1\theta_2 x^{10}}-2 {r_h}) \log ^3({r_h})
   \Biggl(6 {g_s} {N_f} \log (N) \left(9 b^2 {r_h}^2+C_{\theta_1\theta_2 x^{10}}^2\right)+8 \pi
    \biggl[b^2 {r_h}^2 \left(27-8 \left(\sqrt{3}-3 b\right) {g_s}
   {N_f}\right)\nonumber\\
& & \hskip -0.8in+3 C_{\theta_1\theta_2 x^{10}}^2\biggr] -3 {g_s} {N_f} \left(9 b^2
   {r_h}^2+C_{\theta_1\theta_2 x^{10}}^2\right) \log \left(9 b^2C_{\theta_1\theta_2 x^{10}}^4
   {r_h}^2+C_{\theta_1\theta_2 x^{10}}^6\right)\Biggr)^2\Biggr\};\nonumber\\
& & \hskip -0.8in h ((\ref{h-def})\ {\rm near}\ (\ref{alpha_theta_12})) = \frac{2 \sqrt{\pi } \sqrt{{g_s}} \sqrt{N} \left(\frac{3 {g_s} M^2 \log (r) \left(-\frac{{g_s}
   {N_f} \log (N)}{8 \pi }+\frac{3 {g_s} {N_f} \left(\log (C_{\theta_1\theta_2 x^{10}})+\frac{1}{2}\right)}{2 \pi
   }+1\right)}{4 \pi  N}+1\right)}{C_{\theta_1\theta_2 x^{10}}^2},
\end{eqnarray}
}
where 
\begin{eqnarray}
\label{Sigma12-defs}
& &  \Sigma_1\equiv \left(19683 \sqrt{6} \alpha _{\theta _1}^6+6642 \alpha _{\theta _2}^2 \alpha _{\theta _1}^3-40 \sqrt{6}
   \alpha _{\theta _2}^4\right);\nonumber\\
& & \Sigma_2\equiv -{N_f} \log \left(9
   b^2 C_{\theta_1\theta_2 x^{10}}^4 {r_h}^2+C_{\theta_1\theta_2 x^{10}}^6\right)+\frac{8 \pi }{{g_s}}+2 {N_f} \log (N),
\end{eqnarray}
one obtains (\ref{A}).

One may replace $\frac{{\cal O}(10^{-13})\Sigma_1}{\kappa_{A_{\beta^0}'}^3 \epsilon ^4\alpha_{\theta _2}^3}$ in (\ref{zetaoveretabetaLON}) by:
{\footnotesize
\begin{eqnarray}
\label{Sigma1overalphatheta2cubed}
& & \frac{{\cal O}(10^{-13}){\cal P}\int_{S^2_{\rm squashed}\times_w S^3_{\rm squashed}}\sin\theta_1 \sin\theta_2 \Biggl\{ 48213.3 N^{21/10} \left(-\frac{\sin \left(\theta
   _2\right)}{500}+\sin ^6\left(\theta _1\right) \csc
   ^3\left(\theta _2\right)+0.14 \sin ^3\left(\theta
   _1\right) \csc \left(\theta _2\right)\right)\Biggr\}}{\kappa_{A_{\beta^0}'}^3 \epsilon ^4 }\nonumber\\
& & \sim \frac{{\cal O}(10^{-13})\lim_{\epsilon_\beta\rightarrow0^+} \int_0^\pi d\theta_1 \int_{\epsilon_\beta}^{\pi - \epsilon_\beta}d\theta_2 \sin\theta_1 \sin\theta_2
 \Biggl\{ 48213.3 N^{21/10} \left(-\frac{\sin \left(\theta
   _2\right)}{500}+\sin ^6\left(\theta _1\right) \csc
   ^3\left(\theta _2\right)+0.14 \sin ^3\left(\theta
   _1\right) \csc \left(\theta _2\right)\right)\Biggr\}}{\kappa_{A_{\beta^0}'}^3 \epsilon ^4}\nonumber\\
& & = \frac{\lim_{\epsilon_\beta\rightarrow0^+}\frac{N^{21/10}{\cal O}(10^{-13}) \left(88161.5 + 24275\epsilon_\beta\right)}{\epsilon_\beta}}{\kappa_{A_{\beta^0}'}^3 \epsilon ^4 }\nonumber\\
& & \hskip 0.4in\Downarrow N\sim100\nonumber\\
& & \frac{{{\cal O}(10^{-13})} \beta 
   \Sigma_1}{\kappa_{A_{\beta^0}'}^3 \epsilon ^4 
\alpha_{\theta _2}^3}\rightarrow \frac{10^{-7}}{\kappa_{A_{\beta^0}'}^3 \epsilon ^4 \epsilon_\beta}.
\end{eqnarray}
}
to obtain (\ref{PVSigma1overalphatheta2cubed-2}).
\end{itemize}

\section{Co-frames near $\psi=2n\pi,n=0, 1, 2$-patches}
\label{five-fold-coframes}
\setcounter{equation}{0}
\seceqdd

The coframes $e^{2, 3, 4}\in e^{a=1,...,7}$ diagonalizing $M_7=S^1_{\cal M}\times_w\left(S^1_{\rm thermal}\times_w M_5\right)$, $M_5$ being a non-Einsteinian generalization of $T^{1,1}$, near $\psi=2n\pi, n=0, 1, 2$-coordinate patches, are given by the following expressions:
{\scriptsize
\begin{eqnarray}
\label{e23}
& &\hskip -0.87in e^2 = \sqrt{ \kappa^2_{1; \beta^0}\csc ^2\left(\theta _2\right) + \kappa^2_{2, \beta^0}\frac{ r^6 \sin ^{61}\left(\theta _2\right) \sin ^{64}\left(\theta _1\right)}{{g_s}^6 \log N ^3 M^3
   {N_f}^3 \log ^3(r)}
+ \kappa^2_{1; \beta}\frac{ \sqrt[4]{\beta } \sqrt{C_{zz}^{(1)}} {\cal C}_q }{\sin ^{\frac{7}{2}}\left(\theta
   _2\right)}} \nonumber\\
& &\hskip -0.87in \times\Biggl[\frac{d\theta_1 \left(\kappa^2_{\theta_1, 1; \beta^0}\frac{  {g_s}^{7/4} \log N  M {N_f} \left(0.25 a^2-0.06 r^2\right) \sin \left(\theta _1\right) \csc
   \left(\theta _2\right) \log (r)}{r^2} + \kappa^2_{\theta_1; 2; \beta}\frac{\sqrt[4]{\beta }
   \sqrt{C_{zz}^{(1)}} {\cal C}_q  {g_s}^{7/4} \log N  M {N_f} \left({1.8}
   a^2+{2.5} r^2\right) \csc \left(\theta _2\right) \log (r)}{r^2 \sqrt{\sin \left(\theta
   _1\right)}}\right)}{\sqrt[4]{N}}\nonumber\\
& &\hskip -0.87in +d\theta_2 \left(\kappa^2_{\theta_2, 1; \beta^0}\frac{ {g_s}^{7/4} M {N_f} \sin ^2\left(\theta _1\right) \csc ^2\left(\theta _2\right) \log (r) \left(3 a^2 \log (r)+0.08
   r\right)}{\sqrt[4]{N} r} + \kappa^2_{\theta_2; 2; \beta}\frac{\sqrt[4]{\beta } \sqrt{C_{zz}^{(1)}}
   {\cal C}_q   {g_s}^{21/4} M^3 {N_f}^3 \sin ^{\frac{9}{2}}\left(\theta _1\right) \csc ^6\left(\theta _2\right) \log
   ^3(r) \left({a^2 r^2 \log (r)}+{r^3}\right)}{N^{3/4}
   r^3}\right)\nonumber\\
& &\hskip -0.87in +{dx} \left(\kappa^2_{x, 1; \beta^0} \sin ^2\left(\theta _1\right)
   \csc \left(\theta _2\right) + \kappa^2_{x; 2; \beta}\frac{ \sqrt[4]{\beta } \sqrt{C_{zz}^{(1)}} {\cal C}_q  {g_s}^{7/2} M^2
   {N_f}^2 \log ^2(r) \sin ^{\frac{9}{2}}\left({\theta _1}\right) \csc ^5\left({\theta _2}\right) \left({3.6} a^2 r \log (r)+{4.8} r^2\right)}{\sqrt{N} r^2}\right)\nonumber\\
& &\hskip -0.87in +{dy} \left(1-\kappa^2_{y, 1; \beta^0}\frac{ {g_s}^{7/2} M^2 {N_f}^2 \sin ^4\left(\theta _1\right) \csc
   ^4\left(\theta _2\right) \log ^2(r) \left(3 a^2 \log (r)+0.08 r\right)^2}{\sqrt{N} r^2}-\kappa^2_{y; 1; \beta}\frac{
   \sqrt[4]{\beta } \sqrt{C_{zz}^{(1)}} {\cal C}_q }{\sin ^{\frac{3}{2}}\left(\theta _1\right)}\right)\nonumber\\
& &\hskip -0.87in +{dz} \left( \kappa^2_{z, 1; \beta^0}\sin
   \left(\theta _2\right)-\kappa^2_{z; 1; \beta}\frac{ \sqrt[4]{\beta } \sqrt{C_{zz}^{(1)}} {\cal C}_q  \sin
   \left(\theta _2\right)}{\sin ^{\frac{3}{2}}\left(\theta _1\right)}\right)\Biggr]\nonumber\\
& &\hskip -0.87in e^3 = \sqrt{\kappa^3_{1; \beta^0} \csc ^2\left(\theta _2\right) + \kappa^3_{2; \beta^0}\frac{ r^6 \sin ^{61}\left(\theta _2\right) \sin ^{64}\left(\theta _1\right)}{{g_s}^6
   \log N ^3 M^3 N^{311/10} {N_f}^3 \log ^3(r)}-\kappa^3_{1; \beta}{\sqrt[4]{\beta } \sqrt{C_{zz}^{(1)}} {\cal C}_q  \sqrt{\sin \left(\theta _1\right)} \csc ^2\left(\theta
   _2\right)}}\nonumber\\
& &\hskip -0.87in \times \Biggl[\frac{d\theta_1 \left(\kappa^3_{\theta_1; 1; \beta^0}\frac{4 {g_s}^{7/4} \log N  M {N_f} \left(0.2 a^2-0.1 r^2\right) \sin \left(\theta
   _1\right) \csc \left(\theta _2\right) \log (r)}{N^{3/20} r^2} + \kappa^3_{\theta_1; 1; \beta}\frac{16.
   \sqrt[4]{\beta } \sqrt{C_{zz}^{(1)}} {\cal C}_q  {g_s}^{7/4} \log N  M {N_f} \left({1.56} a^2-{0.522} r^2\right) \csc \left(\theta _2\right) \log (r) \sin
   ^{\frac{3}{2}}({\theta_1})}{r^2}\right)}{\sqrt[4]{N}}\nonumber\\
& &\hskip -0.87in +\frac{d\theta_2 \left(\kappa^3_{\theta_2; 1; \beta^0}\frac{{g_s}^{7/4} M {N_f} \sin ^2\left(\theta _1\right) \csc ^2\left(\theta _2\right) \log (r) \left(0.0005
   a^2 \log (r)+0.000014 r\right)}{\sqrt[20]{N} r} + \kappa^3_{\theta_2; 1; \beta}\frac{16\sqrt[4]{\beta }
   \sqrt{C_{zz}^{(1)}} {\cal C}_q   {g_s}^{7/4} M {N_f} \sin ^{\frac{5}{2}}\left(\theta _1\right) \csc ^2\left(\theta
   _2\right) \log (r) \left({2.7} a^2 \log (r)+{0.081}
   r\right)}{r}\right)}{\sqrt[4]{N}}\nonumber\\
& &\hskip -0.87in +{dy} \left(1-\kappa^3_{y; 1; \beta^0}\frac{\frac{{g_s}^{7/2} M^2 {N_f}^2 \sin
   ^4\left(\theta _1\right) \csc ^4\left(\theta _2\right) \log ^2(r) \left(4.32 a^2 \log (r)+0.064 r\right)}{r}-\kappa^3_{y; 1; \beta}\frac{16. \sqrt[4]{\beta }
   \sqrt{C_{zz}^{(1)}} {\cal C}_q  {g_s}^{7/2} M^2 {N_f}^2 \sin ^{\frac{9}{2}}\left(\theta _1\right) \csc ^4\left(\theta
   _2\right) \log ^2(r) \left(-{7 a^2 \log (r)}-{0.1r}\right)}{\sqrt[5]{N}
   r}}{\sqrt{N}}\right)\nonumber\\
& &\hskip -0.87in +{dx} \left(\kappa^3_{x; 1; \beta^0} \sin ^2\left(\theta _1\right) \csc \left(\theta _2\right)-\kappa^3_{x; 1; \beta}
   \sqrt[4]{\beta } \sqrt{C_{zz}^{(1)}} {\cal C}_q  \sin ^{\frac{5}{2}}\left(\theta _1\right) \csc \left(\theta
   _2\right)\right) +{dz} \left(\kappa^3_{z; 1; \beta^0}\sin\theta_2 + {\kappa^3_{z; 1; \beta} \sqrt[4]{\beta } \sqrt{C_{zz}^{(1)}} {\cal C}_q
   \sqrt{\sin{\theta _1}}\sin{\theta _2}}
\right)\Biggr]\nonumber\\
\end{eqnarray}

\begin{eqnarray}
\label{e4}
& &\hskip -0.87in e^4 = \sqrt{\kappa^4_{1;\beta^0} \csc ^2\left(\theta _2\right) -\kappa^4_{2;\beta^0}\frac{ r^6 \sin ^{61}\left(\theta _2\right) \sin ^{64}\left(\theta _1\right)}{{g_s}^6 \log N ^3 M^3
   {N_f}^3 \log ^3(r)} -\kappa^4_{1;\beta}\frac{ \sqrt[4]{\beta } \sqrt{C_{zz}^{(1)}} {\cal C}_q }{\sin ^{\frac{7}{2}}\left(\theta
   _2\right)}}\nonumber\\
& &\hskip -0.87in \times \Biggl[d\theta_1 \sqrt[4]{N} \left(\kappa^4_{\theta_1; 1; \beta^0}\frac{\left(0.006 a^2 r^2-0.002
   r^4\right) \csc \left(\theta _1\right)}{{g_s}^{7/4} \log N  M {N_f} \left(0.02 a^4-0.01 a^2 r^2+0.002 r^4\right) \log
   (r)}+\kappa^4_{\theta_1; 1; \beta}\frac{\sqrt[4]{\beta } \sqrt{C_{zz}^{(1)}} {\cal C}_q   \csc ^3\left(\theta
   _2\right)}{{g_s}^{15/4} \log N ^3 M^2 {N_f}^2 \sqrt{\sin \left(\theta _1\right)} \log ^2(r)}\right)\nonumber\\
& &\hskip -0.87in +d\theta_2
  \left(\kappa^4_{\theta_2; 1; \beta^0}\frac{ \sqrt[4]{\beta } \sqrt{C_{zz}^{(1)}} {\cal C}_q  \sqrt[4]{N} \csc ^2\left(\theta
  _2\right)}{{g_s}^{15/4} \log N ^2 M^2 {N_f}^2 \sin ^{\frac{3}{2}}\left(\theta _1\right) \log ^2(r)}-\kappa^4_{\theta_2; 2; \beta^0}\frac{
 r^{10} \sin ^{\frac{127}{2}}\left(\theta _1\right) \sin ^{59}\left(\theta _2\right)}{\sqrt{C_{zz}^{(1)}} {\cal C}_q  {g_s}^{25/4}
 \log N ^3 M^3 \sqrt[4]{N} {N_f}^3 \left(0.022 a^4-0.015 a^2 r^2+0.002 r^4\right) \log ^3(r)}\right)\nonumber\\
& &\hskip -0.87in +{dx}
 \left(\kappa^4_{x; 1; \beta^0}\frac{\left(0.033 a^4-0.02 a^2 r^2+0.004 r^4\right) \sin ^2\left(\theta _1\right)}{0.022 a^4-0.015 a^2 r^2+0.002
 r^4}-\kappa^4_{x; 1; \beta}\frac{ \sqrt[4]{\beta } \sqrt{C_{zz}^{(1)}} {\cal C}_q  \sqrt{N}}{{g_s}^{7/2}
 \log N ^2 M^2 {N_f}^2 \sin ^{\frac{3}{2}}\left(\theta _1\right) \log ^2(r)}\right)\nonumber\\
& &\hskip -0.87in +{dy} \left(\kappa^4_{y; 1; \beta^0}\frac{\pi ^{5/2}  r^9
   \sin ^{57}\left(\theta _2\right) \sin ^{66}\left(\theta _1\right) \left(9.54 a^2 \log (r)+0.265
   r\right)}{{g_s}^8 \log N ^5 M^4 {N_f}^4 \left(0.022 a^4-0.015 a^2 r^2+0.002 r^4\right) \log ^4(r)}+\kappa^4_{y; 1; \beta}\frac{ \sqrt[4]{\beta } \sqrt{C_{zz}^{(1)}} {\cal C}_q   \sqrt{\sin \left(\theta _1\right)} \csc
   ^4\left(\theta _2\right)}{{g_s}^2 \log N ^2 M {N_f} \log (r)}\right)\nonumber\\
& &\hskip -0.87in +{dz} \left(1-\sqrt{N} \left(\kappa^4_{z; 1; \beta^0}\frac{ r^{20} \sin ^{118}\left(\theta _2\right) \sin ^{128}\left(\theta _1\right)}{{g_s}^{39/2} \log N ^{10} M^{10}
   {N_f}^{10} \left(0.022 a^4-0.015 a^2 r^2+0.002 r^4\right)^2 \log ^{10}(r)}+\kappa^4_{z; 1; \beta}\frac{
   \sqrt[4]{\beta } \sqrt{C_{zz}^{(1)}} {\cal C}_q  \sqrt{\sin \left(\theta _1\right)}}{{g_s}^{7/2} \log N ^2 M^2 {N_f}^2
   \log ^2(r)}\right)\right)\Biggr]\nonumber\\
& & \nonumber\\
\end{eqnarray}
}
In  (\ref{e23}) - (\ref{e4}),  ${\cal C}_{zz}^{(1)}$
is a constant of integration appearing in the solution of the ${\cal O}(R^4)$ correction $f_{zz}$ to the MQGP metric component $g_{zz}$, and ${\cal C}_q$ is a constant that appeared in the Jacobi nome $q = \sum_{j=0}^\infty q_j\left(\frac{L}{2}\right)^{4j+1}, q_j\in\mathbb{Z}^+,\ L\equiv  - 1 + {\cal C}_q\beta^{1/8}\left({\cal C}_{zz}^{(1)}\right)^{1/4}\sin^{1/4}\theta_1,\ {\cal C}_q\in\mathbb{C}$ being a constant,  relevant to the ``Kiepert algorithm'' \cite{Bruce-King-Beyond-Quartic} used to solve a quintic to effect the aforementioned diagonalization in \cite{OR4}. In (\ref{e23}) - (\ref{e4}), $\kappa^{a=1,...,6}_{\theta_{1,2}/x/y/z;1;\beta}\ll1$. Except for $e^4$, however, all the rest have an IR-enhancement factor involving some power of $\log r$ appearing in the contributions picked up from the ${\cal O}(R^4)$ terms. These contributions also receive near-Ouyang-embedding enhancements around small $\theta_{1,2}$ - which also provide the most dominant contributions to all the terms of the action.  Further, $\kappa^a_{1;\beta^0}\gg1$ but are accompanied by IR-suppression factors involving exponents of $r$ along with the aforementioned near-Ouyang-embedding enhancements around small $\theta_{1,2}$.

\end{document}